\documentclass[%
 aip,
 jmp,%
 amsmath,amssymb,
reprint,%
]{revtex4-1}

\usepackage{svg}
\svgsetup{
  inkscapeexe=inkscape,
  inkscapearea=page,
  inkscapelatex=false
}

\usepackage[colorlinks,linkcolor=blue,citecolor=blue,urlcolor=black,bookmarks=false,hypertexnames=true]{hyperref} 

\usepackage{amsmath}
\usepackage{xspace}
\usepackage{multirow}
\usepackage{latexsym}
\usepackage{braket}
\usepackage{enumerate}
\usepackage{mathrsfs}
\usepackage{enumitem}
\usepackage{color}
\usepackage[version=3]{mhchem}
\usepackage{caption,setspace}
\usepackage{graphicx}
\usepackage{subfigure}
\usepackage{threeparttable}
\usepackage{url}
\usepackage{adjustbox}
\usepackage{wrapfig}
\usepackage{array}  
\usepackage{float}
\usepackage{textcomp}
\usepackage{changepage}
\usepackage{natbib}
\usepackage{tabularx}
\usepackage{paracol}
\usepackage{cleveref}
\allowdisplaybreaks
\usepackage{array}
\newcolumntype{L}[1]{>{\raggedright\let\newline\\\arraybackslash\hspace{0pt}}m{#1}}
\newcolumntype{C}[1]{>{\centering\let\newline\\\arraybackslash\hspace{0pt}}m{#1}}
\newcolumntype{R}[1]{>{\raggedleft\let\newline\\\arraybackslash\hspace{0pt}}m{#1}}

\hypersetup{colorlinks,breaklinks,linkcolor=,urlcolor=blue,anchorcolor=blue,citecolor=blue}

\crefname{figure}{Figure}{Figures}
\crefname{table}{Table}{Tables}
\crefname{equation}{Eq.}{Eqs.}
\crefname{section}{Section}{Sections}

\newcommand*{\eh}{\ensuremath{E_\mathrm{h}}\xspace}

\begin{document}

\author{Nicholas~Yiching~Chiang}
\affiliation{Department of Chemistry and Biochemistry, The Ohio State University, Columbus, Ohio 43210, USA}

\author{Rajat~Majumder}
\affiliation{Department of Chemistry and Biochemistry, The Ohio State University, Columbus, Ohio 43210, USA}

\author{Alexander~Yu.~Sokolov}
\email{sokolov.8@osu.edu}
\affiliation{Department of Chemistry and Biochemistry, The Ohio State University, Columbus, Ohio 43210, USA}

\title{
\color{blue}
Molecular \textit{g}-Tensors From Spin--Orbit Quasidegenerate $N$-electron Valence Perturbation Theory: Benchmarks, Intruder-State Mitigation, and Practical Guidelines
\vspace{0.25cm}
}

\begin{abstract}
Accurate prediction of molecular $g$-tensors for open-shell systems requires a balanced treatment of multireference electron correlation and relativistic spin--orbit coupling. 
Here, we develop and benchmark spin--orbit quasidegenerate second-order $N$-electron valence perturbation theory (SO-QDNEVPT2) for $g$-tensor calculations, treating dynamical correlation and spin--orbit effects consistently within a multistate effective Hamiltonian framework. 
Two $g$-tensor approaches are implemented: a spin-free effective Hamiltonian (EH) approach based on second-order response and a Kramers (K) approach that extracts $g$ from spin-mixed SO-QDNEVPT2 states. 
We assess their performance on a benchmark set of 23 molecules spanning diatomics and small polyatomics, low- to high-spin species, and weak to strong spin--orbit coupling. 
Across the dataset, SO-QDNEVPT2 improves agreement with experiment relative to state-averaged complete active-space self-consistent field.
The EH and K formalisms agree for modest $g$-shifts but the Kramers approach becomes essential when the shifts become large. 
We demonstrate that QDNEVPT2 results can be sensitive to intruder-state instabilities that can be effectively mitigated with level-shift or renormalization techniques.
We then analyze the dependence of SO-QDNEVPT2 results on key computational parameters, including active space, number of states, state-averaging weights, gauge origin, and basis set. 
These results establish SO-QDNEVPT2 as a robust framework for computing $g$-tensors in correlated, relativistic open-shell molecules, offering practical guidelines for its applications.
\end{abstract}

\titlepage

\maketitle

\section{Introduction}
\label{sec:intro}

The electronic ${g}$-tensor quantifies the interaction between the spin magnetic moment of an unpaired electron and an external magnetic field. 
As a fundamental parameter in electron paramagnetic resonance (EPR) spectroscopy, it provides direct insight into the local electronic structure of open-shell molecules and materials.\cite{abragamElectronParamagneticResonance2012} 
Beyond spectral interpretation, ${g}$-tensors reveal information on molecular magnetic anisotropy, an essential property for characterizing single-molecule magnets and designing materials for quantum information storage and spintronic applications.\cite{wolfowiczQuantumGuidelinesSolidstate2021,brigantiMagneticAnisotropyTrends2021,gatteschiMolecularNanomagnets2006,kahnMolecularMagnetism1993,hendrickxSweetspotOperationGermanium2024} 
To extract this information, accurate theoretical predictions of ${g}$-tensors are indispensable for assigning experimental spectral features, establishing structure--property relationships that guide materials discovery, and delivering microscopic insight into anisotropy mechanisms and local electronic structure that is often inaccessible from measurements alone.\cite{galiRecentAdvancesInitio2023,rayTheoreticalInvestigationSingleMoleculeMagnet2023,ivadyFirstPrinciplesCalculation2018,guoDiscoveryDysprosiumMetallocene2022,zadroznySlowMagnetizationDynamics2013,bodensteinDevelopmentApplicationComplete2022,birnoschiRelativisticQuantumChemical2024,chengRelativisticExactTwocomponent2023}

Despite their importance, reliable theoretical calculations of ${g}$-tensors remain challenging. 
Two ingredients are essential: an accurate treatment of relativistic spin–orbit coupling\cite{truhlarIntroductionRelativisticElectronic2025,liuIdeasRelativisticQuantum2010} and a balanced description of electron correlation. 
Variational four-component and two-component relativistic methods\cite{ganyushinFullyVariationalSpinorbit2013,jenkinsVariationalRelativisticTwoComponent2019,jorgenaa.jensenRelativisticFourcomponentMulticonfigurational1996,reynoldsZeroFieldSplittingParameters2019,shiozakiRelativisticInternallyContracted2015,tangExactTwoComponentCompleteActive2024,hoyerCorrelatedDiracCoulomb2023,liuRelativisticCoupledclusterEquationofmotion2021,wuRelativisticCoreValence2025,huRelativisticTwoComponentMultireference2020,luExactTwoComponentRelativisticMultireference2022} provide a rigorous framework for incorporating spin–orbit effects, but their computational cost scales steeply when combined with high-level electron correlation treatments.
In contrast, perturbative approaches\cite{singhChallengesMultireferencePerturbation2018,tranDoublehybridDensityFunctional2020,chengPerturbativeTreatmentSpinorbit2014,chengPerturbativeTreatmentSpinorbitcoupling2018,roosRelativisticQuantumChemistry2004} for computing  ${g}$-tensors are far more affordable, yet their reliance on approximate relativistic corrections can limit accuracy, particularly for systems containing heavy elements. 
Striking an effective balance between accuracy and efficiency remains a central difficulty in ${g}$-tensor calculations.

To address this challenge, a state-interaction (SI) approach based on quasidegenerate perturbation theory has been developed, offering a favorable compromise between computational cost and predictive power.\cite{gerlochParamagneticPropertiesUnsymmetricaI1975,malmqvistCASSCFStateInteraction1989,malmqvistRestrictedActiveSpace2002,bolvinAlternativeApproachGMatrix2006,tatchenCalculatingElectronParamagnetic2009,neeseEfficientAccurateApproximations2005,alessioEquationofMotionCoupledClusterProtocol2021,alessioOriginMagneticAnisotropy2023} 
In this strategy, spin-pure scalar-relativistic wavefunctions are computed first, and spin–orbit coupling is subsequently introduced by diagonalizing an effective two-component Hamiltonian in the spin-pure basis.
The SI approach has been combined with a variety of electronic structure methods ranging from density functional theory (DFT)\cite{zhouCalculationZeemanEffect2021,cebreiro-gallardoStateInteractionApproachGMatrix2025, jangidLinearizedPairDensityFunctional2026} and equation-of-motion coupled cluster (EOM-CC)\cite{kahlerStateInteractionApproachEvaluating2023}  to multireference second-order perturbation theories (CASPT2, NEVPT2)\cite{singhChallengesMultireferencePerturbation2018,bolvinAlternativeApproachGMatrix2006,vancoillieCalculationEPRTensors2007,vancoillieMulticonfigurationalTensorCalculations2008,singhChallengesMultireferencePerturbation2018,langSpindependentPropertiesFramework2019,fedorovSpinorbitMultireferenceMultistate2001} and density matrix renormalization group (DMRG).\cite{roemeltSpinOrbitCoupling2015,sayfutyarovaElectronParamagneticResonance2018} 
In particular, the SI-based CASPT2 and NEVPT2 methods have been widely used to compute ${g}$-tensors and simulate magnetic properties for a variety of open-shell molecules with complicated electronic structures.\cite{chibotaruInitioCalculationAnisotropic2012,vancoillieMultireferenceInitioCalculations2010,bokarevSpinDensityDistribution2014,badia-romanoFieldinducedInternalFe2013,charronUnravelingEffectsMagnetic2016,pokhilkoMulticonfigurationalElectronicStructure2025,lunghi2019,blackabyExtremeGTensorAnisotropy2022,roufRelativisticApproximationsParamagnetic2017,sharmaEPRENDORTheoretical2017} 

Despite its successes, the SI framework has significant limitations.
Although exact in the limit of full configuration interaction, it effectively treats spin--orbit coupling as the first-order perturbation and may require expressing the Hamiltonian in a large configuration space to obtain accurate results.
For example, a benchmark of SI-based NEVPT2 method revealed that it systematically overestimates experimental ${g}$-tensors,\cite{singhChallengesMultireferencePerturbation2018} highlighting the need for further methodological developments.

Recently, we developed a spin--orbit quasidegenerate second-order $n$-electron valence perturbation theory (SO-QDNEVPT2) that integrates dynamic electron correlation and spin--orbit coupling within a unified perturbative framework.\cite{majumderSimulatingSpinOrbit2023,majumderConsistentSecondOrderTreatment2024}
In contrast to conventional SI approaches, SO-QDNEVPT2 treats both effects on equal footing, ensuring that their interplay is captured consistently through the second order of multireference perturbation theory. 
When implemented with exact two-component Douglas--Kroll--Hess spin--orbit Hamiltonians (sf-X2C-1e+so-DKH),\cite{liuExactTwocomponentHamiltonians2009,liSpinSeparationAlgebraic2012} SO-QDNEVPT2 shows high accuracy in zero-field splitting for molecules with up to sixth-row elements ($\lesssim$ 2.5 \% error)\cite{majumderConsistentSecondOrderTreatment2024}, while retaining the favorable computational scaling of standard nonrelativistic QDNEVPT2.\cite{angeliQuasidegenerateFormulationSecond2004,angeliNewPerspectivesMultireference2007}

In this work, we extend SO-QDNEVPT2 to the calculation of molecular ${g}$-tensors and present the first comprehensive benchmark of its performance. 
We examine two different strategies for extracting $g$-values from SO-QDNEVPT2 and evaluate how the results depend on computational parameters, including active space selection, the number of states incorporated in the quasidegenerate manifold, state-averaging weights in the CASSCF reference, gauge origin, and basis set.
We also demonstrate that the QDNEVPT2 and SO-QDNEVPT2 calculations may be sensitive to intruder-state problems, which have been largely overlooked in prior studies, and introduce practical approaches for mitigating these instabilities.
Together, this work provides a rigorous evaluation of SO-QDNEVPT2 for ${g}$-tensor calculations and establish clear guidelines for its reliable application in studies of molecular magnetism.

\section{Theory}
\label{sec:theory}

\subsection{Quasidegenerate second-order $N$-electron valence perturbation theory}
\label{sec:qdnevpt2}

We begin by introducing second-order quasidegenerate $N$-electron valence perturbation theory\cite{angeliQuasidegenerateFormulationSecond2004} (QDNEVPT2) for a nonrelativistic Born--Oppenheimer (BO) molecular Hamiltonian ($\hat{H}_{\mathrm{BO}}$).
QDNEVPT2 is an efficient and accurate approach to capture static and dynamic electron correlation in multiple electronic states represented by complete active-space self-consistent field (CASSCF)\cite{hinzeMCSCFMulticonfigurationSelfconsistentfield1973,roosCompleteActiveSpace1980,siegbahnCompleteActiveSpace1981,wernerQuadraticallyConvergentMCSCF1981,wernerSecondOrderMulticonfiguration1985} reference wavefunctions $\ket{\Psi^{(0)}_{I}}$.
Here and throughout, the index $I$ labels the orthonormal CASSCF model states included in the multistate treatment.
Although any orthogonal set of $\ket{\Psi^{(0)}_{I}}$ can be used to perform QDNEVPT2 calculations, it is common to compute them using the state-averaged CASSCF algorithm (SA-CASSCF),\cite{wernerQuadraticallyConvergentMCSCF1981,wernerSecondOrderMulticonfiguration1985} 
where molecular orbitals are optimized for a weighted average of the selected CASSCF state energies.

A distinctive feature of QDNEVPT2\cite{angeliQuasidegenerateFormulationSecond2004} and its single-state NEVPT2 variant\cite{angeliIntroductionElectronValence2001,angeliElectronValenceState2002} is the choice of the zeroth-order Hamiltonian 
\begin{align}
	\hat{H}^{(0)}  = C + \sum_{i}  \epsilon_{i} a_{i}^{\dagger} a_{i} +  \sum_{a}  \epsilon_{a} a_{a}^{\dagger} a_{a} + \hat{H}_{\mathrm{act}} ,
	\label{eq:dyallH0}
\end{align}
known as the Dyall Hamiltonian,\cite{dyallChoiceZerothorderHamiltonian1995,sokolovMultireferencePerturbationTheories2024} where $(i,j,k,l)$, $(w,x,y,z)$, and $(a,b,c,d)$ denote the core, active, and virtual spin-orbitals, respectively, and $a_{p}^{\dagger}$ ($a_p$) creates (annihilates) an electron in spin-orbital $p$.
In all reference CASSCF states $\ket{\Psi^{(0)}_{I}}$, the core orbitals are doubly occupied, the active orbitals can be partially occupied, and the virtual orbitals are empty. 
The quantities $\epsilon_i$ and $\epsilon_a$ are the core and virtual orbital energies computed as the eigenvalues of one-electron generalized Fock matrix.\cite{dyallChoiceZerothorderHamiltonian1995,sokolovMultireferencePerturbationTheories2024}
The $\hat{H}_{\mathrm{act}}$ operator contains all active-space contributions of the BO Hamiltonian $\hat{H}_{\mathrm{BO}}$, i.e., all one- and two-electron terms with indices restricted to $(w,x,y,z)$.
Finally, $C$ collects state-independent constants associated with the core orbitals.

Starting from the Dyall zeroth-order Hamiltonian $\hat{H}^{(0)}$ and the perturbation operator $\hat{V}=\hat{H}_{\mathrm{BO}}-\hat{H}^{(0)}$, each electronic state $I$ is represented by a perturbation expansion, $\ket{\Psi_{I}} = \ket{\Psi^{(0)}_{I}} + \ket{\Psi^{(1)}_{I}} + \ket{\Psi^{(2)}_{I}} + \ldots $, where the zeroth-order term $\ket{\Psi^{(0)}_{I}}$ is the CASSCF model state that captures static correlation within the active space, and the higher-order corrections $\ket{\Psi^{(k)}_{I}}$ ($k>0$) account for dynamic correlation by incorporating excitations that involve non-active spin-orbitals.
Only the first-order wavefunctions $\ket{\Psi^{(1)}_{I}}$ need to be evaluated to compute the QDNEVPT2 electronic energies.
To reduce computational cost, $\ket{\Psi^{(1)}_{I}}$ are expanded in a basis of internally contracted wavefunctions $\ket{\Phi_{\mu,I}}$ called perturbers\cite{angeliIntroductionElectronValence2001,angeliElectronValenceState2002}
\begin{align}
	\ket{\Psi^{(1)}_{I}} \approx \sum_{\mu} t^{(1)}_{\mu,I}\,\hat{\tau}_{{\mu}} \ket{\Psi^{(0)}_{I}}
	\equiv \sum_{\mu} t^{(1)}_{\mu,I}\,\ket{\Phi_{\mu,I}} ,
	\label{eq:psi1}
\end{align}
where $\hat{\tau}_{{\mu}}$ denotes a two-electron excitation operator that generates configurations outside the CASSCF model space and $t^{(1)}_{\mu,I}$ are the associated state-specific first-order amplitudes.

The perturbers $\ket{\Phi_{\mu,I}}$ are partitioned into eight excitation classes according to the number of electrons added to or removed from the active space.
Two internal contraction approximations are commonly used to compute $t^{(1)}_{\mu,I}$:
(i) \emph{strong contraction} (SC), in which a single contracted perturber is retained for each excitation class,\cite{angeliIntroductionElectronValence2001,angeliElectronValenceState2002,angeliThirdorderMultireferencePerturbation2006}
and (ii) \emph{full internal contraction} (FIC, also often referred to as partial contraction), in which multiple perturbers are retained within each class.\cite{kempferEfficientImplementationApproximate2025,angeliQuasidegenerateFormulationSecond2004,parkAnalyticalGradientTheory2020,sharmaQuasidegeneratePerturbationTheory2016}
Although commonly employed in QDNEVPT2 implementations, the SC approach has been shown to exhibit orbital invariance problems leading to instabilities with evaluating molecular response properties.\cite{sivalingamComparisonFullyInternally2016,guoSparseMapsSystematicInfrastructure2016,parkAnalyticalGradientTheory2019}
In this work, we employ the orbitally invariant FIC scheme, which avoids response instabilities and has a small effect on the resulting correlation energies (relative to fully uncontracted QDNEVPT2) while keeping the computational cost tractable.\cite{sokolovTimedependentFormulationMultireference2016,sokolovTimedependentNelectronValence2017}

To account for the interactions between nearly degenerate model states upon including dynamic correlation effects, QDNEVPT2 constructs and diagonalizes an effective Hamiltonian:
\begin{align}
	\rm \mathbf{H}_{eff}\mathbf{Y}=\mathbf{Y}\mathbf{E},
	\label{eq:Heffdiag}
\end{align}
where the columns of $\mathbf{Y}$ are the eigenvectors (state-mixing coefficients) and $\mathbf{E}$ is a diagonal matrix of QDNEVPT2 energies.
In the Hermitian QDNEVPT2 formulation,\cite{sharmaQuasidegeneratePerturbationTheory2016,langCombinationMultipartitioningHamiltonian2020,shavittQuasidegeneratePerturbationTheories1980,certainNewPartitioningPerturbation1970,kirtmanSimultaneousCalculationSeveral1981} 
the matrix elements of $\rm \mathbf{H}_{eff}$ are
\begin{align}
	\braket{\Psi^{(0)}_{I}| \hat{H}_{eff} |\Psi^{(0)}_{J} }
	&= E^{(0)}_{I} \delta_{IJ}
	+ \braket{\Psi^{(0)}_{I}| \hat{V} |\Psi^{(0)}_{J} }
	\nonumber\\
	&\quad + \frac{1}{2} \braket{\Psi^{(0)}_{I}| \hat{V} |\Psi^{(1)}_{J} }
	+ \frac{1}{2} \braket{\Psi^{(1)}_{I}| \hat{V} |\Psi^{(0)}_{J} } ,
	\label{eq:QDNEVPT2_EH}
\end{align}
where $E^{(0)}_{I}$ are the reference CASSCF energies for state $I$.
The diagonal elements of Eq.~\eqref{eq:QDNEVPT2_EH} recover the sum of CASSCF and NEVPT2 correlation energies for each state, whereas the off-diagonal elements describe coupling between different reference states induced by $\hat{V}$ through their first-order wavefunctions $\ket{\Psi^{(1)}_{I}}$.
Diagonalization of $\rm \mathbf{H}_{eff}$ therefore yields a consistent multistate description of dynamical correlation in quasidegenerate electronic states, known as the ``diagonalize--perturb--diagonalize'' strategy.\cite{zaitsevskiiMultipartitioningQuasidegeneratePerturbation1995,shavittQuasidegeneratePerturbationTheories1980} 

\subsection{Spin--orbit QDNEVPT2} 
\label{sec:so-qdnevpt2}

Although conventional QDNEVPT2 provides accurate and efficient treatment of electron correlation in open-shell and strongly correlated molecules, it neglects relativistic effects that are crucial for describing magnetic properties.
The spin--orbit QDNEVPT2 approach (SO-QDNEVPT2)\cite{majumderConsistentSecondOrderTreatment2024} addresses this limitation by efficiently capturing relativity and establishing an accurate multistate framework in which spin--orbit coupling and dynamic correlation are incorporated on equal footing.

SO-QDNEVPT2 is formulated in terms of a two-component relativistic Hamiltonian,
\begin{align}
	\label{eq:2c_hamiltonian}
	\hat{H}_{\mathrm{2c}} = \hat{H}_{\mathrm{SF}} + \hat{H}_{\mathrm{SO}},
\end{align}
obtained from an approximate transformation of the four-component Dirac--Coulomb--Breit Hamiltonian.\cite{pyykkoRelativisticEffectsChemistry2012,liuAdvancesRelativisticMolecular2014}
In \cref{eq:2c_hamiltonian}, the operator $\hat{H}_{\mathrm{SF}}$ incorporates the spin-free (scalar) relativistic effects in its one-electron contributions and has the same two-electron Coulomb repulsion term as the nonrelativistic BO Hamiltonian $\hat{H}_{\mathrm{BO}}$.
The remaining counterpart, $\hat{H}_{\mathrm{SO}}$, collects the spin-dependent relativistic contributions responsible for spin--orbit coupling.

The reference SA-CASSCF model states $\ket{\Psi^{(0)}_{I}}$ are obtained using $\hat{H}_{\mathrm{SF}}$, so that scalar relativistic effects are incorporated variationally in the zeroth-order description of electronic structure.
Consistently, the Dyall Hamiltonian $\hat{H}^{(0)}$ (\cref{eq:dyallH0}) is constructed using the one-electron terms from $\hat{H}_{\mathrm{SF}}$, ensuring that the SA-CASSCF model states remain its eigenfunctions: $\hat{H}^{(0)} \ket{\Psi^{(0)}_{I}} = E^{(0)}_{I} \ket{\Psi^{(0)}_{I}}$.
These definitions of zeroth-order Hamiltonian and reference wavefunctions result in a modified form of the perturbation operator
\begin{align}
	\label{eq:V_SO_def}
	\hat{V} = \hat{H}_{\mathrm{2c}} - \hat{H}^{(0)}
	= \hat{V}_{\mathrm{ee}} + \hat{H}_{\mathrm{SO}},
\end{align}
where $\hat{V}_{\mathrm{ee}} \equiv \hat{H}_{\mathrm{2c}} - \hat{H}^{(0)} - \hat{H}_{\mathrm{SO}}$ captures dynamic correlation stemming from nonrelativistic two-electron interactions involving non-active orbitals.
Importantly, the resulting perturbation expansions for the wavefunctions and energies treat dynamic correlation and spin--orbit coupling on equal footing, incorporating their contributions consistently at each order in perturbation theory.

The SO-QDNEVPT2 energies are computed by diagonalizing an effective Hamiltonian\cite{majumderConsistentSecondOrderTreatment2024}
\begin{align}
	\braket{\Psi^{(0)}_{I}| \hat{H}_{eff}^{\mathrm{SO}} |\Psi^{(0)}_{J} }
	&= E^{(0)}_{I} \delta_{IJ}
	+ \braket{\Psi^{(0)}_{I}| \hat{V}_{\mathrm{ee}}+\hat{H}_{\mathrm{SO}} |\Psi^{(0)}_{J} }
	\nonumber\\
	&\quad + \frac{1}{2} \braket{\Psi^{(0)}_{I}| \hat{V}_{\mathrm{ee}}+\hat{H}_{\mathrm{SO}} |\tilde{\Psi}^{(1)}_{J} }\nonumber\\
	&\quad+ \frac{1}{2} \braket{\tilde{\Psi}^{(1)}_{I}| \hat{V}_{\mathrm{ee}}+\hat{H}_{\mathrm{SO}} |\Psi^{(0)}_{J} } ,
	\label{eq:SO_QDNEVPT2_EH}
\end{align}
where $\ket{\tilde{\Psi}^{(1)}_{I}}$ denotes the first-order wavefunctions obtained using $\hat{V}$ defined in \cref{eq:V_SO_def},
\begin{align}
	\ket{\tilde{\Psi}^{(1)}_{I}} \approx \sum_{\mu} \tilde{t}^{(1)}_{\mu,I}\,\ket{\Phi_{\mu,I}} .
	\label{eq:psi1_tilde}
\end{align}
Compared to \cref{eq:QDNEVPT2_EH}, \cref{eq:SO_QDNEVPT2_EH} contains additional spin--orbit terms from $\hat{H}_{\mathrm{SO}}$ and uses the  $\ket{\tilde{\Psi}^{(1)}_{I}}$ wavefunctions that incorporate dynamic correlation and spin--orbit coupling in first-order perturbation on equal footing.
The resulting ``diagonalize--perturb--diagonalize'' scheme yields spin-mixed energies and wavefunctions that include spin--orbit coupling and dynamic correlation through second order.
In the limit of full configuration interaction, diagonalizing $\mathbf{H_{eff}^{SO}}$ (\cref{eq:SO_QDNEVPT2_EH}) delivers the exact energies and wavefunctions of two-component relativistic Hamiltonian $\hat{H}_{\mathrm{2c}}$ (\cref{eq:2c_hamiltonian}).

The SO-QDNEVPT2 framework is compatible with any two-component relativistic Hamiltonian and its physically motivated partitioning into $\hat{H}_{\mathrm{SF}}$ and $ \hat{H}_{\mathrm{SO}}$ (\cref{eq:2c_hamiltonian}).
In our previous work,\cite{majumderConsistentSecondOrderTreatment2024} we introduced two SO-QDNEVPT2 variants: BP2-QDNEVPT2 and DKH2-QDNEVPT2.
In both methods, $\hat{H}_{\mathrm{SF}}$ is constructed from the spin-free exact two-component one-electron transformation (sf-X2C-1e)\cite{liSpinSeparationAlgebraic2012}, providing an accurate and robust description of scalar relativistic effects.
To treat spin--orbit coupling in BP2-QDNEVPT2, $\hat{H}_{\mathrm{SO}}$ is taken from the Breit--Pauli Hamiltonian,\cite{breitDiracsEquationSpinSpin1932,bearparkSpinorbitInteractionsSelf1993,berningSpinorbitMatrixElements2000} which is attractive because of its compact form but becomes increasingly unreliable for molecules containing fourth-row and heavier elements.
In DKH2-QDNEVPT2, $\hat{H}_{\mathrm{SO}}$ is obtained using the exact two-component second-order Douglass--Kroll--Hess approach developed by Liu and co-workers (sf-X2C-1e+so-DKH2),\cite{liuExactTwocomponentHamiltonians2009,liSpinSeparationAlgebraic2012} which offers improved accuracy and robustness for compounds with heavier elements.\cite{majumderConsistentSecondOrderTreatment2024}
To reduce the computational cost associated with treating spin--orbit two-electron interactions, both BP2-QDNEVPT2 and DKH2-QDNEVPT2 employ the spin--orbit mean-field approximation (SOMF).\cite{hessMeanfieldSpinorbitMethod1996, berningSpinorbitMatrixElements2000}
The explicit working expressions for $\hat{H}_{\mathrm{SF}}$ and $\hat{H}_{\mathrm{SO}}$ used in BP2-QDNEVPT2 and DKH2-QDNEVPT2 are given in Ref.~\citenum{majumderConsistentSecondOrderTreatment2024}.

For comparison, we also report results obtained with conventional state-interaction variants of spin--orbit QDNEVPT2.\cite{majumderSimulatingSpinOrbit2023,majumderConsistentSecondOrderTreatment2024,langCombinationMultipartitioningHamiltonian2020,pokhilkoMulticonfigurationalElectronicStructure2025} 
In these methods, dynamic correlation is treated exactly as in conventional QDNEVPT2, and spin--orbit coupling is introduced only through first-order couplings between the SA-CASSCF reference states.
The corresponding effective Hamiltonian is therefore simplified to
\begin{align}
	\braket{\Psi^{(0)}_{I}|  \hat{H}_{eff}^{\mathrm{SO}} |\Psi^{(0)}_{J} }
	&= E^{(0)}_{I} \delta_{IJ}
	+ \braket{\Psi^{(0)}_{I}| \hat{V}_{\mathrm{ee}}+\hat{H}_{\mathrm{SO}} |\Psi^{(0)}_{J} }
	\nonumber\\
	&\quad + \frac{1}{2} \braket{\Psi^{(0)}_{I}| \hat{V}_{\mathrm{ee}} |\Psi^{(1)}_{J} }
	+ \frac{1}{2} \braket{\Psi^{(1)}_{I}| \hat{V}_{\mathrm{ee}} |\Psi^{(0)}_{J} } ,
	\label{eq:SO_QDNEVPT2_EH_firstorder}
\end{align}
where $\ket{\Psi^{(1)}_{I}}$ is the conventional QDNEVPT2 first-order wavefunction obtained from $\hat{V}_{\mathrm{ee}}$ alone (cf.\ \cref{eq:psi1}).
In Eq.~\eqref{eq:SO_QDNEVPT2_EH_firstorder}, the spin--orbit contribution enters exclusively through the reference-space matrix elements of $\hat{H}_{\mathrm{SO}}$, whereas the second-order terms provide only the dynamic correlation corrections generated by $\hat{V}_{\mathrm{ee}}$.
We consider two first-order (state-interaction) variants:
i) BP1-QDNEVPT2 that employs the Breit--Pauli form of $\hat{H}_{\mathrm{SO}}$ 
and ii) DKH1-QDNEVPT2 that uses $\hat{H}_{\mathrm{SO}}$ derived within the sf-X2C-1e+so-DKH1 approach.\cite{majumderSimulatingSpinOrbit2023,majumderConsistentSecondOrderTreatment2024}
To provide accurate description of scalar relativistic effects, both BP1-QDNEVPT2 and DKH1-QDNEVPT2 employ $\hat{H}_{\mathrm{SF}}$ derived from the sf-X2C-1e transformation.

\subsection{Calculating $g$-tensors using SO-QDNEVPT2}
\label{sec:g_tensor}

In this work, we extend SO-QDNEVPT2 to calculating ${g}$-tensors of open-shell molecules.
The ${g}$-tensor is a key parameter of the effective spin Hamiltonian used to describe the Zeeman response of a low-energy magnetic manifold.\cite{atanasovFirstPrinciplesApproach2015}
In the absence of additional interactions, the spin Hamiltonian can be expressed as
\begin{align}
	\hat{H}_{S} = \mu_{B}\mathbf{B}^\mathrm{T} \mathbf{g} \mathbf{\tilde{S}},
	\label{eq:SH}
\end{align}
where $\mu_{B}$ is the Bohr magneton, $\mathbf{B}$ is the external magnetic field vector, $\mathbf{g}$ is the ${g}$-tensor, and $\mathbf{\tilde{S}}$ is the effective spin operator acting in the target low-energy subspace (with effective spin quantum number $\tilde{S}$).\cite{vancoillieCalculationEPRTensors2007,vancoillieMulticonfigurationalTensorCalculations2008}
The ${g}$-tensor can be computed by mapping the \emph{ab initio} electronic Zeeman Hamiltonian onto Eq.~\eqref{eq:SH}.
In the nonrelativistic form used here, the \emph{ab initio} Zeeman Hamiltonian is
\begin{align}
	\label{eq:Hze}
	\hat{H}_{Ze} = \mu_{B}\mathbf{B}^\mathrm{T} \bigl(\mathbf{L} + g_{e}\mathbf{S}\bigr),
\end{align}
where $\mathbf{L}$ and $\mathbf{S}$ are the orbital and spin angular momentum operators, respectively, and $g_{e}$ is the free-electron $g$-factor.\cite{fanMeasurementElectronMagnetic2023}
Importantly, mapping $\hat{H}_{Ze}$ onto $\hat{H}_{S}$ requires evaluating the matrix elements of the operators in \cref{eq:SH} and \cref{eq:Hze} in a common basis of many-electron wavefunctions provided by the chosen electronic structure method.

In this work, we evaluate the ${g}$-tensor using SO-QDNEVPT2 by employing two complementary strategies that differ in how spin--orbit coupling is incorporated in the working basis.
We refer to these as the effective Hamiltonian (EH) approach and the Kramers (K) approach.

\textit{1) Effective Hamiltonian (EH) approach.}
The EH approach follows the second-order treatment of spin--orbit and Zeeman interactions introduced by McWeeny,\cite{mcweenyOriginSpinHamiltonianParameters1965}
in which both effects are included perturbatively within a scalar-relativistic, spin-free basis.
Let $\ket{\Psi^{SM}_{0}}$ denote a component ($M$) of the spin multiplet of interest with total spin $S$, obtained as an eigenstate of the spin-free Hamiltonian.
Defining
\begin{align}
	\hat{H}_{1} = \hat{H}_{\mathrm{SO}} + \hat{H}_{Ze},
\end{align}
the second-order effective electronic Hamiltonian within the $(2S+1)$-dimensional spin manifold has matrix elements
\begin{align}
	H_{MM'}
	&= \delta_{MM'} E_{0}
	+ \bra{\Psi^{SM}_{0}}\hat{H}_{1}\ket{\Psi^{SM'}_{0}}
	\nonumber\\
	&\quad
	- \sum_{S',M'',I>0}
	\frac{
		\bra{\Psi^{SM}_{0}}\hat{H}_{1}\ket{\Psi^{S'M''}_{I}}\,
		\bra{\Psi^{S'M''}_{I}}\hat{H}_{1}\ket{\Psi^{SM'}_{0}}
	}{
		E_{I}-E_{0}
	},
	\label{eq:geff}
\end{align}
where $\ket{\Psi^{S'M''}_{I}}$ are excited eigenstates of the spin-free Hamiltonian with energies $E_I$ and spin quantum numbers $S',M''$, while $E_0$ is the energy of the reference multiplet.
Assuming that the effective spin in \cref{eq:SH} is identified with the spin of the spin-free basis used in \cref{eq:geff}, the spin Hamiltonian matrix elements can be constructed in the same $\ket{\Psi^{SM}}$ basis, and the ${g}$-tensor is obtained by matching the resulting Zeeman response.\cite{atanasovFirstPrinciplesApproach2015}

A commonly used working expression for the ${g}$-tensor elements derived from this mapping is\cite{neeseCalculationZeroFieldSplittings1998,singhChallengesMultireferencePerturbation2018}
	\begin{equation}
		\begin{aligned}
			\label{eq:g_Neese}
			g_{kl} = g_{e}\delta_{kl}
			&-\frac{1}{S} \sum_{I>0}  \frac{ \bra{\Psi^{SS}_{0}} \hat{L}_k \ket{\Psi^{SS}_{I}} \bra{\Psi^{SS}_{I}}\sum_{pq} F^{l}_{pq}\hat{D}^{z}_{pq}\ket{\Psi^{SS}_{0}}}{E_{I}-E_{0}} \\
			&-\frac{1}{S}\sum_{I>0}\frac{ \bra{\Psi^{SS}_{0}} \sum_{pq} F^{k}_{pq}\hat{D}^{z}_{pq}\ket{\Psi^{SS}_{I}}\bra{\Psi^{SS}_{I}} \hat{L}_l \ket{\Psi^{SS}_{0}} }{E_{I}-E_{0}} ,
		\end{aligned}
	\end{equation}
where $k,l\in\{x,y,z\}$, $\delta_{kl}$ is the Kronecker delta, $\hat{L}_k$ is the $k$-component of orbital angular momentum operator, $F^{k}_{pq}$ are the spin--orbit mean-field matrix elements along Cartesian axis $k$ ($\hat{H}_{\mathrm{SO}} = \sum_{lpq} F^{l}_{pq}\hat{D}^{l}_{pq}$), $\hat{D}^{z}_{pq}$ is the $z$-component of spin-density operator in the spin-orbital basis, and the sum over $I$ runs over excited spin-free electronic states.

The SO-QDNEVPT2 treatment described in \cref{sec:so-qdnevpt2} yields spin-mixed states by incorporating $\hat{H}_{\mathrm{SO}}$ directly in the effective Hamiltonian (cf.\ \cref{eq:SO_QDNEVPT2_EH}), which is incompatible with the spin-free assumptions underlying \cref{eq:g_Neese}.
Therefore, within the EH approach we perform a scalar-relativistic QDNEVPT2 calculation \emph{without} spin--orbit coupling to obtain spin-free wavefunctions and excitation energies, and then evaluate \cref{eq:g_Neese} using these spin-free quantities.
Because the mapping is performed through an effective Hamiltonian constructed in a spin-free basis, we refer to this strategy as the EH approach.

\textit{2) Kramers (K) approach.}
To enable ${g}$-tensor calculations using wavefunctions that already include spin--orbit coupling, we also employ a second strategy that starts from the spin-mixed SO-QDNEVPT2 eigenstates.
In this approach, the Zeeman response of the spin-mixed manifold is mapped directly onto the effective spin Hamiltonian by equating the operators in Eqs.~\eqref{eq:SH} and \eqref{eq:Hze}\cite{cebreiro-gallardoEfficientStateinteractionApproach2025,tatchenCalculatingElectronParamagnetic2009,chibotaruInitioCalculationAnisotropic2012}
\begin{align}
	\begin{bmatrix}
		g_{xx} \mathbf{\tilde{S}}_{x} + g_{xy} \mathbf{\tilde{S}}_{y} + g_{xz} \mathbf{\tilde{S}}_{z}\\
		g_{yx} \mathbf{\tilde{S}}_{x} + g_{yy} \mathbf{\tilde{S}}_{y} + g_{yz} \mathbf{\tilde{S}}_{z}\\
		g_{zx} \mathbf{\tilde{S}}_{x} + g_{zy} \mathbf{\tilde{S}}_{y} + g_{zz} \mathbf{\tilde{S}}_{z}
	\end{bmatrix}
	=
	\begin{bmatrix}
		\mathbf{L}_{x}+ g_{e} \mathbf{S}_{x} \\
		\mathbf{L}_{y}+ g_{e} \mathbf{S}_{y} \\
		\mathbf{L}_{z}+ g_{e} \mathbf{S}_{z}
	\end{bmatrix}
	\equiv
	\begin{bmatrix}
		\mathbf{J}_{x}\\
		\mathbf{J}_{y}\\
		\mathbf{J}_{z}
	\end{bmatrix}
	\equiv {\mathbf{J}} , 
	\label{eq:ME_g}
\end{align}
where ${\mathbf{J}_{x}}$, ${\mathbf{J}_{y}}$ and ${\mathbf{J}_{z}}$  are the magnetic moment operators and $g_{lk}$ ($l,k=x,y,z$) are the elements of the ${g}$-tensor.
The effective spin operators $\mathbf{\tilde{S}}_{x}$, $\mathbf{\tilde{S}}_{y}$, and $\mathbf{\tilde{S}}_{z}$ are represented by the conventional angular-momentum matrices for the chosen reference multiplet spin quantum number $\tilde{S}$:
\begin{align}
\label{eq:S_matrix}
(\mathbf{\tilde{S}}_{x})_{\tilde{M}\tilde{M}'}
&=\frac{1}{2}\sqrt{(\tilde{S}+\tilde{M})(\tilde{S}-\tilde{M}+1)}\delta_{\tilde{M}',\tilde{M}-1} \notag \\
&+ \frac{1}{2}\sqrt{(\tilde{S}-\tilde{M})(\tilde{S}+\tilde{M}+1)}\delta_{\tilde{M}',\tilde{M}+1} , \\ 
(\mathbf{\tilde{S}}_{y})_{\tilde{M}\tilde{M}'}
&= \frac{i}{2}\sqrt{(\tilde{S}-\tilde{M})(\tilde{S}+\tilde{M}+1)}\delta_{\tilde{M}',\tilde{M}+1} \notag \\
&-\frac{i}{2}\sqrt{(\tilde{S}+\tilde{M})(\tilde{S}-\tilde{M}+1)}\delta_{\tilde{M}',\tilde{M}-1} , \\ 
(\mathbf{\tilde{S}}_{z})_{\tilde{M}\tilde{M}'}&=\tilde{M}\delta_{\tilde{M}\tilde{M}'} .
\end{align}  
To determine the $g$-tensor elements in \cref{eq:ME_g}, we diagonalize the SO-QDNEVPT2 effective Hamiltonian in Eqs.~\eqref{eq:SO_QDNEVPT2_EH} or \eqref{eq:SO_QDNEVPT2_EH_firstorder} and compute the $ (2\tilde{S}+1)\times(2\tilde{S}+1)$ magnetic moment matrices $\mathbf{J}_{x}$, $\mathbf{J}_{y}$, and $\mathbf{J}_{z}$ for the reference multiplet in the spin-mixed SO-QDNEVPT2 eigenstate basis.
We then transform these matrices to the eigenbasis of $\mathbf{J}_{z}$, which provides a convenient convention for defining $\tilde{M}$ and the matrix representation of $\mathbf{\tilde{S}}$.
The ${g}$-tensor elements are obtained by matching the matrix elements on the left- and right-hand sides of \cref{eq:ME_g}, leading to the working equations:
\begin{equation}
	\begin{aligned}
		&g_{kx} = \sqrt{\frac{2}{\tilde{S}}}\operatorname{Re}((\mathbf{J}_k)_{\tilde{S},\tilde{S}-1}) ,
		\\ &
		g_{ky}  = -\sqrt{\frac{2}{\tilde{S}}}\operatorname{Im}((\mathbf{J}_k)_{\tilde{S},\tilde{S}-1}) ,
		\\ &
		g_{kz}  = \frac{1}{\tilde{S}}(\mathbf{J}_k)_{\tilde{S}\tilde{S}} ,
	\end{aligned}  
	\label{eq:g_working}
\end{equation}
where $k\in\{x,y,z\}$
This procedure is closely related to the approach of Chibotaru and Ungur,\cite{chibotaruInitioCalculationAnisotropic2012}
where the ${g}$-tensor is extracted by comparing Zeeman splittings predicted by the effective spin Hamiltonian and by the \emph{ab initio} Zeeman operator.

For $\tilde{S}=1/2$, Kramers' theorem implies a doubly degenerate ground-state manifold in the absence of $\hat{H}_{Ze}$, forming a Kramers pair $(\Phi,\tilde{\Phi})$.
In this case, \cref{eq:ME_g} leads to the well-known relation\cite{gerlochParamagneticPropertiesUnsymmetricaI1975}
\begin{align}
	G_{kl}
	=
	2 \sum_{u=\Phi,\tilde{\Phi}}\sum_{v=\Phi,\tilde{\Phi}}
	\bra{u} \hat{L}_{k} + g_{e} \hat{S}_{k} \ket{v}\,
	\bra{v} \hat{L}_{l} + g_{e} \hat{S}_{l} \ket{u},
	\label{eq:Kramers}
\end{align}
where $\mathbf{G}=\mathbf{g}\mathbf{g}^{T}$.
This relation, first implemented by Gerloch and McMeeking,\cite{gerlochParamagneticPropertiesUnsymmetricaI1975}
is widely used to compute $g$-tensors using multireference electronic structure methods such as 
CASSCF,\cite{ganyushinFullyVariationalSpinorbit2013}
CASPT2,\cite{vancoillieCalculationEPRTensors2007,bolvinAlternativeApproachGMatrix2006}
NEVPT2,\cite{ganyushinFullyVariationalSpinorbit2013,singhChallengesMultireferencePerturbation2018}
and DMRG.\cite{sayfutyarovaElectronParamagneticResonance2018}
Although many studies refer to this procedure as a ``quasi-degenerate perturbation theory (QDPT)'' approach,\cite{singhChallengesMultireferencePerturbation2018,lanMolecularTensorsAnalytical2015}
we refer to this method as the Kramers (K) approach to avoid confusion with QDNEVPT2.

\textit{3) Method abbreviation.}
Throughout this work, we denote each computational protocol for calculating $g$-tensors by A-B-C, where A specifies the spin--orbit treatment (BP1, BP2, DKH1, or DKH2), B specifies the electronic structure level (QDNEVPT2 or CASSCF), and C specifies the ${g}$-tensor evaluation strategy (K or EH).
For example, DKH2-QDNEVPT2-K denotes calculations in which spin--orbit coupling is incorporated via the second-order DKH2 operator within SO-QDNEVPT2 and the ${g}$-tensor is obtained using the Kramers approach.
Since the EH calculations are performed using the scalar-relativistic basis without incorporating spin--orbit coupling, BP1-QDNEVPT2-EH and BP2-QDNEVPT2-EH yield the same results and will be denoted as BP-QDNEVPT2-EH.

\subsection{Mitigating intruder states}
\label{sec:intruder_states}

\begin{figure*}[t!]
	\centering
	\includegraphics[width=\textwidth]{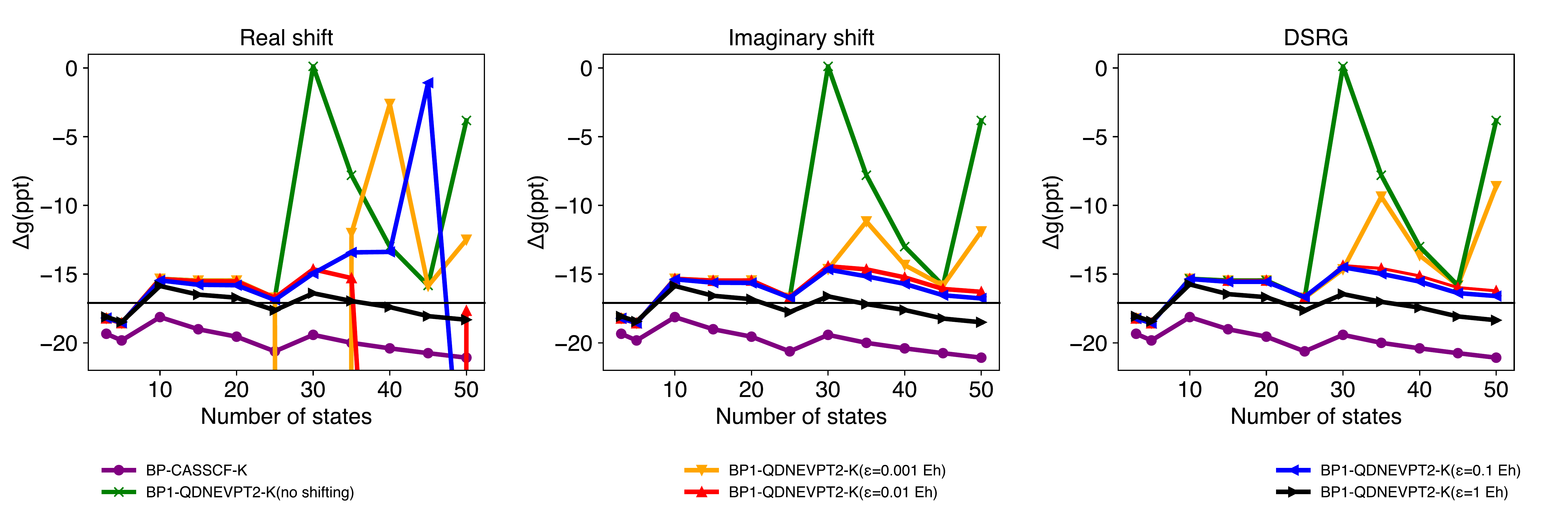}
	\caption{Perpendicular $g$-shift of ZnH in parts per thousand (ppt) relative to the free-electon $g$-value computed using BP1-CASSCF-K and BP1-QDNEVPT2-K, the ANO-RCC basis set, and the (3e, 5o) active space as a function of the number of states.
	For BP1-QDNEVPT2-K, results are computed with three level shifting techniques and different values of shift parameter ($\varepsilon$, $E_h$).
	The experimental $g$-value ($-$17.1 ppt) is indicated by the horizontal black line.\cite{weltnerMagneticAtomsMolecules1983}
	}
	\label{fig:ZnH_intruder}
\end{figure*}

\begin{figure*}[t!]
	\centering
	\includegraphics[width=\textwidth]{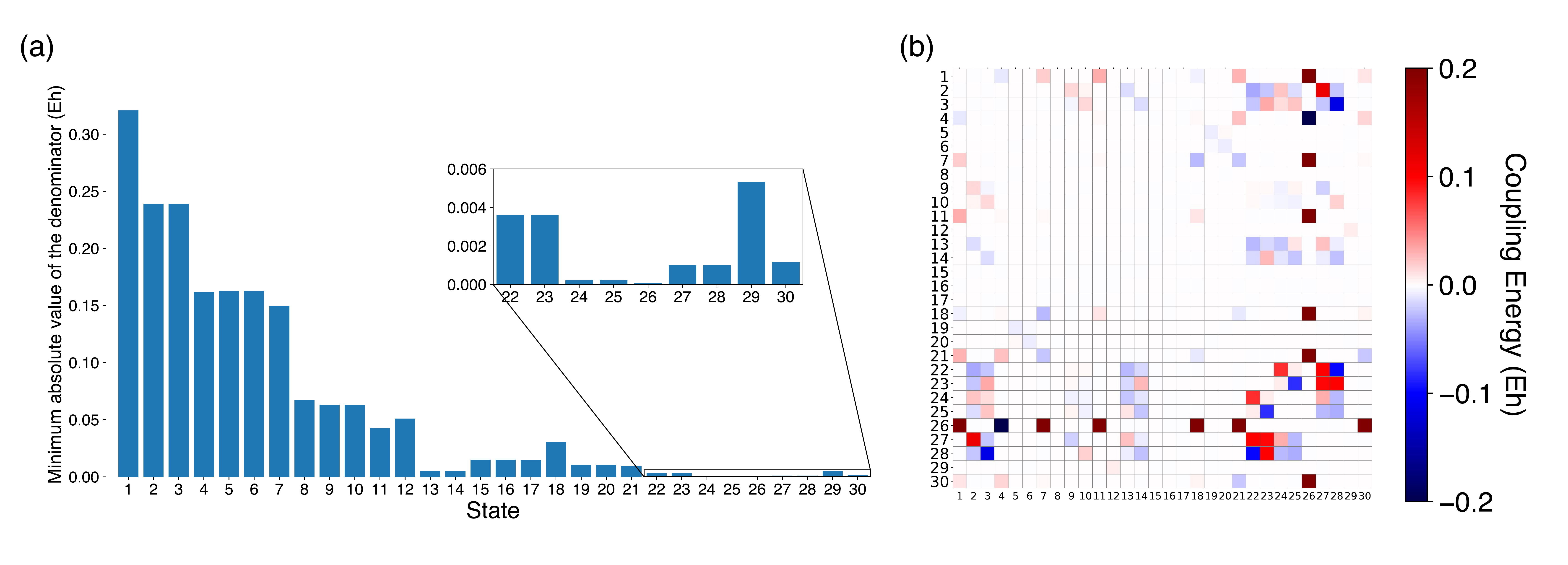}
	\caption{(a) Magnitude of the smallest [$-1'$] correlation energy denominator (\eh, \cref{eq:intruder_amp}) computed for 30 electronic states in ZnH using QDNEVPT2, the ANO-RCC basis set, and the (3e, 5o) active space.
	(b) Heat map indicating the magnitude of QDNEVPT2 effective Hamiltonian matrix elements for the calculation in (a).
	}
	\label{fig:Heff_intruder}
\end{figure*}

\begin{figure*}[t!]
	\centering
	\includegraphics[width=\textwidth]{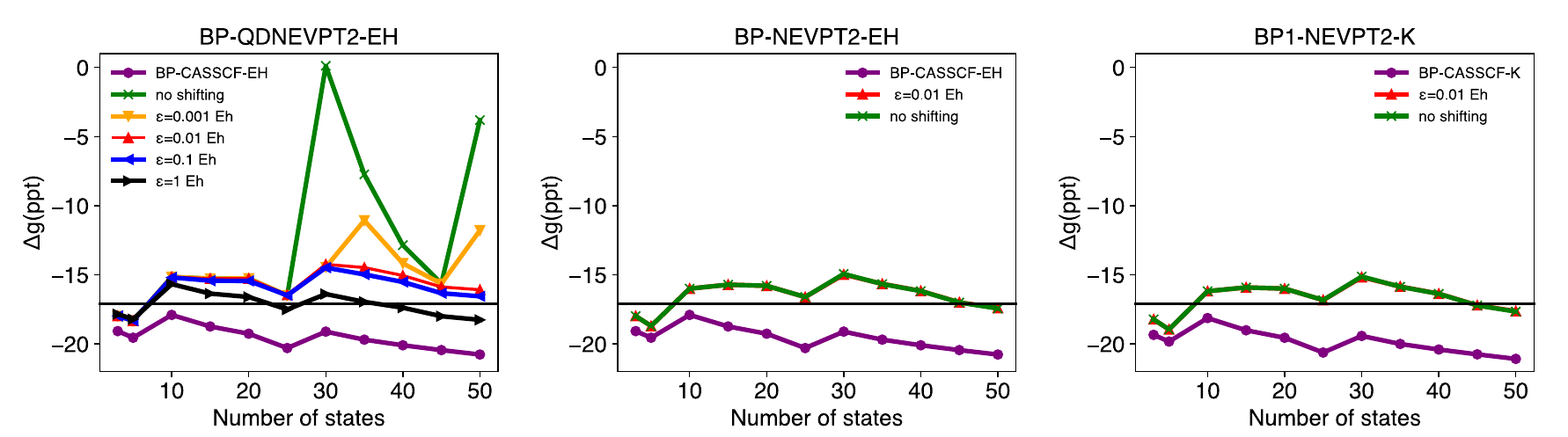}
	\caption{Perpendicular $g$-shift of ZnH in parts per thousand (ppt) relative to the free-electron $g$-value computed using CASSCF, QDNEVPT2, and non-QD multistate NEVPT2 with the ANO-RCC basis set and the (3e, 5o) active space as a function of the number of states.
	The Breit--Pauli Hamiltonian (BP) was used to describe spin--orbit coupling in all calculations. 
	For QDNEVPT2 and NEVPT2, results are computed with imaginary level shifting and different values of shift parameter ($\varepsilon$, $E_h$).
	The experimental $g$-value ($-$17.1 ppt) is indicated by the horizontal black line.\cite{weltnerMagneticAtomsMolecules1983}
	}
	\label{fig:ZnH_intruder_nevpt2}
\end{figure*}

Although multireference theories are effective in describing electron correlation and spin--orbit coupling, they are known to suffer from intruder-state problems, which arise due to unphysical near-degeneracies between the zeroth-order wavefunctions $\ket{\Psi_I^{(0)}}$ and excitations out of the model space.\cite{choeIdentifyingRemovingIntruder2001,forsbergMulticonfigurationPerturbationTheory1997} 
Thanks to the Dyall form of zeroth-order Hamiltonian $\hat{H}^{(0)}$ (\cref{eq:dyallH0}), which incorporates two-electron interactions already at zeroth order, NEVPT2 and QDNEVPT2 are resilient to intruder-state problems for the ground and low-lying excited electronic states.\cite{}
However, these methods can still exhibit intruder-like behavior for sufficiently high-lying excited states, especially when their excitation energies approach (or exceed) the first ionization potential.\cite{hayashiQuasidegenerateExtensionLocal2024,liApplicationDyallHamiltonianbased2025,pokhilkoMulticonfigurationalElectronicStructure2025}

To illustrate the appearance of intruders, we consider the linear system of equations that defines the SO-QDNEVPT2 amplitudes in \cref{eq:psi1_tilde}:
\begin{align}
	\label{eq:soqdnevpt_amplitudes}
	\sum_{\nu} K_{\mu \nu, I}\, \tilde{t}_{\nu, I}^{(1)}
	&= -\bra{\Phi_{\mu, I}} \hat{V} \ket{\Psi_I^{(0)}},
	\\
	\label{eq:koopmans_matrix}
	K_{\mu \nu, I}
	&= \bra{\Phi_{\mu, I}} \bigl(\hat{H}^{(0)} - E_I^{(0)}\bigr)\ket{\Phi_{\nu, I}},
\end{align}
where $K_{\mu\nu,I}$ is the matrix representation of the shifted Dyall Hamiltonian in the perturber basis for state $I$, known as the generalized Koopmans matrix.
Choosing $\hat{V} = \hat{V}_{\mathrm{ee}}$ reduces \cref{eq:soqdnevpt_amplitudes} to the expression defining the conventional QDNEVPT2 amplitudes $t_{\nu, I}^{(1)}$ (\cref{eq:psi1}).
For both QDNEVPT2 and SO-QDNEVPT2, $\hat{V}$ can be expressed as a sum of contributions ($\hat{V}^{[k]}$) from eight excitation classes that are labeled by the number of electrons added to or removed from the active space upon excitation ($k$ $=$ $0$, $\pm1$, $\pm2$, $0'$, and $\pm1'$).\cite{dyallChoiceZerothorderHamiltonian1995,sokolovMultireferencePerturbationTheories2024}
Calculating $\tilde{t}_{\nu, I}^{(1)}$ requires diagonalizing and inverting the generalized Koopmans matrices $K_{\mu\nu,I}$ for each state $I$ and excitation class $[k]$.

As an example, we consider the [$-1'$] excitation class.
In the eigenbasis of the corresponding generalized Koopmans matrix ($\ket{\Psi^{[-1']}_{\rho a,I}}$), the associated amplitudes have the following form:
\begin{align}
	\tilde{t}^{[-1']}_{\rho a,I}
	&=
	\frac{\bra{\Psi^{[-1']}_{\rho a,I}}\hat{V}\ket{\Psi^{(0)}_{I}}}
	{\bra{\Psi^{[-1']}_{\rho a,I}}\bigl(\hat{H}^{(0)}-E^{(0)}_{I}\bigr)\ket{\Psi^{[-1']}_{\rho a,I}}} \notag \\
	&=
	\frac{\bra{\Psi^{[-1']}_{\rho a,I}}\hat{V}\ket{\Psi^{(0)}_{I}}}
	{E^{[-1']}_{\rho a,I}-E^{(0)}_{I}}
	=
	\frac{\bra{\Psi^{[-1']}_{\rho a,I}}\hat{V}\ket{\Psi^{(0)}_{I}}}
	{\epsilon_{a}+e_{\rho,I}},
	\label{eq:intruder_amp}
\end{align}
where $(E^{[-1']}_{\rho a,I} - E^{(0)}_{I})$ is the generalized Koopmans eigenvalue, $\epsilon_{a}$ is the virtual orbital energy entering $\hat{H}^{(0)}$ (\cref{eq:dyallH0}), and $e_{\rho,I}$ is an active-space ionization energy for channel $\rho$ of state $I$.
For typical active spaces, $\epsilon_{a}$ is nonnegative ($\epsilon_{a}\gtrsim0$), so small denominators arise when $e_{\rho,I}$ becomes small or negative ($e_{\rho,I}\lesssim0$).
This situation is most likely for highly excited reference states with excitation energies approaching (or exceeding) the first ionization potential, where some active-space ionization channels can satisfy $e_{\rho,I}\approx -\epsilon_{a}$.
When such roots are included in the multistate treatment, the denominator in \cref{eq:intruder_amp} can become anomalously small, producing unphysically large amplitudes.
These amplified amplitudes can, in turn, generate spuriously strong couplings in $\mathbf{H}_{\mathrm{eff}}$ and lead to erratic or nonconvergent molecular properties.

We demonstrate this behavior in \cref{fig:ZnH_intruder}, which plots the perpendicular $g$-shift $\Delta g_{\perp}$ of the ZnH molecule computed using BP1-CASSCF-K and BP1-QDNEVPT2-K, 3 electrons and 5 orbitals in the active space [(3e, 5o)], and the ANO-RCC basis set.
The $g$-shift is reported in parts per thousand, $\Delta g_{\perp} = (g_{\perp} - g_e)\times10^3$, where $g_e$ is the free-electron value.
To illustrate emerging intruder-state behavior in BP1-QDNEVPT2-K, the $\Delta g_{\perp}$ values (green line with crosses) are plotted as a function of the number of states included in the calculation ($N_{\mathrm{state}}$) and are compared to experimental value ($-$17.1 ptt) indicated by the horizontal black line.
Increasing $N_{\mathrm{state}}$ from 25 to 30 gives rise to a steep increase in the BP1-QDNEVPT2-K $\Delta g_{\perp}$ value (from $-$16.724 to 0.115 ppt), in stark contrast to the $g$-shift from BP1-CASSCF-K that varies much more smoothly.
This instability also arises in the conventional (spin-free) QDNEVPT2 calculations that produce unphysical excitation energies with $N_{\mathrm{states}}>25$, indicating that this behavior originates from the perturbative treatment of electron correlation rather than from the inclusion of spin--orbit coupling.

To diagnose the source of the instability, in \cref{fig:Heff_intruder}a we plot the magnitude of the smallest [$-1'$] amplitude denominator (\cref{eq:intruder_amp}) for each state in the QDNEVPT2 calculation with $N_{\mathrm{state}} = 30$.
States with the smallest denominators (24 to 28, 30) are precisely those that introduce exceptionally large off-diagonal couplings in the QDNEVPT2 effective Hamiltonian, as illustrated in \cref{fig:Heff_intruder}b.
These couplings distort the multistate diagonalization, yielding unphysical excitation energies and ${g}$-shifts.
As demonstrated in \cref{fig:ZnH_intruder_nevpt2}, neglecting the couplings between electronic states by performing a non-QD multistate NEVPT2 calculation eliminates intruder-state problems for the ground-state $g$-tensor.

Our numerical tests indicate that, out of eight excitation classes, the [$-1'$] class is the most likely to cause the intruder-state behavior, although problems are occasionally observed for the [$+1'$] class as well.
Such issues are more likely to occur in small active spaces (e.g., $\lesssim 6$ active orbitals) where states near the ionization continuum can be reached with relatively modest $N_{\mathrm{state}}$, or when performing calculations with diffuse atom-centered basis sets where the virtual orbital energies tend to have small values ($\epsilon_{a}\approx0$).
Expanding the active space by including additional occupied and/or virtual orbitals generally reduces the incidence of intruder-like behavior by improving the zeroth-order description and shifting problematic channels away from near-degeneracy.

To mitigate intruder-state problems, we tested three approaches\cite{changChoiceOptimalShift2012,camachoIntruderStatesMultireference2009}: real level shift\cite{roosMulticonfigurationalPerturbationTheory1995,parkAnalyticalDerivativeCoupling2017} ($\hat{H}^{(0)} \;\to\; \hat{H}^{(0)} + \varepsilon$), imaginary level shift ($\hat{H}^{(0)} \;\to\; \hat{H}^{(0)} + i\varepsilon$)\cite{forsbergMulticonfigurationPerturbationTheory1997,parkImaginaryShiftCASPT22019}, and exponential amplitude damping inspired by the driven similarity renormalization group (DSRG) approach.\cite{liDrivenSimilarityRenormalization2017}
Defining the amplitude denominator as $\Delta$, the application of these techniques involves modifying the amplitude equations as follows:
\begin{align}
\label{eq:im}
&	\frac{1}{\Delta}
	\;\to\;
	\frac{1}{\Delta+\varepsilon}  \quad \mathrm{(real~shift)} ,
\\ &
\label{eq:re}
\frac{1}{\Delta}
	\;\to\;
	\frac{\Delta}{\Delta^{2}+\varepsilon^{2}}  \quad \mathrm{(imaginary~shift)} ,
\\ &
\label{eq:DSRG}
\frac{1}{\Delta}
	\;\to\;
	\frac{1-e^{-\Delta^{2}/\varepsilon^{2}}}{\Delta}  \quad \mathrm{(DSRG~shift)} ,
\end{align} 
where $\varepsilon$ is a real shift constant expressed in atomic energy units (\eh).

\cref{fig:ZnH_intruder} compares the performance of three intruder-state mitigation approaches for the calculation of ZnH $\Delta g_{\perp}$ using BP1-QDNEVPT2-K. 
The real level shift technique translates the denominators along the real energy axis, shifting the near-singular values away from zero but introducing new instabilities.
As a result, applying real shift is the least reliable of three approaches, requiring large shift values ($\varepsilon \ge 1$ \eh) to achieve stable results. 
In contrast, the imaginary shift and DSRG approaches produce intruder-free results with much smaller energy shift parameters ($\varepsilon \ge 0.01$ \eh), showing similar performance.
Importantly, with modest $\varepsilon$ values (0.01 to 0.1 \eh), the imaginary shift and DSRG techniques completely eliminate the intruder-state problems while leaving the data essentially unchanged for the low-lying states unaffected by intruders.
\cref{fig:ZnH_intruder_nevpt2} demonstrates that level shifting is also effective for the BP-QDNEVPT2-EH calculations. 

In practice, the emergence of intruder-state behavior can be detected by checking the magnitude of correlation amplitudes and the coupling matrix elements of QDNEVPT2 effective Hamiltonian.
In this work, we adopt a conservative approach where we apply level shift only for molecules and excitation classes, which exhibit intruder-like problems.
Since the imaginary shift and DSRG approaches show similar performance, we only report results obtained by applying an imaginary level shift.

\section{Computational Details}
\label{sec:comp_details}

\begin{table*}[t!]
	\captionsetup{justification=raggedright,singlelinecheck=false,font=footnotesize}
	\caption{
		Ground-state spin multiplicity ($2S+1$), structural parameters, and experimental $g$-shifts (in parts per thousand, ppt) for the 23 molecules included in the benchmark study (\cref{sec:results_and_discussion}). 
		Bond lengths are in \AA\ and angles are in degrees.
		The set is partitioned into two subsets according to the dominant orbital character of the central atom’s unpaired electron(s): $p$-shell or $d$-shell.
		See \cref{sec:comp_details} and Table S1 for computational details.
	}
	\label{tab:geometry}
	\setlength{\extrarowheight}{2pt}
	\setstretch{1}
	\centering
	\hspace*{-0.8cm}
	\begin{threeparttable}
		\begin{tabularx}{\textwidth}{>{\centering\arraybackslash}X
				>{\centering\arraybackslash}X
				>{\centering\arraybackslash}X>{\centering\arraybackslash}X}
			\hline\hline
                       & $2S+1$  & Structural parameters & Experimental $g$-shift \\ \hline
    $p$-shell: &  &     &  \\
    ZnH   & 2     & 1.595\tnote{a} & $-$17.1\tnote{i,j} \\
    CdH   & 2     & 1.781\tnote{a}  & $-$49.9\tnote{i,j} \\
    HgH   & 2     &  1.766\tnote{a}  & $-$174.3\tnote{i,j} \\
    ZnF   & 2     &  1.799\tnote{b} & $-$6.3\tnote{i,j} \\
    CdF   & 2     &  2.014\tnote{b} & $-$17.3\tnote{i,j} \\
    HgF   & 2     & 2.077\tnote{b} & $-$41.3\tnote{i,j} \\
    NCl   & 3     & 1.643\tnote{c} & 5.4\tnote{i,k}  \\
    NBr   & 3     & 1.808\tnote{c}   & 19.3\tnote{i,k} \\
    NI    & 3     & 2.007\tnote{c}  & 31.0\tnote{i,k} \\ 
    Ge\(_{2}^{+}\)  & 4     &  2.461\tnote{c} & $-$63.3\tnote{i,k}  \\ \hline
    $d$-shell: &    &   &  \\
    CaH   & 2     & 2.0025\tnote{d} & $-$5.7\tnote{i,j} \\
    SrH   & 2     & 2.1456\tnote{d} & $-$15.8\tnote{i,j}   \\
    BaH   & 2     & 2.2319\tnote{e} & $-$27.7\tnote{i,j} \\
    PdH   & 2     & 1.529\tnote{a} & 290.6\tnote{i,j} \\
    RhH\(_2\)   & 2     & 1.510 (Rh--H), \qquad\qquad 84.00 (HRhH)\tnote{f} & $-$52.3, 678.6, 860.6\tnote{l} \\
    IrH\(_2\) & 2     & 1.540 (Ir--H), \qquad\qquad\qquad 91.90 (HIrH)\tnote{g}  &  $-$454.2, 661.6, 1712.6\tnote{l} \\
    CuCl\(_{4}^{2-}\)(\(D_{4h}\))  & 2     & 2.291 (Cu--Cl)\tnote{h}  & 46.7, 46.7, 229.7\tnote{m} \\
    Cu(NH\(_{3}\))\(_{4}^{2+}\) & 2     & Ref.~\citenum{singhChallengesMultireferencePerturbation2018}   &  44.7, 44.7, 238.7\tnote{n}    \\
    TiF\(_{3}\)  (\(D_{3h}\))  & 2     & 1.774 (Ti--F)\tnote{h}  & $-$111.3, $-$111.3, $-$11.1\tnote{o}   \\
    MnF   & 7     & 1.833\tnote{c} & $-$1.3\tnote{i,k} \\
    MnCl  & 7     & 2.241\tnote{c}  & $-$7.3\tnote{i,k} \\
    MnBr  & 7     & 2.395\tnote{c} &  $-$9.3\tnote{i,k} \\
    MnI   & 7     & 2.610\tnote{c} & $-$9.3\tnote{i,k} \\ \hline\hline
		\end{tabularx}
	\end{threeparttable}
	\begin{tablenotes}
		\scriptsize
		\begin{minipage}[t]{0.48\textwidth}
			\item[a] $^{a}$ Experimental value from Ref.\citenum{radzigReferenceDataAtoms1985}.
			\item[b] $^{b}$ Optimized value from Ref.\citenum{belanzoniEvaluationDensityFunctional2001}.
			\item[c] $^{c}$ Optimized value in Ref.\citenum{patchkovskiiCalculationEPRGTensors2001}.
			\item[d] $^{d}$ Experimental value from Ref.\citenum{NISTWebBook}.
			\item[e] $^{e}$ Experimental value from Ref.\citenum{RAM201318}.
			\item[f] $^{f}$ Optimized value from Ref.\citenum{balasubramanianElectronicStatesPotential1988}.
			\item[g] $^{g}$ Optimized values from Ref.\citenum{balasubramanianPotentialEnergySurfaces1990}.
			\item[h] $^{h}$ Optimized values  from Ref.\citenum{vancoillieCalculationEPRTensors2007}.
		\end{minipage}\hfill
		\begin{minipage}[t]{0.48\textwidth}
			\item[i] $^{i}$ The $\Delta g_{\perp}$ value.
			\item[j] $^{j}$ Experimental value from Ref.\citenum{weltnerMagneticAtomsMolecules1983}.
			\item[k] $^{k}$ Experimental value from Ref.\citenum{patchkovskiiCalculationEPRGTensors2001}.
			\item[l] $^{l}$ Experimental values from Ref.\citenum{vanzeeElectronspinResonanceCo1992}.
			\item[m] $^{m}$ Experimental values from Ref.\citenum{chowElectronSpinResonance1973}.
			\item[n] $^{n}$ Experimental values from Ref.\citenum{schollESRENDORCopperII1992}.
			\item[o] $^{o}$ Experimental value from Ref.\citenum{devoreTitaniumDifluorideTitanium1977}.
		\end{minipage}
	\end{tablenotes}
\end{table*}	

The effective Hamiltonian (EH) and Kramers (K) approaches for evaluating $g$-tensors within SO-QDNEVPT2 (\cref{sec:g_tensor}) were implemented in a developer branch of \textsc{Prism}.\cite{MouraSokolov2025Prism}
The SO-QDNEVPT2 code is fully internally contracted and designed to preserve the exact degeneracies of spin--orbit--coupled microstates, following the strategy introduced in our earlier work.\cite{majumderSimulatingSpinOrbit2023}
One- and two-electron integrals, together with the reference SA-CASSCF wavefunctions, were obtained by interfacing \textsc{Prism} with \textsc{PySCF}.\cite{pyscf}
Scalar relativistic effects were treated at the spin-free X2C one-electron (sf-X2C-1e) level\cite{liuExactTwocomponentHamiltonians2009,liSpinSeparationAlgebraic2012} (\cref{sec:so-qdnevpt2}) using the \textsc{PySCF} implementation.
Spin--orbit Hamiltonian matrix elements at the sf-X2C-1e+so-DKH1 and sf-X2C-1e+so-DKH2 levels of theory were generated with \textsc{SOCUTILS},\cite{Wang2022Socutils} interfaced to \textsc{Prism}.
The DKH2 scalar relativistic contributions arising from the transformation of spin-dependent second-order Hamiltonian due to the picture change effect\cite{majumderConsistentSecondOrderTreatment2024} were found to have a very small influence on the computed $g$-shifts ($\lesssim$ 1 \%) and were neglected in our calculations.

To assess the accuracy of SO-QDNEVPT2 for $g$-tensor calculations, we assembled a benchmark set of 23 small molecules with experimentally reported $g$-values.
\Cref{tab:geometry} summarizes the corresponding structural parameters, ground-state spin states, and reference experimental $g$-values.
For analysis, the dataset is partitioned into two subsets according to the dominant orbital character of the unpaired electron(s) on the central atom, namely $p$-shell and $d$-shell species.

An overview of the ${g}$-shift benchmark results is provided in \cref{sec:results_and_discussion:benchmark}.
The parameters of SO-QDNEVPT2 calculations (active space, basis set, number of states, and imaginary level shift) are reported in Table S1 of the Supplementary Material. 
Whenever feasible, multiple active spaces were explored.
Starting orbitals were generated from density functional theory using the BP86 functional,\cite{perdewDensityfunctionalApproximationCorrelation1986,beckeDensityfunctionalExchangeenergyApproximation1988} and the selected active-space orbitals are documented in the Supplementary Material.
The ANO-RCC basis sets\cite{roosMainGroupAtoms2004} were used in most of the calculations reported in \cref{sec:results_and_discussion:benchmark} except for a few polyatomic molecules where the def2-TZVP basis\cite{weigendBalancedBasisSets2005} set was used (Table S1). 
State averaging typically included 50 roots, but this number was reduced for a subset of molecules to alleviate convergence difficulties in the SA-CASSCF calculations (Table~S1).
For all molecules with a spin-doublet ($S$ = 1/2) ground state, all spin states were included. 
For the high-spin molecules in \cref{tab:Highspin}, only the states with $S \ge S’$ where $S’$ is the ground-state spin were incorporated.  
Unless stated otherwise, SA-CASSCF orbitals were optimized with equal weights over all included states, and the coordinate origin was placed at the center of nuclear charge.
Following the benchmark overview, \cref{sec:results_and_discussion:analysis} examines the sensitivity of results to the SO-QDNEVPT2 parameters, including the dependence of computed $g$ values on the active space and number of states (\cref{sec:results_and_discussion:analysis:active_space}), state-averaging weights (\cref{sec:results_and_discussion:analysis:weights}), choice of coordinate origin (\cref{sec:results_and_discussion:analysis:gauge}), and basis set (\cref{sec:results_and_discussion:analysis:basis}).

As discussed in \cref{sec:intruder_states}, QDNEVPT2 calculations can exhibit intruder-state behavior, which may be controlled using level shifts.
Throughout \cref{sec:results_and_discussion:benchmark,sec:results_and_discussion:analysis}, we adopt a conservative protocol in which a small imaginary shift ($\varepsilon=0.01~\eh$) is applied only to problematic amplitudes, identified by excitation norms exceeding 1.0 and off-diagonal effective Hamiltonian couplings larger than $0.05~\eh$.
In the vast majority of affected cases, the problematic contributions originate from the $[-1']$ excitation class.
The use of an imaginary shift is explicitly indicated in the corresponding tables and figures.

For each molecule, the computed $g$-tensor is reduced to three principal $g$-factors, reported as $g$-shifts in parts per thousand (ppt) relative to the free-electron value, $\Delta g_i = (g_i - g_e)\times 10^3$, where $i$ labels a principal axis.
In the Kramers formalism (K), the resulting $g$ tensor need not be symmetric.\cite{sharmaGfactorSymmetryTopology2024}
Accordingly, Kramers principal values are obtained by taking the square roots of the eigenvalues of the symmetric matrix $\mathbf{G}=\mathbf{g}\mathbf{g}^{T}$ where $\mathbf{g}$ is obtained via \cref{eq:g_working}, rather than by directly diagonalizing $\mathbf{g}$.
Since the perpendicular $g$-shift of diatomic molecules is significantly larger than their parallel $g$-component, only the perpendicular values are included in our benchmark.
All numerical results are provided in the Supplementary Material.

\section{Results and Discussion}
\label{sec:results_and_discussion}

\subsection{Benchmark of SO-QDNEVPT2 for ${g}$-tensor calculations}
\label{sec:results_and_discussion:benchmark}

\begin{figure*}[t!]
	\centering
	\includegraphics[width=\textwidth]{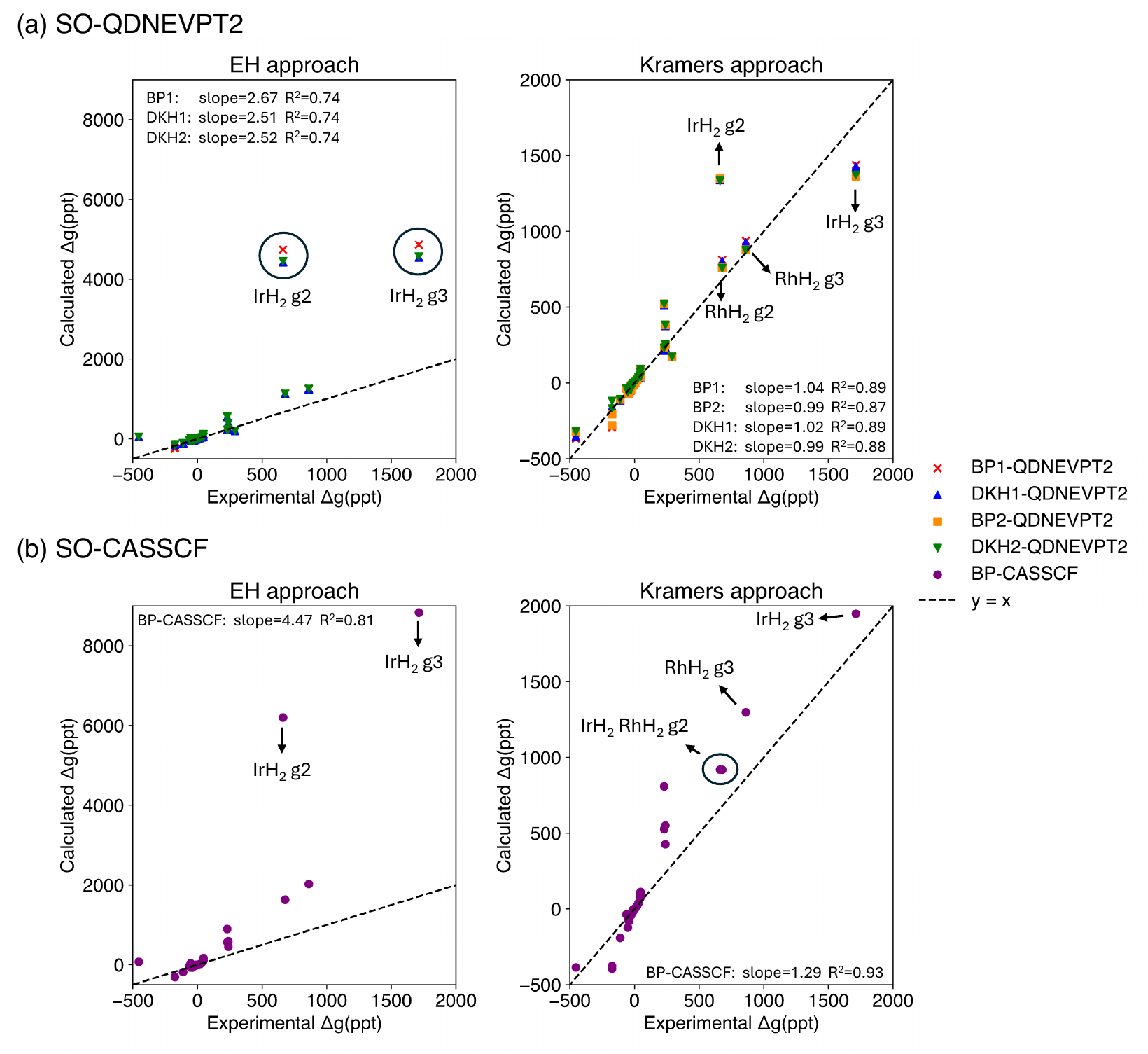}
	\caption{Correlation between experimental and computed $g$-shifts (in parts per thousand, ppt) for the full benchmark set, comparing the effective Hamiltonian (EH) and Kramers (K) $g$-tensor formalisms combined with SO-QDNEVPT2 (a) and SA-CASSCF (b). 
		See the Supplementary Material for numerical values.
	}
	\label{fig:overall}
\end{figure*}

We begin with an overview of SO-QDNEVPT2 benchmark results by comparing the performance of this method between two $g$-tensor formalisms (K and EH) and four levels of spin--orbit treatment (BP1, BP2, DKH1, and DKH2).
As discussed in \cref{sec:comp_details}, our benchmark set contains 23 diatomic and polyatomic molecules (\cref{tab:geometry}) spanning $g$-shifts from near zero to several hundred parts per thousand ($\Delta g_i = (g_i - g_e)\times 10^3$, ppt) and ground-state spin values from 1/2 to 3.

\cref{fig:overall} compares experimental $g$-shifts from \cref{tab:geometry} with the EH and K predictions from SO-QDNEVPT2 (panel a) and SA-CASSCF (panel b).
Overall, SA-CASSCF systematically overestimates the magnitude of the $g$-shifts across the benchmark. 
Incorporating dynamical correlation via SO-QDNEVPT2 markedly reduces these systematic errors and brings the computed trends into substantially better agreement with experiment for the majority of molecules.

Within the SO-QDNEVPT2 results, the EH and K formalisms yield very similar predictions for modest $g$-shifts ($|\Delta g|\lesssim 100$~ppt).
In contrast, for large shifts, which are indicative of strong spin--orbit mixing, the EH results deteriorate systematically.
This behavior is expected because EH treats spin--orbit effects only through second order and becomes unreliable when the interaction is strong (\cref{sec:g_tensor}).\cite{vancoillieCalculationEPRTensors2007,lanMolecularTensorsAnalytical2015}

\begin{figure*}[t!]
    \centering
    \includegraphics[width=\textwidth]{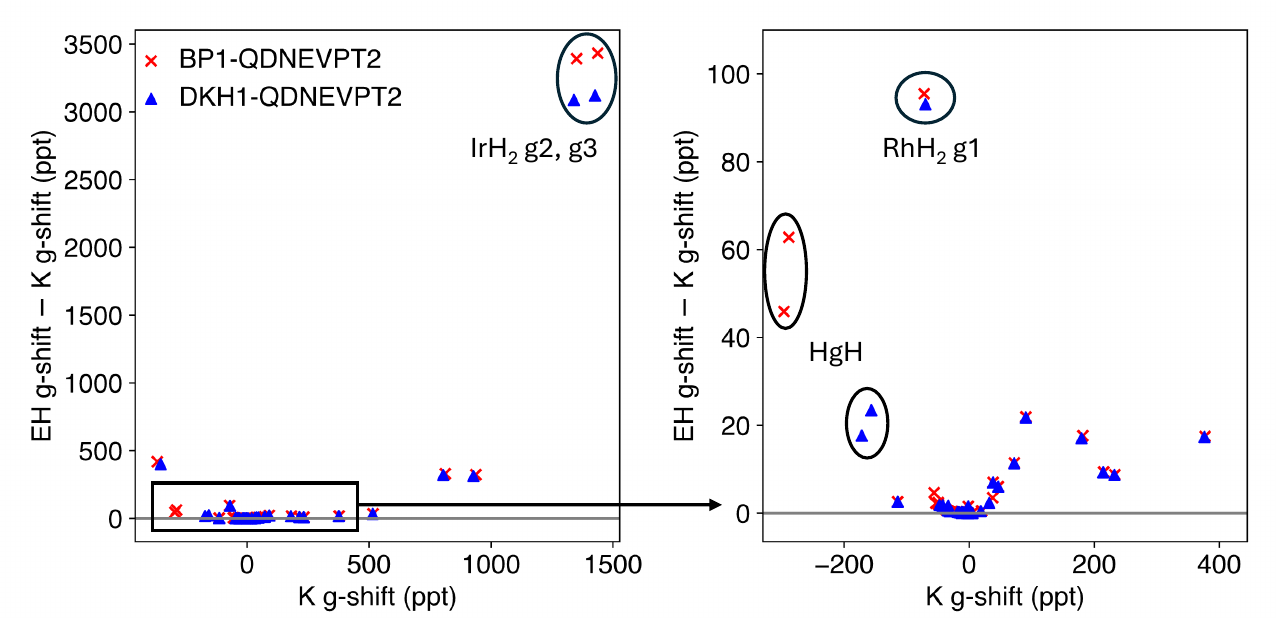}
    \caption{
    	Difference in $g$-shift predictions between the EH and K formalisms plotted as a function of K $g$-shift across the benchmark set (in parts per thousand, ppt).
    }
\label{fig:diff_K_EH}
\end{figure*}

The magnitude dependence of this effect is quantified in \cref{fig:diff_K_EH}, which plots the difference between the EH and K $g$-values as a function of the K $g$-shift.
Over the $-50$ to $50$~ppt range, the deviation in $g$-values computed using the two approaches remains within 5~ppt, confirming that EH is an adequate approximation when spin--orbit effects are weak.
Beyond this regime the discrepancy grows with $|\Delta g|$, and the Kramers formalism becomes essential.
Accordingly, we use the Kramers results as the primary reference when comparing spin--orbit Hamiltonians below.

\begin{figure*}[t!]
    \centering
    \includegraphics[width=\textwidth]{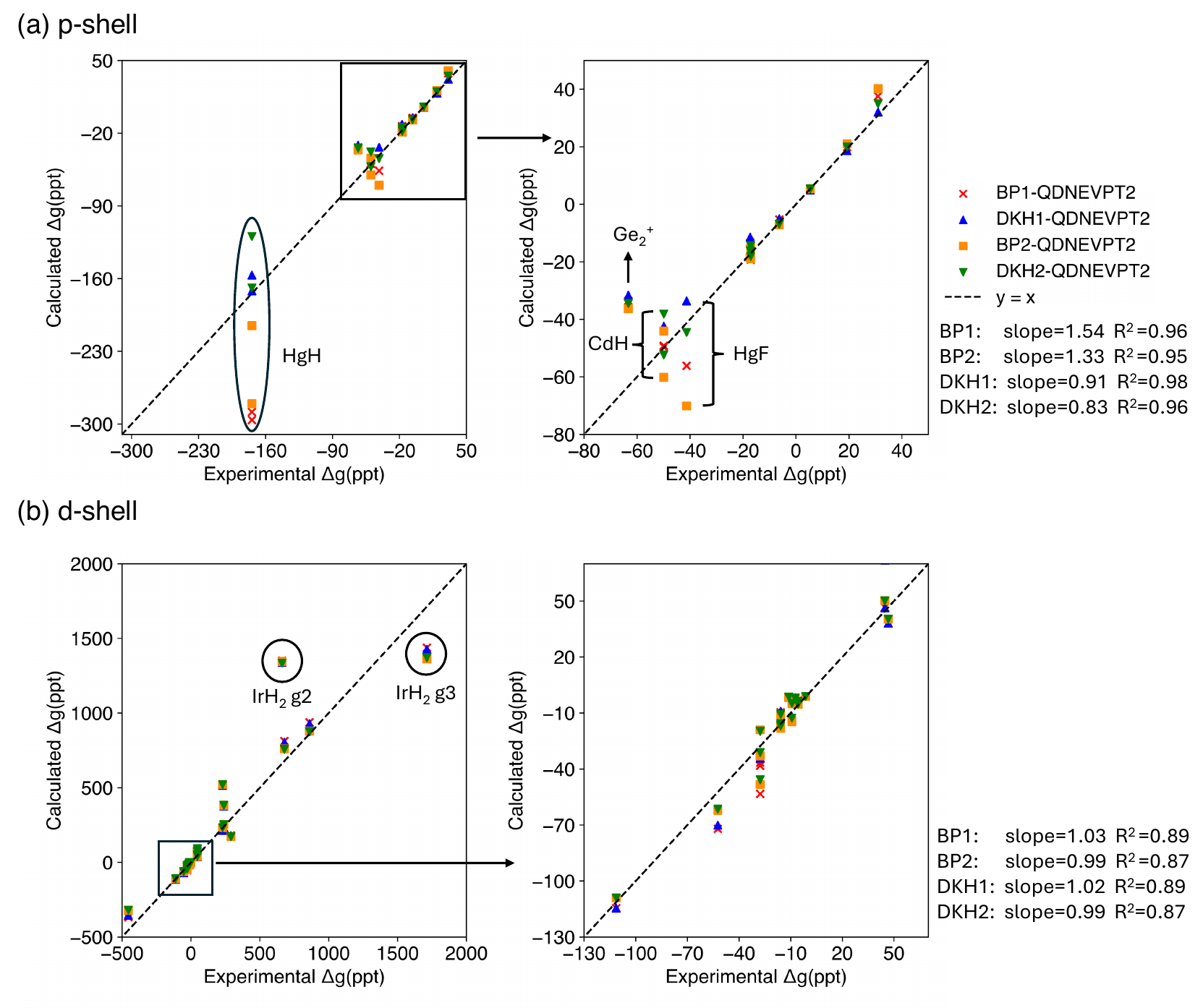}
    \caption{
    	Correlation between experimental and computed $g$-shifts (in parts per thousand, ppt) for molecules with unpaired electrons localized on either the $p$- (a) or $d$-shell (b).
    }
\label{fig:toc_pd}
\end{figure*}

Restricting our analysis to the Kramers approach, we next assess the impact of the spin--orbit Hamiltonian.
To rationalize trends, we separate the benchmark into subsets where the dominant spin--orbit mixing responsible for the $g$-shift derives from either $p$-shell or $d$-shell orbitals.
For the $p$-shell subset (\cref{fig:toc_pd}a), the DKH-based treatments track experiment substantially better than the BP-based approximations as $|\Delta g|$ increases.
This is reflected in the linear-regression slopes: DKH results are closer to the ideal unit slope (slopes $\approx 0.91$ and $0.83$ for DKH1 and DKH2, respectively), whereas BP exhibits pronounced systematic error (slopes $\approx 1.54$ and $1.33$ for BP1 and BP2).
Including second-order spin--orbit contributions in the effective Hamiltonian (BP2 and DKH2) has only a minor impact for molecules with $|\Delta g|<40$~ppt.
In contrast, for $p$-shell species exhibiting larger $|\Delta g|$ (e.g., HgF, CdH, and HgH), the BP2/BP1 and DKH2/DKH1 differences become more pronounced, consistent with stronger spin--orbit mixing in this subset.
The most significant effect of second-order spin--orbit coupling contributions is observed for HgH computed using the (13e, 10o) active space where the BP2 -- BP1 and DKH2 -- DKH1 $g$-shift differences amount to 83.1 ppt and 37.2 ppt, respectively.

For the $d$-shell subset (\cref{fig:toc_pd}b), BP and DKH perform comparably, even in systems containing heavier elements.
This can be understood from the weaker effective spin--orbit coupling experienced by more diffuse $d$ electrons (relative to $p$ electrons), which reduces the sensitivity to the specific one-electron spin--orbit treatment.\cite{lanMolecularTensorsAnalytical2015}
Including second-order spin--orbit terms in the effective Hamiltonian (BP2 and DKH2) produces only modest changes relative to the dominant difference between BP and DKH in this subset.

\begin{table*}[t!]
\begin{adjustwidth}{-2cm}{-2cm}
	\caption{ Perpendicular $g$-shift $\Delta g_{\perp}$ (in parts per thousand, ppt) of low-spin ($S = 1/2$) diatomic molecules computed using SO-QDNEVPT2-K with the ANO-RCC basis set}
	\label{tab:Lowspin}
	\setlength{\tabcolsep}{10pt}
	\setstretch{1}
	\centering
	\begin{threeparttable}
\begin{tabular}{cccccccc}
\hline\hline
          & Active space & $N_\mathrm{state}$\tnote{a} &  Experiment\tnote{b}  & BP1   & BP2   & DKH1  & DKH2 \\ \hline
    $p$-shell &       &       &       &       &       &       &  \\
    ZnH   & (3e, 5o) & 50    & $-$17.1 & $-$16.3 & $-$19.0 & $-$15.5 & $-$18.2 \\
          & (13e, 10o) & 50    & $-$17.1 & $-$19.3 & $-$15.1 & $-$18.3 & $-$14.5 \\
    CdH   & (3e, 5o) & 50    & $-$49.9 & $-$49.2 & $-$60.2 & $-$42.4 & $-$52.4 \\
          & (13e, 10o) & 50    & $-$49.9 & $-$49.6 & $-$44.1 & $-$42.4 & $-$38.2 \\
    HgH   & (3e, 5o) & 50    & $-$174.3 & $-$296.2 & $-$280.5 & $-$171.9 & $-$169.0 \\
          & (13e, 10o) & 35    & $-$174.3 & $-$288.5 & $-$205.3 & $-$156.6 & $-$119.5 \\
    ZnF   & (7e, 10o) & 50    & $-$6.3  & $-$5.3  & $-$7.2  & $-$5.1  & $-$6.9 \\
    CdF   & (7e, 10o) & 45    & $-$17.3 & $-$13.3 & $-$18.5 & $-$11.4 & $-$16.2 \\
    HgF   & (7e, 10o) & 30    & $-$41.3 & $-$56.2 & $-$70.1 & $-$33.6 & $-$44.5 \\
 \hline
    $d$-shell &       &       &       &       &       &       &  \\
    CaH   & (3e, 5o) & 50    & $-$5.7  & $-$3.5  & $-$5.3  & $-$3.5  & $-$5.3 \\
          & (3e, 7o) & 50    & $-$5.7  & $-$3.9  & $-$5.2  & $-$3.8  & $-$5.1 \\
          & (3e, 9o) & 50    & $-$5.7  & $-$3.2  & $-$4.0  & $-$3.1  & $-$3.9 \\
    SrH   & (3e, 5o) & 50    & $-$15.8 & $-$12.8 & $-$18.1 & $-$11.8 & $-$17.0 \\
          & (3e, 7o) & 50    & $-$15.8 & $-$13.1 & $-$17.0 & $-$12.0 & $-$15.7 \\
          & (3e, 9o) & 50    & $-$15.8 & $-$9.7  & $-$11.4 & $-$9.0  & $-$10.7 \\
    BaH   & (3e, 5o) & 50    & $-$27.7 & $-$53.3 & $-$48.2 & $-$47.5 & $-$45.7 \\
          & (3e, 7o) & 50    & $-$27.7 & $-$38.3 & $-$33.0 & $-$34.1 & $-$31.2 \\
          & (3e, 9o) & 50    & $-$27.7 & $-$36.6 & $-$19.0 & $-$34.2 & $-$19.8 \\
    PdH   & (11e, 10o) & 50    & 290.6 & 181.7 & 172.4 & 179.6 & 171.4 \\
    \hline \hline 
\end{tabular} 
	\end{threeparttable}
		\begin{tablenotes}
	 \item[a] \scriptsize {$^{a}$ Number of states}
        \item[b] \scriptsize {$^{b}$ Experimental values from Ref.~\citenum{weltnerMagneticAtomsMolecules1983}.}
		\end{tablenotes}
\end{adjustwidth}
\end{table*}

\begin{table*}[t!]
\begin{adjustwidth}{-2cm}{-2cm}
	\caption{Perpendicular $g$-shift $\Delta g_{\perp}$ (in parts per thousand, ppt) of high-spin molecules computed with SO-QDNEVPT2-K using the ANO-RCC basis set for NCl, NBr, and NI, and the ANO-RCC-VTZP basis for MnF, MnCl, MnBr, and MnI.
	}
	\label{tab:Highspin}
	\setlength{\tabcolsep}{10pt}
	\setstretch{1}
	\centering
	\begin{threeparttable}
\begin{tabular}{ccccccccc}
\hline\hline
          & $2S+1$  & Active space & $N_\mathrm{state}$\tnote{a} &  Experiment\tnote{b}  & BP1   & BP2   & DKH1  & DKH2 \\ \hline
    $p$-shell &       &       &       &       &       &       &       &  \\
    NCl   & 3     & (12e, 8o) & 50    & 5.4   & 5.0   & 5.4   & 5.0   & 5.3 \\
    NBr   & 3     & (12e, 8o) & 50    & 19.3  & 19.9  & 21.1  & 18.6  & 19.9 \\
    NI    & 3     & (12e, 8o) & 50    & 31    & 37.6  & 40.2  & 32.1  & 34.9 \\ 
        Ge\(_{2}^{+}\)  &    4   & (7e, 7o) & 6     & $-$63.3 & $-$33.6 & $-$36.3 & $-$31.6 & $-$34.5 \\\hline
    $d$-shell &       &       &       &       &       &       &       &  \\
    MnF   & 7     & (12e, 11o) & 3     & $-$1.3  & $-$1.1  & $-$1.1  & $-$1.0  & $-$1.1 \\
    MnCl  & 7     & (12e, 11o) & 3     & $-$7.3  & $-$1.8  & $-$2.0  & $-$1.8  & $-$2.0 \\
    MnBr  & 7     & (12e, 11o) & 3     & $-$9.3  & $-$4.3  & $-$5.1  & $-$4.1  & $-$4.9 \\
    MnI   & 7     & (12e, 11o) & 3     & $-$9.3  & $-$13.9 & $-$14.6 & $-$11.9 & $-$12.8 \\
    \hline \hline 
\end{tabular} 
	\end{threeparttable}
		\begin{tablenotes}
	 \item[a] \scriptsize {$^{a}$ Number of states}
        \item[b] \scriptsize {$^{b}$ Experimental values from Ref.~\citenum{patchkovskiiCalculationEPRGTensors2001}.}
		\end{tablenotes}
\end{adjustwidth}
\end{table*}

The trends established in \cref{fig:overall,fig:diff_K_EH,fig:toc_pd} hold across both the low-spin ($S=1/2$, \cref{tab:Lowspin}) and high-spin ($S>1/2$, \cref{tab:Highspin}) subsets. 
As shown in \cref{tab:Highspin}, the $d$-shell MnX (X = F, Cl, Br, I; $S=3$) molecules exhibit small $g$-shifts, which depend little on the level of spin--orbit coupling treatment. 
As a result, the BP and DKH results agree closely throughout the MnX series, with only a modest increase in the BP--DKH difference for MnI relative to MnF--MnBr due to a larger nuclear charge. 
In contrast, the $p$-shell NX (X = Cl, Br, I; $S=1$) radicals and Ge$_2^{+}$ ($S=3/2$) show a noticeably stronger dependence on both the choice of spin--orbit Hamiltonian and the inclusion of second-order terms.

\begin{table*}[t!]
\begin{adjustwidth}{-2cm}{-2cm}
	\caption{The $g$-shifts of polyatomic molecules (in parts per thousand, ppt) computed using SO-QDNEVPT2-K.
	For RhH\(_2\) and IrH\(_2\), the ANO-RCC basis set was used, while the def2-TZVP basis was employed for CuCl\(_{4}^{2-}\), Cu(NH\(_{3}\))\(_{4}^{2+}\), and TiF\(_{3}\). }
	\label{tab:polyatomic}
	\setlength{\tabcolsep}{10pt}
	\setstretch{1}
	\centering
	\begin{threeparttable}
    \begin{tabular}{ccccccccc}
\hline\hline
          & Active space & $N_\mathrm{state}$\tnote{a} & $\Delta g$ &  Experiment\tnote{b}  & BP1   & BP2   & DKH1  & DKH2 \\ \hline
    CuCl\(_{4}^{2-}\) & (9e,5o) & 5     & $\parallel$ & 46.7  & 90.8  & 91.2  & 90.6  & 91.1 \\
          &       &       & $\perp$ & 229.7 & 516.6 & 521.8 & 515.2 & 520.7 \\
          & (11e,6o) & 6\tnote{c} & $\parallel$ &   46.7    & 38.2  & 40.3  & 38.1  & 40.3 \\
          &       &       & $\perp$ &  229.7     & 215.0 & 232.3 & 214.3 & 232.0 \\
    Cu(NH\(_{3}\))\(_{4}^{2+}\) & (9e,5o) & 5     & $\parallel$ & 44.7  & 72.2  & 73.2  & 72.0  & 73.1 \\
          &       &       & $\perp$ & 238.7 & 377.3 & 383.2 & 376.2 & 382.3 \\
          & (11e,6o) & 6\tnote{c} & $\parallel$ &   44.7    & 46.5  & 50.2  & 46.4  & 50.1 \\
          &       &       & $\perp$ &  238.7     & 232.9 & 253.8 & 232.2 & 253.2 \\
    TiF\(_{3}\)  & (1e,5o) & 5     & $\parallel$ & $-$111.3 & $-$114.5 & $-$109.0 & $-$114.5 & $-$109.1 \\
          &       &       & $\perp$ & $-$11.1 & $-$1.5  & $-$1.4  & $-$1.5  & $-$1.4 \\
    RhH\(_2\)   & (11e,8o) & 45    & 1     & $-$52.3 & $-$72.1 & $-$62.2 & $-$70.0 & $-$61.5 \\
&       &       & 2     & 678.6 & 812.4 & 760.4 & 803.9 & 756.8 \\
&       &       & 3     & 860.6 & 937.6 & 879.0 & 927.8 & 875.3 \\
IrH\(_2\) & (11e,8o) & 40    & 1     & $-$454.2 & $-$368.1 & $-$321.6 & $-$353.7 & $-$320.7 \\
&       &       & 2     & 661.6 & 1350.8 & 1345.6 & 1340.0 & 1333.3 \\
&       &       & 3     & 1712.6 & 1437.4 & 1362.3 & 1426.5 & 1370.6 \\
          \hline\hline
\end{tabular}

	\end{threeparttable}
		\begin{tablenotes}
 \item[a] \scriptsize {$^{a}$ Number of states}
        
\item[b] \scriptsize {$^{b}$  Experimental values from Ref.~\citenum{vanzeeElectronspinResonanceCo1992} for RhH\(_2\) and IrH\(_2\), Ref.~\citenum{chowElectronSpinResonance1973} for CuCl\(_{4}^{2-}\),
Ref.~\citenum{schollESRENDORCopperII1992} for Cu(NH$_3$)$_4^{2+}$, and Ref.~\citenum{devoreTitaniumDifluorideTitanium1977} for TiF$_3$.
}

\item[c] \scriptsize {$^{c}$ The LMCT state (fifth excited state) is assigned five times the weight of the remaining states in the SA-CASSCF reference.}
		\end{tablenotes}
\end{adjustwidth}
\end{table*}

The polyatomic complexes in \cref{tab:polyatomic} exhibit the same qualitative behavior as the diatomics while showing some dependence on the parameters of reference SA-CASSCF calculation.
For the copper(II) benchmarks, $\mathrm{[CuCl_4]^{2-}}$ and $\mathrm{[Cu(NH_3)_4]^{2+}}$, the minimal (9e, 5o) active space with five states reproduces the trends reported previously\cite{singhChallengesMultireferencePerturbation2018} but yields sizable absolute errors.
Expanding the active space to (11e, 6o) by including the metal--ligand bonding $\sigma_{d_{x^2-y^2}}$ orbital and increasing the state-averaging weight on the ligand-to-metal charge-transfer (LMCT) state (Table~\ref{tab:polyatomic}) markedly improves agreement with experiment and aligns the SO-QDNEVPT2 results with prior multireference studies.\cite{langSpindependentPropertiesFramework2019,langCombinationMultipartitioningHamiltonian2020}
Since in these copper complexes the $g$ shifts originate from the unpaired electrons in the $d$-shell, BP and DKH give similar principal values, and the difference between first- and second-order spin--orbit variants (e.g., BP1 vs.\ BP2, DKH1 vs.\ DKH2) is typically only a few ppt.

For TiF\(_3\), the compact (1e, 5o) active space with five states yields a \(g_{\parallel}\) shift close to experiment (errors of only a few ppt) and shows essentially no BP--DKH separation, consistent with earlier observations for \(d\)-shell molecules. 
In contrast, the perpendicular shift is strongly underestimated in magnitude (\(\Delta g_{\perp}\approx -1.5\) ppt vs.\ \(-11.1\) ppt), suggesting that additional excited-state contributions relevant to \(\Delta g_{\perp}\) are not fully captured using this minimal active space. 
For RhH\(_2\) and IrH\(_2\), the shifts computed using the (11e, 8o) SA-CASSCF reference span from moderate to very large values, and the \(d\)-shell character again leads to similar BP and DKH predictions. 
At the same time, the second-order variants (BP2/DKH2) systematically reduce the deviations relative to the first-order treatments for RhH\(_2\), improving all three principal shifts. 
IrH\(_2\) is markedly more challenging: all four SO-QDNEVPT2-K variants remain comparable to one another yet show substantial residual discrepancies for the largest experimental components (see \cref{sec:results_and_discussion:analysis:active_space} for additional discussion).

Taken together, these benchmarks establish three overarching trends.
(i) The K formalism is generally more reliable than EH and becomes essential when $|\Delta g|$ is large.
(ii) For $p$-shell $g$ shifts, DKH-based spin--orbit Hamiltonians reduce systematic errors relative to BP with significant differences between the first-order (state-interaction) BP1/DKH1 and second-order BP2/DKH2 variants.
(iii) For $d$-orbital systems, BP and DKH perform similarly, and higher-order spin--orbit approximations typically produce only minor changes in $g$-shift.
In the following section, we analyze how these conclusions depend on technical parameters of SO-QDNEVPT2 calculations, namely: active space, number of states and state-averaging weights, basis set, and coordinate system origin.

\subsection{Dependence of $g$-shifts on SO-QDNEVPT2 parameters}
\label{sec:results_and_discussion:analysis}

In this section, we analyze how the choice of key SO-QDNEVPT2 parameters affects computed $g$-shifts, focusing on (1) the orbital active space and the number of states included in the QDNEVPT2 effective Hamiltonian, (2) SA-CASSCF state-averaging weights, (3) the coordinate origin (gauge), and (4) the one-electron basis set.
The focus of this study to investigate convergence behavior and practical parameter choices.

\subsubsection{Active space size and number of states}
\label{sec:results_and_discussion:analysis:active_space}

\begin{figure*}[t!]
   \centering
    \includegraphics[width=\textwidth]{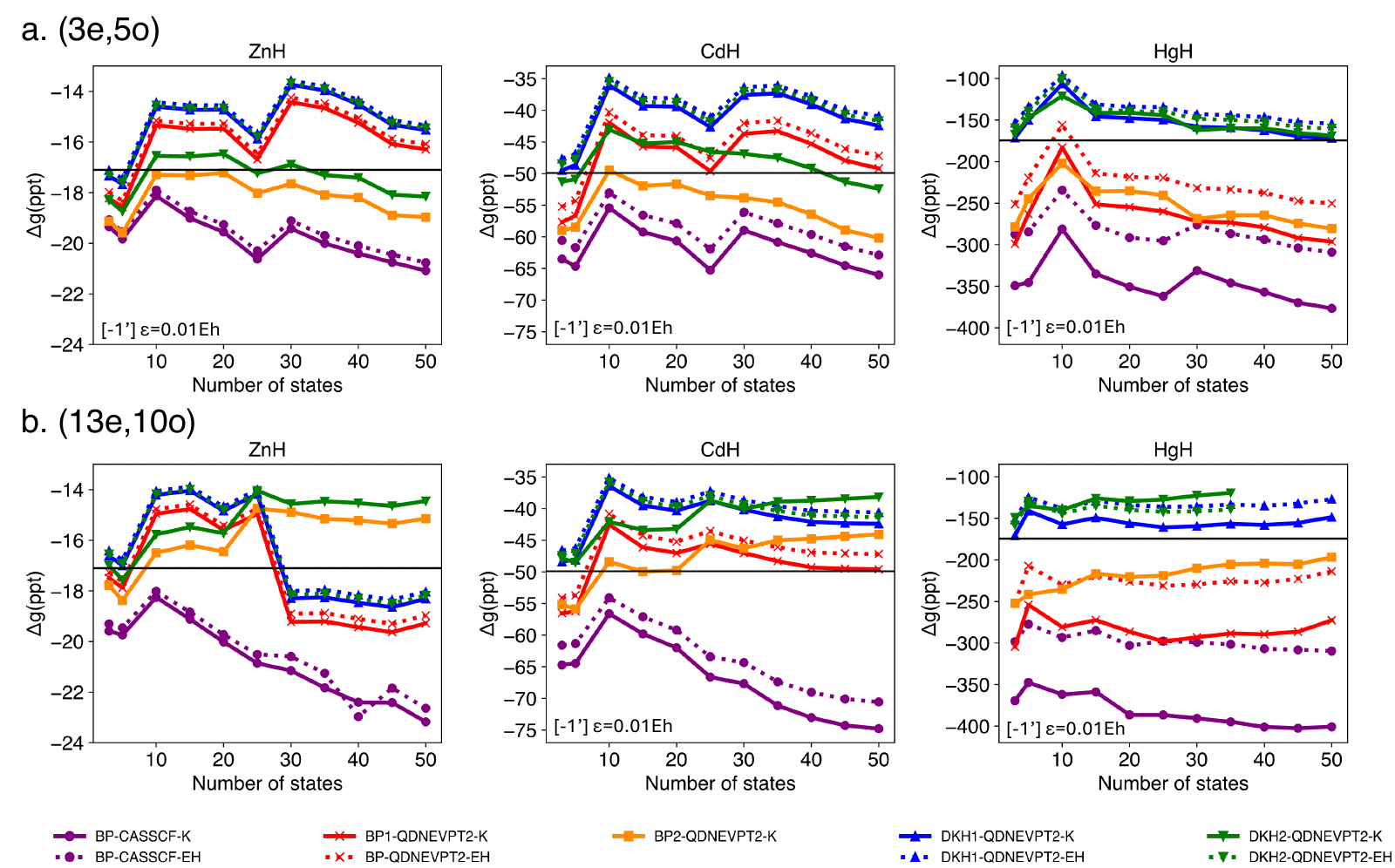}
    \caption{
    Perpendicular $g$-shift $\Delta g_{\perp}$ (in parts per thousand, ppt) for ZnH, CdH, and HgH as a function of the number of states computed using (a) (3e, 5o) and (b) (13e, 10o) active space.
    Horizontal lines denote the experimental $g$-shift values for comparison (Ref.~\citenum{weltnerMagneticAtomsMolecules1983}).
    An imaginary level shift $\varepsilon=0.01~\eh$ is applied to the $[-1']$ amplitudes for all calculations except for the ZnH (13e, 10o) data set.
    }
    \label{fig:ZnHCdHHgH_351310}
\end{figure*}

We first assess the sensitivity of $\Delta g$ to (i) the definition of the active space and (ii) the number of states included in the QDNEVPT2 effective Hamiltonian.
Representative sets include $p$-shell diatomics with $S = 1/2$ (ZnH, CdH, HgH; ZnF, CdF, HgF; CaH, SrH, BaH), $d$-shell transition metal hydrides (RhH$_2$, IrH$_2$, PdH), and diatomic molecules with $S = 1$ (NCl, NBr, NI).
All diatomic molecules possess the $\Sigma$ ground electronic state.
The RhH$_2$ and IrH$_2$ molecules have the $^2A_1$ ground states.

\textit{1) ZnH, CdH, and HgH.}
\Cref{fig:ZnHCdHHgH_351310}a reports $\Delta g_{\perp}$ for ZnH, CdH, and HgH computed using a small (3e, 5o) active space that incorporates the frontier $s$-orbitals of both atoms and three $p$-orbitals of the metal atom.
For each molecule, the $g$-shifts are plotted as a function of the number of states included in the SO-QDNEVPT2 Hamiltonian, up to the maximum (50) permitted by this active space.
All three molecules exhibit similar trends with increasing $N_\mathrm{state}$, indicating a common underlying electronic mechanism for the $g$-shift.
The curves show pronounced changes when $N_\mathrm{state}$ increases from 5 to 10 and from 25 to 30 with more gradual convergence in other regions of this parameter.

When comparing the performance of different SO-QDNEVPT2 approximations, the results follow the trends observed in \cref{sec:results_and_discussion:benchmark} for $p$-shell molecules.
The computed $g$-shifts are quite sensitive to the choice of the spin--orbit Hamiltonian and the order of SO-QDNEVPT2 approximation (i.e., BP1 vs.\ BP2, DKH1 vs.\ DKH2), particularly for the heavier molecules in this series (CdH, HgH).
For ZnH, all SO-QDNEVPT2 methods predict $g$-shifts within $\sim$ 4 ppt of each other.
As the nuclear charge increases (Zn $<$ Cd $<$ Hg), the BP and DKH results separate, consistent with increasing importance of spin--orbit effects.
For CdH and HgH, the best overall agreement with experiment is shown by DKH2-QDNEVPT2-K.
Notably, the $g$-shifts computed using DKH2-QDNEVPT2-K and the (3e, 5o) active space are most accurate either with very few states ($N_\mathrm{state}\!\approx\!5$) or with the full set of states ($N_\mathrm{state}=50$), whereas intermediate choices show larger errors.

To analyze how computed $g$-shifts depend on the choice of active orbitals, we performed calculations with a larger (13e, 10o) active space that additionally incorporates the occupied $d$-shell of the metal atom.
\cref{fig:ZnHCdHHgH_351310}b shows results including up to 50 reference states.
The DKH2-QDNEVPT2-K data for HgH is presented up to 35 states due to current limitations of our implementation.
We note that for this active space $N_\mathrm{state}$ = 50 no longer represents a complete set of states.
Compared to the (3e, 5o) results, the (13e, 10o) $\Delta g_{\perp}$ curves are generally smoother as a function of $N_\mathrm{state}$, possibly due to a higher density of states.
A notable exception is ZnH, which shows a more pronounced change in the computed $g$-shift value from 25 to 30 states when using a larger active space and the first-order (BP1, DKH1) SO-QDNEVPT2 methods.
Our numerical tests indicate that this sudden change originates from a strong coupling between the ground state and higher-lying configurations describing spin--orbit and dynamic correlation effects.
This analysis is further supported by the results of second-order (BP2, DKH2) methods that treat this coupling at a higher-level of theory and show a smoother dependence on $N_\mathrm{state}$.

For $N_\mathrm{state}\!\approx\!5$, the SO-QDNEVPT2 results depend weakly on the active space, changing by only a few ppt when going from (3e, 5o) to (13e, 10o).
As $N_\mathrm{state}$ increases, the calculations become markedly more sensitive: for most methods, strong active-space dependence is observed when $N_\mathrm{state}\gtrsim 20$.
In this regime, enlarging the active space does not necessarily improve accuracy.
This is most evident for DKH2-QDNEVPT2-K, which performs best with the compact (3e, 5o) space, yet for (13e, 10o) and $N_\mathrm{state}>20$ it can be substantially worse than DKH1-QDNEVPT2-K.
Overall, these trends reflect a known limitation of extensive state averaging: including many states can degrade the SA-CASSCF orbitals and thereby reduce the reliability of subsequent multireference perturbation treatments.\cite{ganyushinFullyVariationalSpinorbit2013,majumderSimulatingSpinOrbit2023}

\begin{figure*}[t!]
   \centering
    \includegraphics[width=\textwidth]{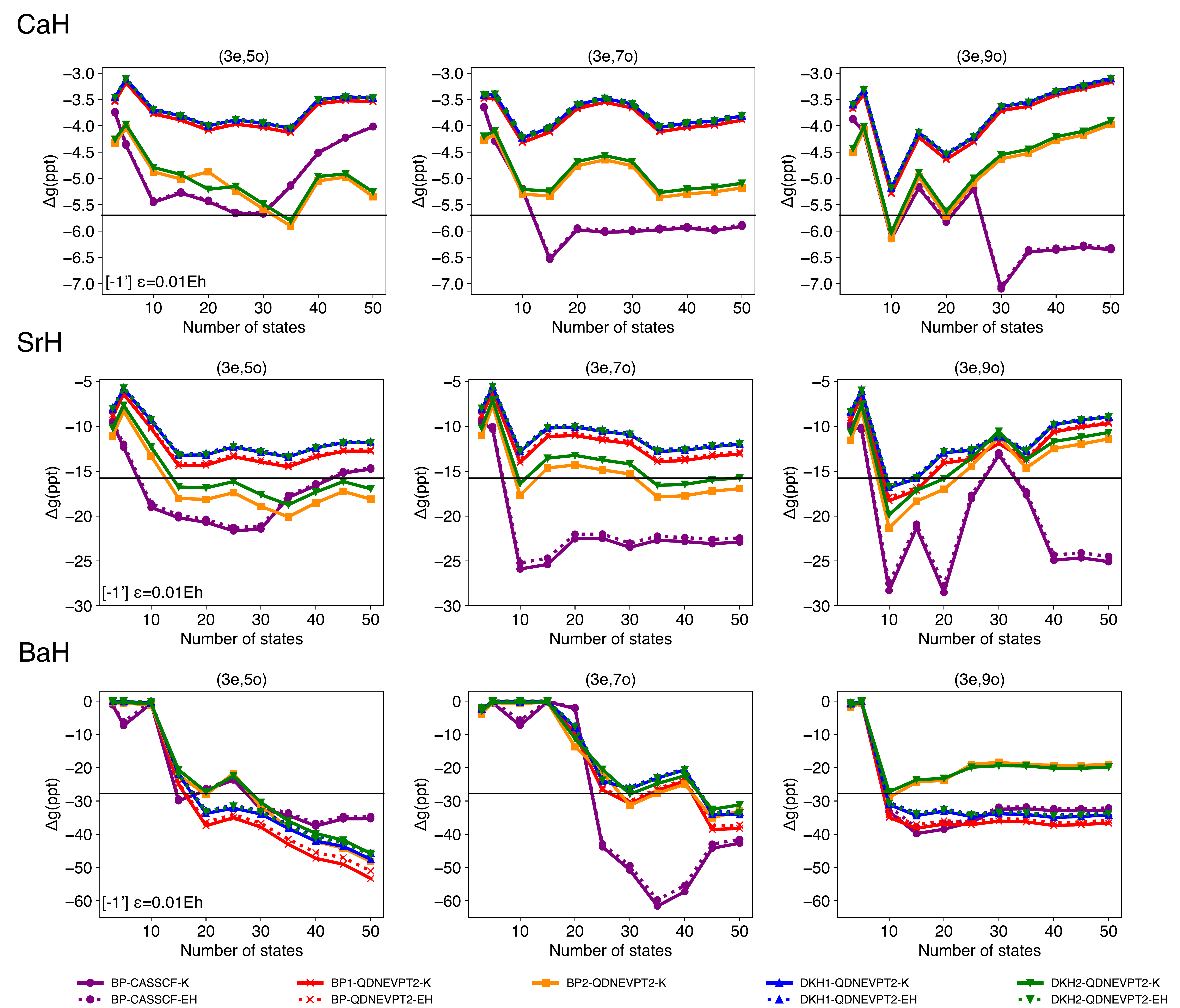}
    \caption{
    	Perpendicular $g$-shift $\Delta g_{\perp}$ (in parts per thousand, ppt) for CaH, SrH, and BaH as a function of the number of states and active space size.
    	Horizontal lines denote the experimental $g$-shift values for comparison (Ref.~\citenum{weltnerMagneticAtomsMolecules1983}).
    	An imaginary level shift $\varepsilon=0.01~\eh$ is applied to the $[-1']$ amplitudes for the (3e, 5o) calculations.
}
\label{fig:CaSrBaH_3579}
\end{figure*}

\textit{2) CaH, SrH, and BaH.}
For these molecules, we examined three active spaces: 
(i) (3e, 5o), comprising three $\sigma$ and two $\pi$ orbitals with dominant contributions from the atomic $s$-functions on both centers and from the metal $p$- and $d_{z^2}$-functions;
(ii) (3e, 7o), obtained by adding two low-lying virtual $\delta_d$-orbitals; and
(iii) (3e, 9o), which further includes two virtual $\pi_d$-orbitals.
Although Ca, Sr, and Ba are formally $s$-block elements, the unpaired electron in their hydrides resides in a $\sigma$-orbital with pronounced metal $d_{z^2}$-character, which makes their $g$-shifts resemble the $d$-shell subset discussed in \cref{sec:results_and_discussion:benchmark}.

The resulting $\Delta g_\perp$ trends are summarized in \cref{fig:CaSrBaH_3579}.
Consistent with $d$-shell behavior, the DKH and BP Hamiltonians yield very similar $g$-shifts for this series, whereas inclusion of second-order spin--orbit terms has a noticeable impact and generally improves agreement for both Hamiltonians.

CaH and SrH exhibit closely matching dependence on the number of included states across all three active spaces, indicating similar low-energy electronic structure.
Unlike the ZnH/CdH/HgH series, using only a few states leads to appreciable errors in $\Delta g_\perp$.
For both CaH and SrH, at least 10 states are needed to improve agreement with experiment.
Increasing $N_\mathrm{state}$ further and enlarging the active space have overall a detrimental effect on the accuracy of computed $g$-shifts.

BaH behaves differently.
With only a few states, $\Delta g_\perp$ is predicted to be nearly zero for all active spaces, producing a large error of $\sim$25~ppt relative to experiment\cite{weltnerMagneticAtomsMolecules1983}.
Convergence is strongly active-space dependent at intermediate $N_\mathrm{state}$: for (3e, 5o) and (3e, 7o), roughly $\sim$20 and $\sim$30 states, respectively, are required to obtain an accurate shift.
In contrast, the largest space (3e, 9o) reaches a stable value already at $\sim$10 states and changes much less upon further increasing $N_\mathrm{state}$.
Overall, as in the ZnH/CdH/HgH series, the choice of active space primarily affects the results in the intermediate range of $N_\mathrm{state}$ (approximately $10 \lesssim N_\mathrm{state} \lesssim 40$).

\begin{figure*} [t!]
   \centering
    \includegraphics[width=\textwidth]{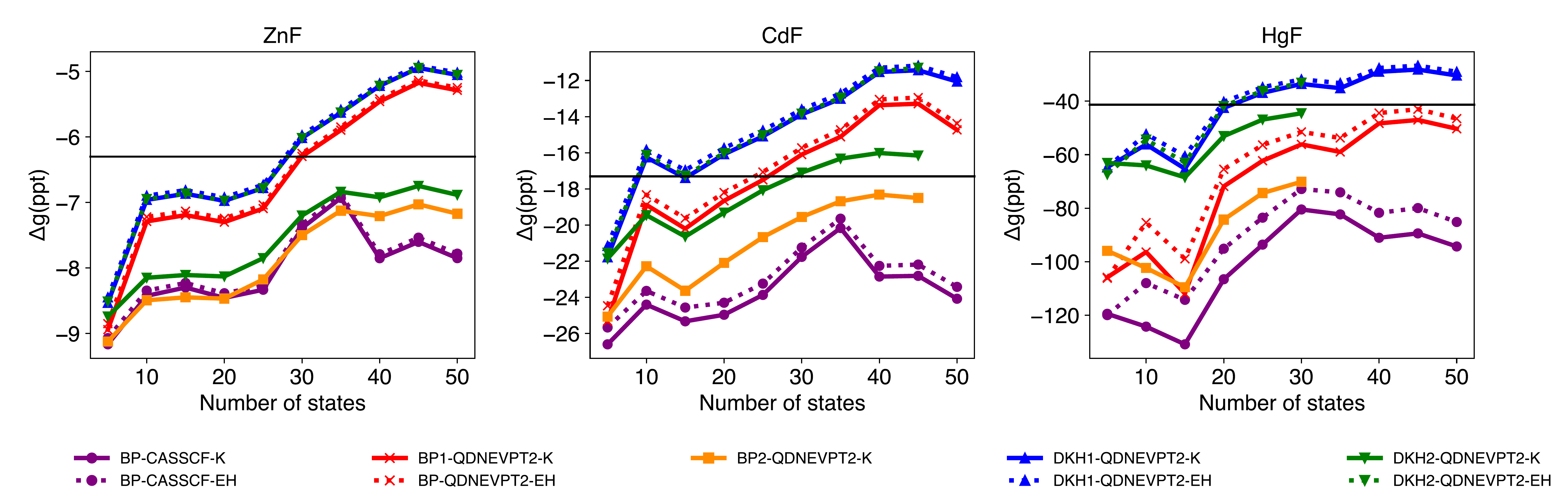}
    \caption{
    	Perpendicular $g$-shift $\Delta g_{\perp}$ (in parts per thousand, ppt) for ZnF, CdF, and HgF as a function of the number of states computed using SO-QDNEVPT2 with the (7e, 10o) active space.
    	Horizontal lines denote the experimental $g$-shift values for comparison (Ref.~\citenum{weltnerMagneticAtomsMolecules1983}).
}
\label{fig:ZnFCdFHgF}
\end{figure*}

\textit{3) ZnF, CdF, HgF.}
Next, we investigate the results for the Zn--Hg fluoride series (\cref{fig:ZnFCdFHgF}) that were computed using the (7e, 10o) active space incorporating the occupied $\sigma$-bond, two F $p$-lone pairs, the singly occupied $\sigma$-antibonding orbital, as well as two $\sigma$ and two $\pi$ virtual orbitals.
As for the hydrides, the $g$-shifts originate from the $p$-shell configurations, and the BP--DKH separation grows with nuclear charge.
However, convergence with respect to $N_\mathrm{state}$ is systematically slower than for ZnH/CdH/HgH: $\Delta g_{\perp}$ decreases more gradually as additional states are included.
This behavior is consistent with a larger number of energetically accessible configurations contributing to the spin--orbit response when both metal and ligand $p$ manifolds are involved.

In contrast to ZnH/CdH/HgH, using only a few states in the effective Hamiltonian leads to substantial errors for all SO-QDNEVPT2 variants in this series. 
The most reliable agreement with experiment is achieved with DKH2-QDNEVPT2-K once a sufficiently large number of states is included ($N_\mathrm{state}\gtrsim 30$).
The BP1-QDNEVPT2-K method also shows a good agreement with experiment for large $N_\mathrm{state}$.
However, for HgF the BP2-QDNEVPT2-K approximation exhibits a larger deviation from experiment, indicating that the apparent accuracy of BP1-QDNEVPT2-K in this series likely arises from fortuitous error cancellation.

\begin{figure*}[t!]
   \centering
    \includegraphics[width=\textwidth]{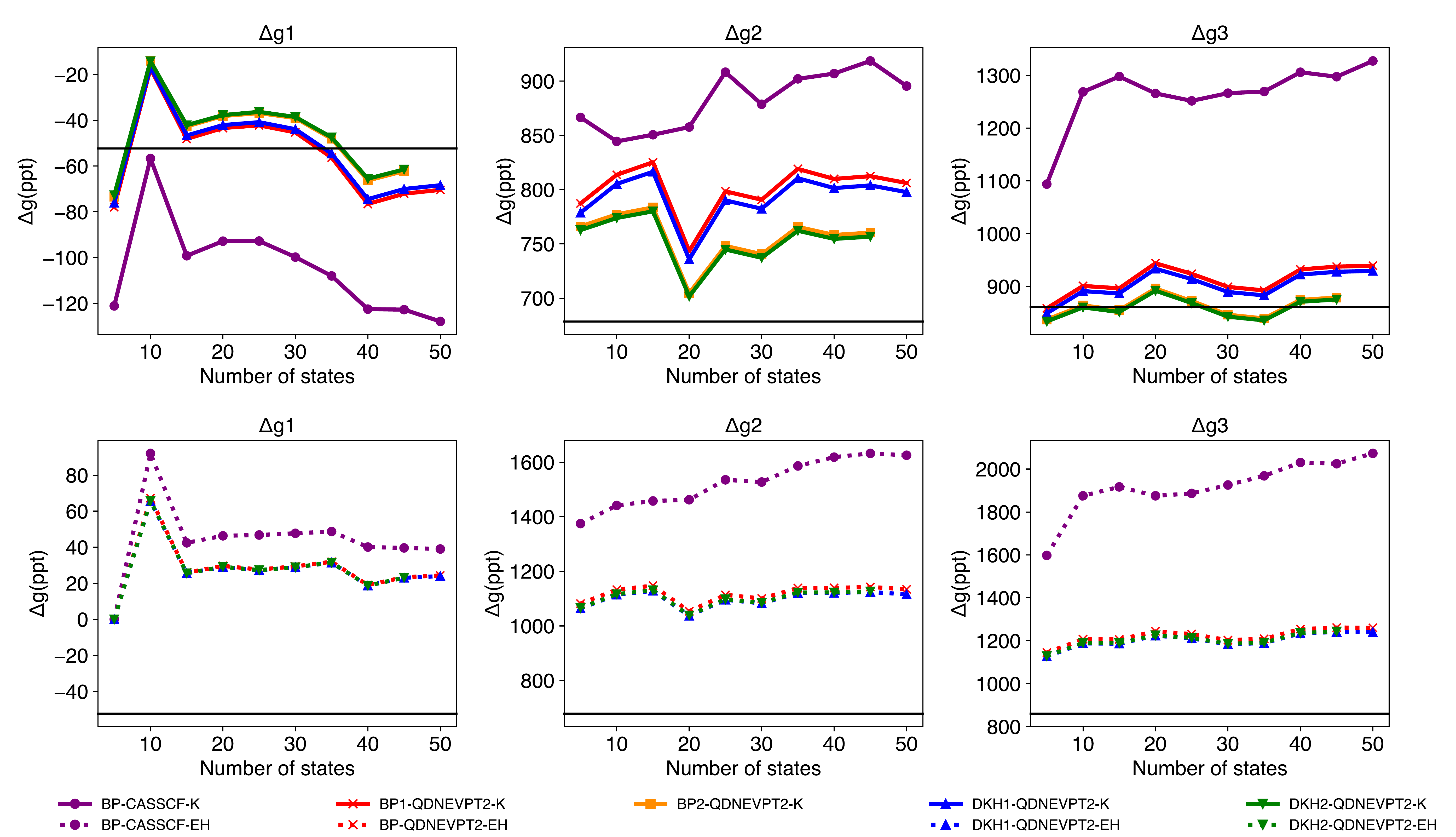}
    \caption{RhH$_2$ $g$-shifts (in parts per thousand, ppt) as functions of the number of states.
    	Horizontal lines denote the experimental $g$-shift values for comparison (Ref.~\citenum{vanzeeElectronspinResonanceCo1992}). 
}
\label{fig:RhH2}
\end{figure*}

\begin{figure*}[t!]
   \centering
    \includegraphics[width=10cm]{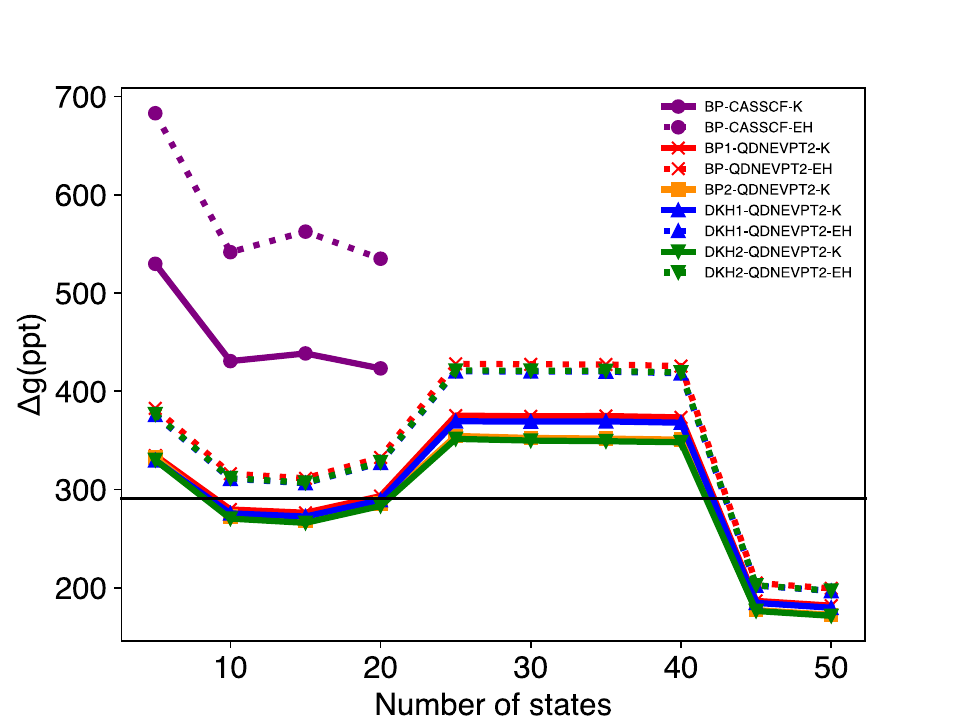}
    \caption{Perpendicular $g$-shift $\Delta g_{\perp}$ (in parts per thousand, ppt) for PdH  as a function of the number of states.
    Horizontal lines denote the experimental $g$-shift values for comparison (Ref.~\citenum{weltnerMagneticAtomsMolecules1983}).
}
\label{fig:PdH}
\end{figure*}

\begin{figure*}[t!]
	\centering
	\includegraphics[width=\textwidth]{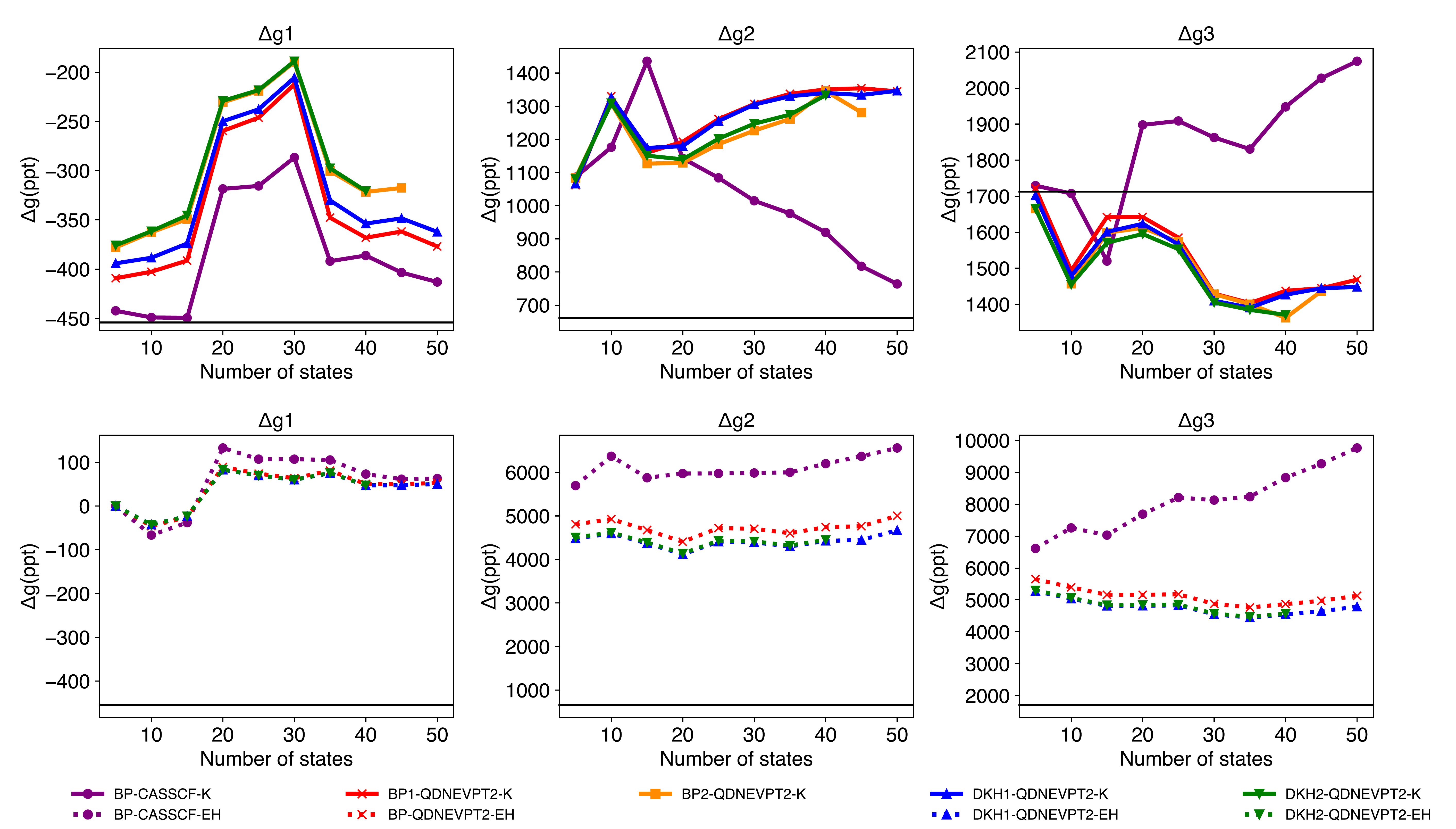}
	\caption{IrH$_2$ $g$-shifts (in parts per thousand, ppt) as functions of the number of states.
		Horizontal lines denote the experimental $g$-shift values for comparison (Ref.~\citenum{vanzeeElectronspinResonanceCo1992}).
	}
	\label{fig:IrH2}
\end{figure*}

\textit{4) RhH$_2$, IrH$_2$, and PdH.} 
We next analyze the transition metal hydrides RhH$_2$, IrH$_2$, and PdH (\cref{fig:RhH2,fig:PdH,fig:IrH2}) with $d$-block elements near the middle of transition metal series.
The calculations employ (11e, 8o) active spaces for RhH$_2$ and IrH$_2$ and (11e, 10o) for PdH (see the Supplementary Material).
In this subset, the dominant spin--orbit response is controlled by metal $d$ orbitals.
Consequently, the Kramers results obtained with BP and DKH are nearly indistinguishable, even for Ir ($Z=77$).
This behavior contrasts with HgH ($Z=80$), where $p$-shell spin--orbit coupling produces a much larger BP--DKH separation, consistent with earlier CASSCF-level analyses.\cite{lanMolecularTensorsAnalytical2015}

As already noted in \cref{sec:results_and_discussion:benchmark}, the EH $g$-tensor approximation fails for RhH$_2$ and IrH$_2$ (\cref{fig:RhH2,fig:IrH2}), producing substantially larger deviations from experiment than the Kramers formalism.
This is expected because $\Delta g$ in these molecules is governed primarily by spin--orbit mixing among excited states rather than by simple ground--excited couplings.
Such excited-state interaction terms are not fully included in the EH expression (\cref{eq:g_Neese}).\cite{lanMolecularTensorsAnalytical2015}

For RhH$_2$ (\cref{fig:RhH2}), all SO-QDNEVPT2-K variants show a broadly similar and relatively smooth dependence on $N_\mathrm{state}$, in contrast to SA-CASSCF-K that exhibits qualitatively different trends as additional states are included.
Incorporating dynamical correlation via QDNEVPT2 significantly improves the agreement with experiment.
The most accurate results are obtained with the second-order variants (BP2- and DKH2-QDNEVPT2-K), whereas the first-order treatments (BP1 and DKH1) exhibit noticeably larger errors in $\Delta g_2$ and $\Delta g_3$ (by $\sim$50~ppt).

For PdH (\cref{fig:PdH}), the SA-CASSCF calculations with 20 or fewer states overestimate $\Delta g_\perp$ by $\sim$ 100 ppt and converge to an incorrect ground state when $N_\mathrm{state}$ $>$ 20.
SO-QDNEVPT2 corrects this qualitative failure and yields the best agreement with experiment for $N_\mathrm{state}\lesssim 20$.
However, including substantially more states produces abrupt changes in $\Delta g_\perp$, consistent with the increasing influence of very high-lying states on orbital optimization and on the structure of the effective Hamiltonian.

IrH$_2$ (\cref{fig:IrH2}) is the heaviest system in this subset and displays the strongest state dependence at the SA-CASSCF level.
Interestingly, SA-CASSCF yields the closest agreement with experiment among the methods tested, while SO-QDNEVPT2 reduces the sensitivity to $N_\mathrm{state}$ but increases the residual error.
All four Kramers SO-QDNEVPT2 variants (BP1, BP2, DKH1, DKH2) perform similarly, suggesting that the remaining discrepancy in $g$-shift is dominated by the treatment of electron correlation rather than spin--orbit coupling.
The fact that SA-CASSCF appears more accurate in this case, yet differs systematically from SO-QDNEVPT2, further points to a significant role of error cancellation in the SA-CASSCF results.

\begin{figure*}[t!]
   \centering
    \includegraphics[width=\textwidth]{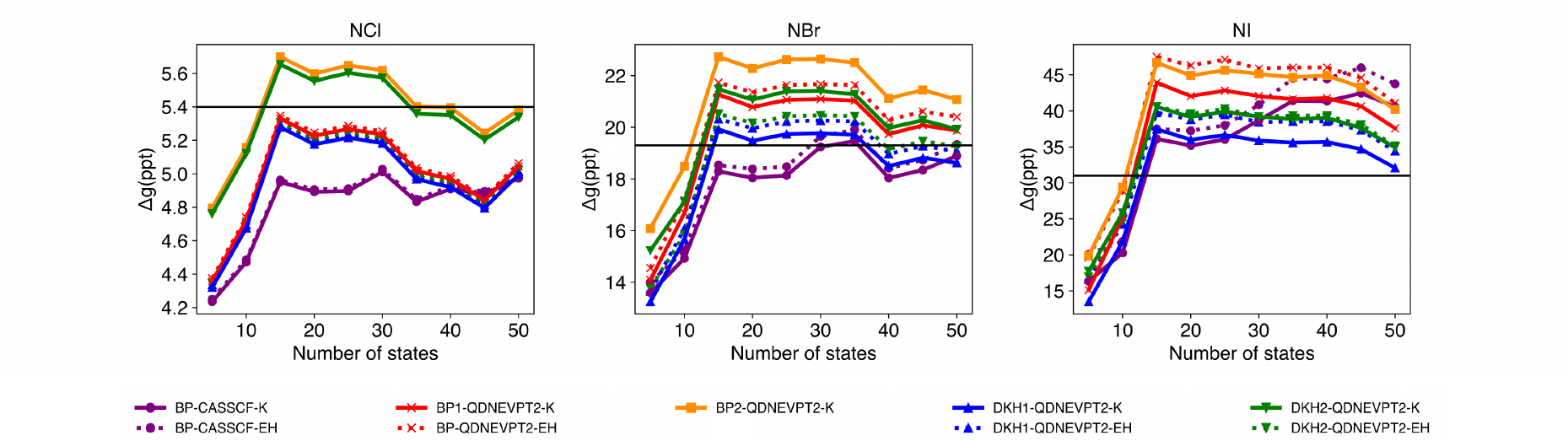}
    \caption{Perpendicular $g$-shift $\Delta g_{\perp}$ (in parts per thousand, ppt) for the triplet radicals NCl, NBr, and NI versus the number of states included in the effective Hamiltonian (active space (12e, 8o), the ANO-RCC basis set).
    Horizontal lines denote the experimental $g$-shift values for comparison (Ref.~\citenum{patchkovskiiCalculationEPRGTensors2001}).
}
\label{fig:triplet}
\end{figure*}

\textit{5) NCl, NBr, and NI.} 
Finally, we consider the diatomic triplet radicals NCl, NBr, and NI, for which $\Delta g_{\perp}$ was computed using the (12e, 8o) active space (\cref{fig:triplet}).
As observed for other $p$-shell molecules in \cref{sec:results_and_discussion:benchmark}, the BP--DKH separation grows with increasing nuclear charge (NCl $<$ NBr $<$ NI).
Using fewer than $\sim$10 states leads to substantial errors in $\Delta g_{\perp}$, whereas the results become essentially stable for $N_\mathrm{state}\gtrsim 15$.
In this converged regime, DKH1- and DKH2-QDNEVPT2-K provide the closest overall agreement with experiment.
Despite the different spin multiplicity, the qualitative behavior across SO-QDNEVPT2 resembles that observed for the ZnF/CdF/HgF series (\cref{fig:ZnFCdFHgF}).

\textit{Summary.}
Reliable calculations of $g$-shifts using SO-QDNEVPT2 require an active space that includes all orbitals essential to the dominant spin--orbit response while avoiding redundant orbitals that destabilize convergence.
In practice, expanding the active space with extra virtual orbitals most often increases the sensitivity to $N_\mathrm{state}$, whereas including chemically important occupied orbitals can reduce this sensitivity and yield smoother trends.
Moreover, taking $N_\mathrm{state}$ too large may introduce diffuse, high-lying states that distort the SA-CASSCF orbital optimization and the resulting effective Hamiltonian, leading to nonmonotonic or noisy $\Delta g$.
Consequently, the optimal balance between active-space size and $N_\mathrm{state}$ is strongly molecule dependent and can vary substantially across different types of electronic structure.

\subsubsection{State-averaging weights}
\label{sec:results_and_discussion:analysis:weights}

\begin{figure*}[t!]
   \centering
    \includegraphics[width=\textwidth]{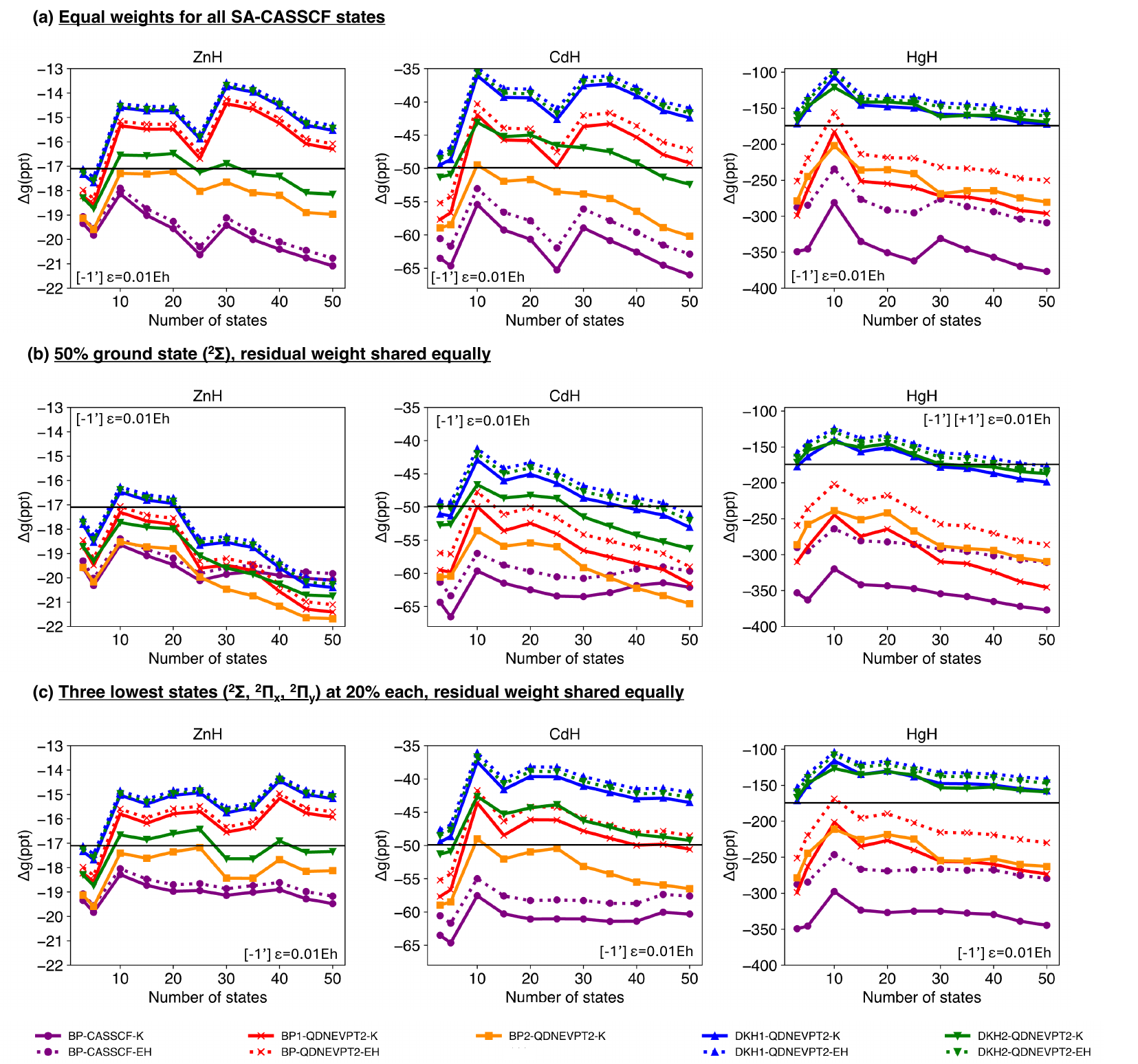}
    \caption{
    	Effect of SA-CASSCF weighting on $\Delta g_{\perp}$ (in parts per thousand, ppt) for ZnH, CdH, and HgH with the (3e, 5o) active space.
    	(a) Equal weights for all states.
    	(b) Ground state ($^{2}\Sigma$) fixed at 50\% weight; remaining states share the residual weight equally.
    	(c) The first three states ($^{2}\Sigma$, $^{2}\Pi_x$, $^{2}\Pi_y$) fixed at 20\% weight each; remaining states weighted equally.
    	Horizontal lines denote the experimental $g$-shift values for comparison (Ref.~\citenum{weltnerMagneticAtomsMolecules1983}).
    	An imaginary shift $\varepsilon=0.01~\eh$ is applied to the $[-1']$ amplitudes in all calculations.
    	For HgH in panel (b), the shift is also applied to the $[+1']$ amplitudes.
    	}
\label{fig:weight}
\end{figure*}

We next investigate the sensitivity of SO-QDNEVPT2 $g$-values to the SA-CASSCF reference weights by focusing on ZnH, CdH, and HgH (\cref{fig:weight}) and the (3e, 5o) active space.
Relative to equal-weight averaging (\cref{fig:weight}a), fixing 50\% of the weight on the $^{2}\Sigma$ ground state and distributing the residual weight among the excited states (\cref{fig:weight}b) stabilizes the convergence of $g$-shifts at the SA-CASSCF level of theory.
However, for SO-QDNEVPT2 this choice leads to a slower convergence with the number of states and less accurate results when all states are included ($N_\mathrm{state}$ = 50), indicating that an overly ground-state-biased reference is unbalanced for the $g$-tensor calculations.
This trend is consistent with the strong contribution of the lowest $^{2}\Pi$ manifold to $\Delta g_{\perp}$ in transition-metal hydrides,\cite{bolvinAlternativeApproachGMatrix2006} which can be significantly reduced by state averaging heavily weighted on the ground state.

\cref{fig:weight}c shows results from a more physically motivated state averaging, which assigns 20\% weight to the ground state and 20\% to each $^{2}\Pi$ component, distributing the remaining weight uniformly.
This scheme yields smoother, more stable behavior for both SA-CASSCF and SO-QDNEVPT2, suggesting that state-averaging should be weighted toward the states that dominate $\Delta g_{\perp}$, especially when a strong dependence on the number of states is observed.

\subsubsection{Gauge-origin error} 
\label{sec:results_and_discussion:analysis:gauge}

\begin{figure*}[t!]
   \centering
    \includegraphics[width=\textwidth]{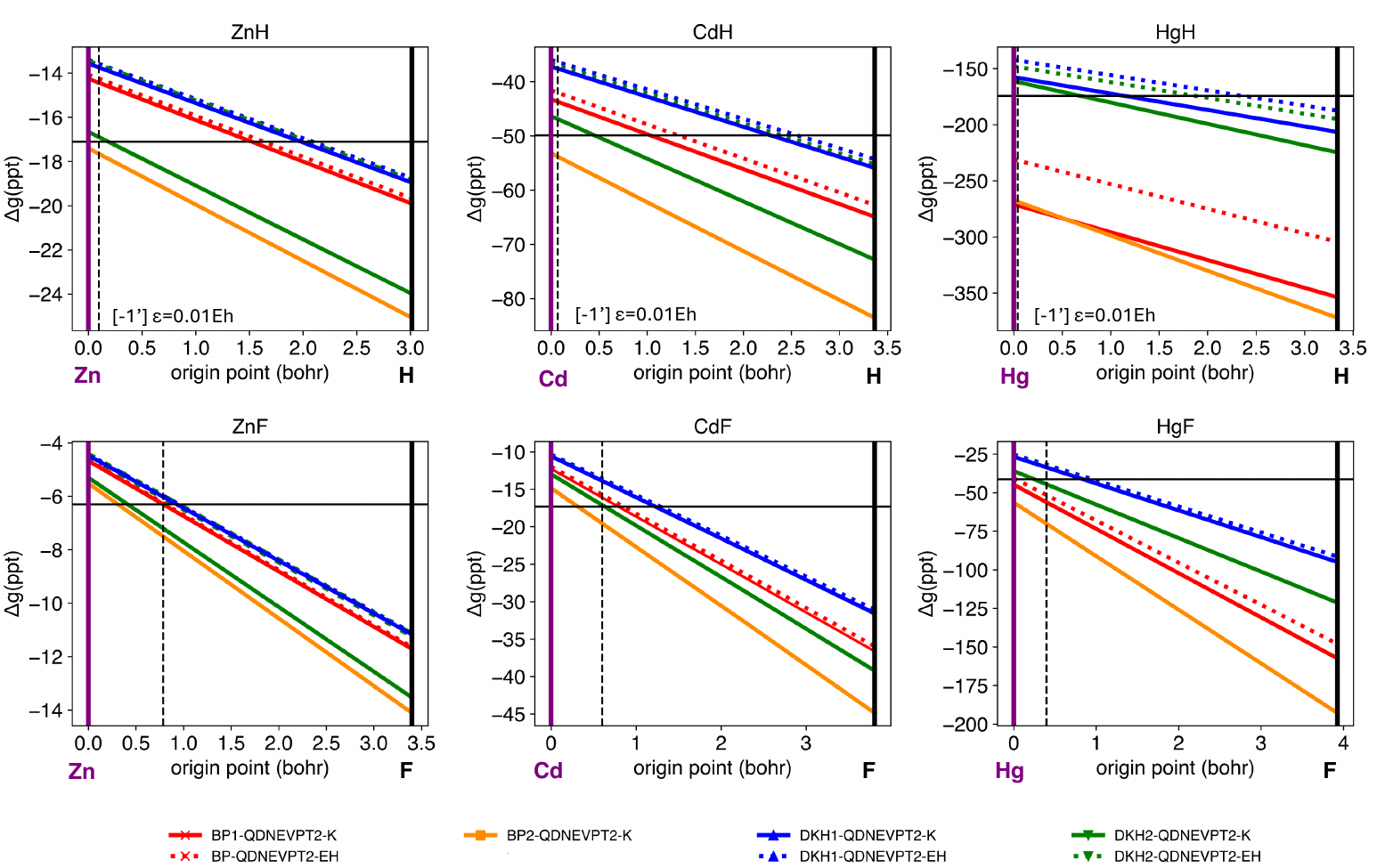}
    \caption{
Coordinate system origin dependence of $\Delta g_{\perp}$ (in parts per thousand, ppt) for XH and XF (X = Zn, Cd, and Hg) computed using (3e, 5o) and (7e, 10o) active space, respectively (ANO-RCC basis set, 30 equally-averaged states).
The origin is shifted along the bond axis from the metal atom (set to 0 $a_0$) toward the ligand. 
The vertical dashed line marks the center of nuclear charge.
Horizontal lines denote the experimental $g$-shift values for comparison (Ref.~\citenum{weltnerMagneticAtomsMolecules1983}).
An imaginary shift $\varepsilon=0.01~\eh$ is applied to the $[-1']$ amplitudes for ZnH, CdH, and HgH.
}
\label{fig:gauge}
\end{figure*}

With finite basis sets, computed $g$-tensors can retain a residual dependence on the coordinate system origin.
To quantify this effect, we shift the origin along the molecular axis for XH and XF (X = Zn, Cd, Hg), holding the metal atom at position 0, and compute the $\Delta g_{\perp}$-values using the ANO-RCC basis set (\cref{fig:gauge}).
We employ the (3e, 5o) and (7e, 10o) active spaces for XH and XF, respectively, and 30 equally-averaged states. 
Placing the origin on the metal systematically reduces $\Delta g_{\perp}$, whereas moving it toward the ligand increases $\Delta g_{\perp}$.

When employing standard Gaussian basis sets, using the center of nuclear charge is common in $g$-tensor calculations.\cite{luzanovGaugeinvariantCalculationsMagnetic1994,bolvinAlternativeApproachGMatrix2006,vancoillieCalculationEPRTensors2007}
\cref{fig:gauge} demonstrates that the center of nuclear charge (vertical dashed line) is indeed a convenient choice, yielding expected convergence with the level of theory and consistent performance across the series of six molecules.
Using the highest-level SO-QDNEVPT2 approximation, DKH2-QDNEVPT2-K, the $g$-values computed within 0.5 $a_0$ of the center of nuclear charge show the best agreement with experimental results.

\subsubsection{Basis set effects} 
\label{sec:results_and_discussion:analysis:basis}

\begin{figure*}[t!]
	\centering
	\includegraphics[width=0.97\textwidth]{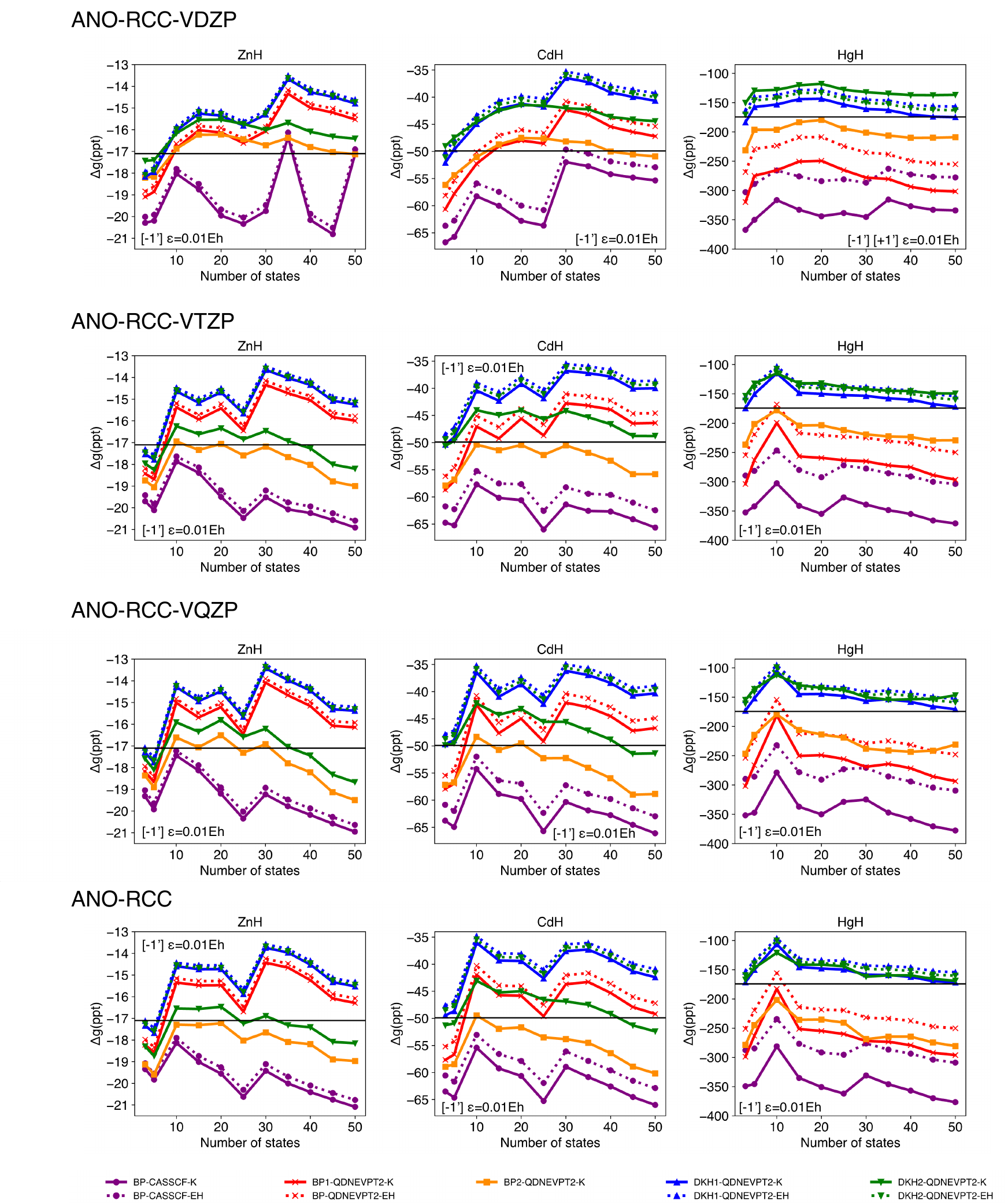}
	\caption{
	The perpendicular $g$-shifts $\Delta g_{\perp }$ (in parts per thousand, ppt) of ZnH, CdH, and HgH computed using the ANO-RCC-VXZP (X = D, T, Q) and uncontracted ANO-RCC basis sets as a function of the number of states.
	All calculations employed the (3e, 5o) active space.
	Horizontal lines denote experimental $g$-shift values for comparison (Ref.~\citenum{weltnerMagneticAtomsMolecules1983}).
	An imaginary shift $\varepsilon=0.01~\eh$ is applied to the $[-1']$ amplitudes for all molecules. 
	For HgH with ANO-RCC-VDZP, the shift is also applied to the $[+1']$ amplitudes.
	}
	\label{fig:ANO_basis}
\end{figure*}

We finally assess the basis set dependence of SO-QDNEVPT2 $g$-factors.
\Cref{fig:ANO_basis} compares ZnH, CdH, and HgH results obtained with the (3e, 5o) active space using the full ANO-RCC basis and the truncated ANO-RCC-VXZP (X = D, T, Q) sets.
We use the ANO-RCC calculations as a reference to quantify errors introduced by basis set truncation in ANO-RCC-VXZP.

As expected, ANO-RCC-VDZP shows the largest deviations from a more complete ANO-RCC.
These differences are modest when only a few states are included ($N_\mathrm{state}\le 10$) but grow systematically as additional states are incorporated, indicating that the truncated double-$\zeta$ basis becomes increasingly inadequate for describing the higher-lying configurations that contribute to $\Delta g_{\perp}$ at larger $N_\mathrm{state}$.
Moving to ANO-RCC-VTZP substantially improves agreement for all three molecules.
For ZnH and CdH, the VTZP curves closely track the ANO-RCC reference across the full $N_\mathrm{state}$ range.
For HgH, however, noticeable basis set errors remain at the triple-$\zeta$ level, reflecting the more stringent basis set requirements imposed by heavy-element spin--orbit coupling.

Further increasing the basis to ANO-RCC-VQZP brings ZnH and CdH results essentially to the uncontracted limit, but small yet discernible differences persist for HgH, most clearly for the second-order spin--orbit treatments (BP2 and DKH2).
The weaker basis sensitivity of the first-order variants (BP1 and DKH1) suggests that the more pronounced basis set dependence in BP2 and DKH2 likely originates from the second-order spin--orbit coupling contributions that are most important for heavier elements.

Overall, our results indicate that ANO-RCC-VTZP is generally sufficient for reliable $g$-tensor predictions in molecules with lighter members.
For heavy-element compounds, at least quadruple-$\zeta$ quality or, ideally, the full ANO-RCC basis should be used, especially with the second-order (BP2/DKH2) SO-QDNEVPT2 approximations.

\section{Conclusions}
\label{sec:conclusions}

In this work, we developed and benchmarked spin--orbit quasidegenerate second-order N-electron valence perturbation theory (SO-QDNEVPT2) for computing $g$-tensors of open-shell chemical systems. 
The SO-QDNEVPT2 framework provides accurate and efficient description of electron correlation and relativistic effects, consistently incorporating dynamic correlation and spin--orbit coupling to multireference second order using Breit--Pauli (BP) and exact two-component Douglas--Kroll--Hess (DKH) two-component Hamiltonians.
We implemented two strategies for computing $g$-tensors using SO-QDNEVPT2, namely: a spin-free effective-Hamiltonian (EH) treatment based on second-order response expressions and a Kramers (K) treatment that extracts $g$ directly from spin-mixed SO-QDNEVPT2 states (commonly referred to as quasidegenerate perturbation theory). 
To assess the accuracy of SO-QDNEVPT2 $g$-tensors, we assembled a benchmark set of 23 molecules spanning from low- to high-spin species, diatomics to small polyatomics, and $g$-shifts from near zero to the strong spin--orbit regime.
Our results demonstrate that spin-free and spin--orbit QDNEVPT2 calculations with many states included are susceptible to intruder-state problems, which can be effectively mitigated using level-shift or renormalization group techniques.

Across the benchmark, SO-QDNEVPT2 significantly improves agreement with experiment relative to state-averaged CASSCF while preserving the qualitative trends dictated by the dominant orbital character of the spin--orbit response. 
For modest $g$-shifts ($|\Delta g|\lesssim 100$~parts per thousand), the EH and K formalisms yield similar results.
However, as $|\Delta g|$ grows, the EH approximation deteriorates systematically, and the Kramers approach becomes essential. 
The choice of spin--orbit Hamiltonian is most consequential for molecules with unpaired electrons in predominantly $p$-shell orbitals: exact two-component DKH-based treatments track experimental trends substantially better than BP as the nuclear charge and $|\Delta g|$ increase. 
In contrast, for $d$-open-shell molecules, BP and DKH produce comparable $g$-values even for heavy elements, consistent with a reduced sensitivity of diffuse $d$-orbitals to spin--orbit coupling. 
The impact of including second-order spin--orbit contributions is generally modest for weak-to-intermediate coupling but becomes significant for strongly spin--orbit-mixed $p$-shell molecules, providing substantial improvements over traditional state-interaction first-order treatments.

To provide practical guidance for routine SO-QDNEVPT2 $g$-tensor calculations, we performed a thorough analysis of how the computed $g$-values depend on active space, number of states, state-averaging weights, coordinate system origin, and basis set. 
Our results demonstrate that all of these parameters can significantly affect the accuracy of SO-QDNEVPT2 $g$-values.
Reliable $g$-predictions usually require an active space that captures the orbitals responsible for the dominant spin--orbit response while avoiding redundant orbitals that destabilize state averaging. 
Including the low-energy states that substantially contribute to $\Delta g$ improves the accuracy of SO-QDNEVPT2 predictions.
However, incorporating highly excited, diffuse states distort state-averaged orbital optimization and the effective Hamiltonian, producing noisy or nonmonotonic trends affected by intruder-state problems. 
Increasing the state-average weights on the first few low-lying states that dominate the magnetic response (rather than heavily favoring the ground state alone) can help stabilize convergence with the number of states in pathological cases.
Finally, using the basis sets of at least triple-$\zeta$ quality and placing the origin of coordinate system near the center of nuclear charge is necessary to reduce the basis set incompleteness and gauge origin error.
For heavier elements (6th period or above), uncontracted or quadruple-$\zeta$ basis sets should be ideally used.

Our results establish SO-QDNEVPT2 as a practical and accurate framework for computing $g$-tensors in open-shell systems where electron correlation and relativistic effects are both important.
Looking forward, several extensions will further strengthen the predictive power and scope of this approach. 
Immediate priorities include improving the efficiency and robustness of SO-QDNEVPT2 implementation for larger active spaces, reducing residual origin sensitivity by adopting gauge-invariant atomic orbitals, and automating active space and state selection for complex chemical systems. 
From a methodological perspective, incorporating higher-level spin--orbit and dynamic correlation treatments and providing access to a wider range of magnetic properties would provide a more complete toolkit for simulating modern spectroscopies. 
Together, these developments will enable routine and accurate simulations of magnetic observables in a wide range of chemical systems.

\section{Supplementary Material}
See the Supplementary Material for tables with all $g$-shift data and plots depicting molecular orbitals selected in the active space for each molecule.

\section{Acknowledgements}
This work was supported by the National Science Foundation under Grant No. CHE-2044648.
Computations were performed at the Ohio Supercomputer Center under projects PAS1963.\cite{osc1987} 
The authors thank James D.\@ Serna for insightful discussion.

\section{Data availability}
The data that supports the findings of this study are available within the article and its supplementary material. 
Additional data can be made available upon reasonable request.


\begin{thebibliography}{133}%
\makeatletter
\providecommand \@ifxundefined [1]{%
 \@ifx{#1\undefined}
}%
\providecommand \@ifnum [1]{%
 \ifnum #1\expandafter \@firstoftwo
 \else \expandafter \@secondoftwo
 \fi
}%
\providecommand \@ifx [1]{%
 \ifx #1\expandafter \@firstoftwo
 \else \expandafter \@secondoftwo
 \fi
}%
\providecommand \natexlab [1]{#1}%
\providecommand \enquote  [1]{``#1''}%
\providecommand \bibnamefont  [1]{#1}%
\providecommand \bibfnamefont [1]{#1}%
\providecommand \citenamefont [1]{#1}%
\providecommand \href@noop [0]{\@secondoftwo}%
\providecommand \href [0]{\begingroup \@sanitize@url \@href}%
\providecommand \@href[1]{\@@startlink{#1}\@@href}%
\providecommand \@@href[1]{\endgroup#1\@@endlink}%
\providecommand \@sanitize@url [0]{\catcode `\\12\catcode `\$12\catcode
  `\&12\catcode `\#12\catcode `\^12\catcode `\_12\catcode `\%12\relax}%
\providecommand \@@startlink[1]{}%
\providecommand \@@endlink[0]{}%
\providecommand \url  [0]{\begingroup\@sanitize@url \@url }%
\providecommand \@url [1]{\endgroup\@href {#1}{\urlprefix }}%
\providecommand \urlprefix  [0]{URL }%
\providecommand \Eprint [0]{\href }%
\providecommand \doibase [0]{https://doi.org/}%
\providecommand \selectlanguage [0]{\@gobble}%
\providecommand \bibinfo  [0]{\@secondoftwo}%
\providecommand \bibfield  [0]{\@secondoftwo}%
\providecommand \translation [1]{[#1]}%
\providecommand \BibitemOpen [0]{}%
\providecommand \bibitemStop [0]{}%
\providecommand \bibitemNoStop [0]{.\EOS\space}%
\providecommand \EOS [0]{\spacefactor3000\relax}%
\providecommand \BibitemShut  [1]{\csname bibitem#1\endcsname}%
\let\auto@bib@innerbib\@empty
\bibitem [{\citenamefont {Abragam}\ and\ \citenamefont
  {Bleaney}(2012)}]{abragamElectronParamagneticResonance2012}%
  \BibitemOpen
  \bibfield  {author} {\bibinfo {author} {\bibfnamefont {A.}~\bibnamefont
  {Abragam}}\ and\ \bibinfo {author} {\bibfnamefont {B.}~\bibnamefont
  {Bleaney}},\ }\href@noop {} {\emph {\bibinfo {title} {Electron Paramagnetic
  Resonance of Transition Ions}}}\ (\bibinfo  {publisher} {Oxford University
  Press},\ \bibinfo {address} {Oxford},\ \bibinfo {year} {2012})\BibitemShut
  {NoStop}%
\bibitem [{\citenamefont {Wolfowicz}\ \emph {et~al.}(2021)\citenamefont
  {Wolfowicz}, \citenamefont {Heremans}, \citenamefont {Anderson},
  \citenamefont {Kanai}, \citenamefont {Seo}, \citenamefont {Gali},
  \citenamefont {Galli},\ and\ \citenamefont
  {Awschalom}}]{wolfowiczQuantumGuidelinesSolidstate2021}%
  \BibitemOpen
  \bibfield  {author} {\bibinfo {author} {\bibfnamefont {G.}~\bibnamefont
  {Wolfowicz}}, \bibinfo {author} {\bibfnamefont {F.~J.}\ \bibnamefont
  {Heremans}}, \bibinfo {author} {\bibfnamefont {C.~P.}\ \bibnamefont
  {Anderson}}, \bibinfo {author} {\bibfnamefont {S.}~\bibnamefont {Kanai}},
  \bibinfo {author} {\bibfnamefont {H.}~\bibnamefont {Seo}}, \bibinfo {author}
  {\bibfnamefont {A.}~\bibnamefont {Gali}}, \bibinfo {author} {\bibfnamefont
  {G.}~\bibnamefont {Galli}},\ and\ \bibinfo {author} {\bibfnamefont {D.~D.}\
  \bibnamefont {Awschalom}},\ }\href
  {https://doi.org/10.1038/s41578-021-00306-y} {\bibfield  {journal} {\bibinfo
  {journal} {Nat Rev Mater}\ }\textbf {\bibinfo {volume} {6}},\ \bibinfo
  {pages} {906} (\bibinfo {year} {2021})}\BibitemShut {NoStop}%
\bibitem [{\citenamefont {Briganti}\ \emph {et~al.}(2021)\citenamefont
  {Briganti}, \citenamefont {Lucaccini}, \citenamefont {Chelazzi},
  \citenamefont {Ciattini}, \citenamefont {Sorace}, \citenamefont {Sessoli},
  \citenamefont {Totti},\ and\ \citenamefont
  {Perfetti}}]{brigantiMagneticAnisotropyTrends2021}%
  \BibitemOpen
  \bibfield  {author} {\bibinfo {author} {\bibfnamefont {M.}~\bibnamefont
  {Briganti}}, \bibinfo {author} {\bibfnamefont {E.}~\bibnamefont {Lucaccini}},
  \bibinfo {author} {\bibfnamefont {L.}~\bibnamefont {Chelazzi}}, \bibinfo
  {author} {\bibfnamefont {S.}~\bibnamefont {Ciattini}}, \bibinfo {author}
  {\bibfnamefont {L.}~\bibnamefont {Sorace}}, \bibinfo {author} {\bibfnamefont
  {R.}~\bibnamefont {Sessoli}}, \bibinfo {author} {\bibfnamefont
  {F.}~\bibnamefont {Totti}},\ and\ \bibinfo {author} {\bibfnamefont
  {M.}~\bibnamefont {Perfetti}},\ }\href {https://doi.org/10.1021/jacs.1c02502}
  {\bibfield  {journal} {\bibinfo  {journal} {J. Am. Chem. Soc.}\ }\textbf
  {\bibinfo {volume} {143}},\ \bibinfo {pages} {8108} (\bibinfo {year}
  {2021})}\BibitemShut {NoStop}%
\bibitem [{\citenamefont {Gatteschi}, \citenamefont {Sessoli},\ and\
  \citenamefont {Villain}(2006)}]{gatteschiMolecularNanomagnets2006}%
  \BibitemOpen
  \bibfield  {author} {\bibinfo {author} {\bibfnamefont {D.}~\bibnamefont
  {Gatteschi}}, \bibinfo {author} {\bibfnamefont {R.}~\bibnamefont {Sessoli}},\
  and\ \bibinfo {author} {\bibfnamefont {J.}~\bibnamefont {Villain}},\
  }\href@noop {} {\emph {\bibinfo {title} {Molecular Nanomagnets}}},\ \bibinfo
  {number} {5}\ (\bibinfo  {publisher} {Oxford University Press},\ \bibinfo
  {address} {Oxford ; New York},\ \bibinfo {year} {2006})\BibitemShut {NoStop}%
\bibitem [{\citenamefont {Kahn}(1993)}]{kahnMolecularMagnetism1993}%
  \BibitemOpen
  \bibfield  {author} {\bibinfo {author} {\bibfnamefont {O.}~\bibnamefont
  {Kahn}},\ }\href@noop {} {\emph {\bibinfo {title} {Molecular Magnetism}}}\
  (\bibinfo  {publisher} {VCH},\ \bibinfo {address} {New York},\ \bibinfo
  {year} {1993})\BibitemShut {NoStop}%
\bibitem [{\citenamefont {Hendrickx}\ \emph {et~al.}(2024)\citenamefont
  {Hendrickx}, \citenamefont {Massai}, \citenamefont {Mergenthaler},
  \citenamefont {Schupp}, \citenamefont {Paredes}, \citenamefont {Bedell},
  \citenamefont {Salis},\ and\ \citenamefont
  {Fuhrer}}]{hendrickxSweetspotOperationGermanium2024}%
  \BibitemOpen
  \bibfield  {author} {\bibinfo {author} {\bibfnamefont {N.~W.}\ \bibnamefont
  {Hendrickx}}, \bibinfo {author} {\bibfnamefont {L.}~\bibnamefont {Massai}},
  \bibinfo {author} {\bibfnamefont {M.}~\bibnamefont {Mergenthaler}}, \bibinfo
  {author} {\bibfnamefont {F.~J.}\ \bibnamefont {Schupp}}, \bibinfo {author}
  {\bibfnamefont {S.}~\bibnamefont {Paredes}}, \bibinfo {author} {\bibfnamefont
  {S.~W.}\ \bibnamefont {Bedell}}, \bibinfo {author} {\bibfnamefont
  {G.}~\bibnamefont {Salis}},\ and\ \bibinfo {author} {\bibfnamefont
  {A.}~\bibnamefont {Fuhrer}},\ }\href
  {https://doi.org/10.1038/s41563-024-01857-5} {\bibfield  {journal} {\bibinfo
  {journal} {Nat. Mater.}\ }\textbf {\bibinfo {volume} {23}},\ \bibinfo {pages}
  {920} (\bibinfo {year} {2024})}\BibitemShut {NoStop}%
\bibitem [{\citenamefont {Gali}(2023)}]{galiRecentAdvancesInitio2023}%
  \BibitemOpen
  \bibfield  {author} {\bibinfo {author} {\bibfnamefont {{\'A}.}~\bibnamefont
  {Gali}},\ }\href {https://doi.org/10.1515/nanoph-2022-0723} {\bibfield
  {journal} {\bibinfo  {journal} {Nanophotonics}\ }\textbf {\bibinfo {volume}
  {12}},\ \bibinfo {pages} {359} (\bibinfo {year} {2023})}\BibitemShut
  {NoStop}%
\bibitem [{\citenamefont {Ray}\ \emph {et~al.}(2023)\citenamefont {Ray},
  \citenamefont {Oakley}, \citenamefont {Sarkar}, \citenamefont {Bai},\ and\
  \citenamefont
  {Gagliardi}}]{rayTheoreticalInvestigationSingleMoleculeMagnet2023}%
  \BibitemOpen
  \bibfield  {author} {\bibinfo {author} {\bibfnamefont {D.}~\bibnamefont
  {Ray}}, \bibinfo {author} {\bibfnamefont {M.~S.}\ \bibnamefont {Oakley}},
  \bibinfo {author} {\bibfnamefont {A.}~\bibnamefont {Sarkar}}, \bibinfo
  {author} {\bibfnamefont {X.}~\bibnamefont {Bai}},\ and\ \bibinfo {author}
  {\bibfnamefont {L.}~\bibnamefont {Gagliardi}},\ }\href
  {https://doi.org/10.1021/acs.inorgchem.2c04013} {\bibfield  {journal}
  {\bibinfo  {journal} {Inorg. Chem.}\ }\textbf {\bibinfo {volume} {62}},\
  \bibinfo {pages} {1649} (\bibinfo {year} {2023})}\BibitemShut {NoStop}%
\bibitem [{\citenamefont {Iv{\'a}dy}, \citenamefont {Abrikosov},\ and\
  \citenamefont {Gali}(2018)}]{ivadyFirstPrinciplesCalculation2018}%
  \BibitemOpen
  \bibfield  {author} {\bibinfo {author} {\bibfnamefont {V.}~\bibnamefont
  {Iv{\'a}dy}}, \bibinfo {author} {\bibfnamefont {I.~A.}\ \bibnamefont
  {Abrikosov}},\ and\ \bibinfo {author} {\bibfnamefont {A.}~\bibnamefont
  {Gali}},\ }\href {https://doi.org/10.1038/s41524-018-0132-5} {\bibfield
  {journal} {\bibinfo  {journal} {Comput. Mater.}\ }\textbf {\bibinfo {volume}
  {4}},\ \bibinfo {pages} {76} (\bibinfo {year} {2018})}\BibitemShut {NoStop}%
\bibitem [{\citenamefont {Guo}\ \emph {et~al.}(2022)\citenamefont {Guo},
  \citenamefont {He}, \citenamefont {Huang}, \citenamefont {Giblin},
  \citenamefont {Billington}, \citenamefont {Heinemann}, \citenamefont {Tong},
  \citenamefont {Mansikkam{\"a}ki},\ and\ \citenamefont
  {Layfield}}]{guoDiscoveryDysprosiumMetallocene2022}%
  \BibitemOpen
  \bibfield  {author} {\bibinfo {author} {\bibfnamefont {F.-S.}\ \bibnamefont
  {Guo}}, \bibinfo {author} {\bibfnamefont {M.}~\bibnamefont {He}}, \bibinfo
  {author} {\bibfnamefont {G.-Z.}\ \bibnamefont {Huang}}, \bibinfo {author}
  {\bibfnamefont {S.~R.}\ \bibnamefont {Giblin}}, \bibinfo {author}
  {\bibfnamefont {D.}~\bibnamefont {Billington}}, \bibinfo {author}
  {\bibfnamefont {F.~W.}\ \bibnamefont {Heinemann}}, \bibinfo {author}
  {\bibfnamefont {M.-L.}\ \bibnamefont {Tong}}, \bibinfo {author}
  {\bibfnamefont {A.}~\bibnamefont {Mansikkam{\"a}ki}},\ and\ \bibinfo {author}
  {\bibfnamefont {R.~A.}\ \bibnamefont {Layfield}},\ }\href
  {https://doi.org/10.1021/acs.inorgchem.1c03980} {\bibfield  {journal}
  {\bibinfo  {journal} {Inorg. Chem.}\ }\textbf {\bibinfo {volume} {61}},\
  \bibinfo {pages} {6017} (\bibinfo {year} {2022})}\BibitemShut {NoStop}%
\bibitem [{\citenamefont {Zadrozny}\ \emph {et~al.}(2013)\citenamefont
  {Zadrozny}, \citenamefont {Atanasov}, \citenamefont {Bryan}, \citenamefont
  {Lin}, \citenamefont {Rekken}, \citenamefont {Power}, \citenamefont {Neese},\
  and\ \citenamefont {Long}}]{zadroznySlowMagnetizationDynamics2013}%
  \BibitemOpen
  \bibfield  {author} {\bibinfo {author} {\bibfnamefont {J.~M.}\ \bibnamefont
  {Zadrozny}}, \bibinfo {author} {\bibfnamefont {M.}~\bibnamefont {Atanasov}},
  \bibinfo {author} {\bibfnamefont {A.~M.}\ \bibnamefont {Bryan}}, \bibinfo
  {author} {\bibfnamefont {C.-Y.}\ \bibnamefont {Lin}}, \bibinfo {author}
  {\bibfnamefont {B.~D.}\ \bibnamefont {Rekken}}, \bibinfo {author}
  {\bibfnamefont {P.~P.}\ \bibnamefont {Power}}, \bibinfo {author}
  {\bibfnamefont {F.}~\bibnamefont {Neese}},\ and\ \bibinfo {author}
  {\bibfnamefont {J.~R.}\ \bibnamefont {Long}},\ }\href
  {https://doi.org/10.1039/C2SC20801F} {\bibfield  {journal} {\bibinfo
  {journal} {Chem. Sci.}\ }\textbf {\bibinfo {volume} {4}},\ \bibinfo {pages}
  {125} (\bibinfo {year} {2013})}\BibitemShut {NoStop}%
\bibitem [{\citenamefont {Bodenstein}\ \emph {et~al.}(2022)\citenamefont
  {Bodenstein}, \citenamefont {Heimermann}, \citenamefont {Fink},\ and\
  \citenamefont
  {Van~W{\"u}llen}}]{bodensteinDevelopmentApplicationComplete2022}%
  \BibitemOpen
  \bibfield  {author} {\bibinfo {author} {\bibfnamefont {T.}~\bibnamefont
  {Bodenstein}}, \bibinfo {author} {\bibfnamefont {A.}~\bibnamefont
  {Heimermann}}, \bibinfo {author} {\bibfnamefont {K.}~\bibnamefont {Fink}},\
  and\ \bibinfo {author} {\bibfnamefont {C.}~\bibnamefont {Van~W{\"u}llen}},\
  }\href {https://doi.org/10.1002/cphc.202100648} {\bibfield  {journal}
  {\bibinfo  {journal} {ChemPhysChem}\ }\textbf {\bibinfo {volume} {23}},\
  \bibinfo {pages} {e202100648} (\bibinfo {year} {2022})}\BibitemShut {NoStop}%
\bibitem [{\citenamefont {Birnoschi}\ \emph {et~al.}(2024)\citenamefont
  {Birnoschi}, \citenamefont {Oakley}, \citenamefont {McInnes},\ and\
  \citenamefont {Chilton}}]{birnoschiRelativisticQuantumChemical2024}%
  \BibitemOpen
  \bibfield  {author} {\bibinfo {author} {\bibfnamefont {L.}~\bibnamefont
  {Birnoschi}}, \bibinfo {author} {\bibfnamefont {M.~S.}\ \bibnamefont
  {Oakley}}, \bibinfo {author} {\bibfnamefont {E.~J.~L.}\ \bibnamefont
  {McInnes}},\ and\ \bibinfo {author} {\bibfnamefont {N.~F.}\ \bibnamefont
  {Chilton}},\ }\href {https://doi.org/10.1021/jacs.4c01930} {\bibfield
  {journal} {\bibinfo  {journal} {J. Am. Chem. Soc.}\ }\textbf {\bibinfo
  {volume} {146}},\ \bibinfo {pages} {14660} (\bibinfo {year}
  {2024})}\BibitemShut {NoStop}%
\bibitem [{\citenamefont
  {Cheng}(2023)}]{chengRelativisticExactTwocomponent2023}%
  \BibitemOpen
  \bibfield  {author} {\bibinfo {author} {\bibfnamefont {L.}~\bibnamefont
  {Cheng}},\ }\href {https://doi.org/10.1080/00268976.2022.2113567} {\bibfield
  {journal} {\bibinfo  {journal} {Mol. Phys.}\ }\textbf {\bibinfo {volume}
  {121}},\ \bibinfo {pages} {e2113567} (\bibinfo {year} {2023})}\BibitemShut
  {NoStop}%
\bibitem [{\citenamefont {Truhlar}\ and\ \citenamefont
  {Li}(2025)}]{truhlarIntroductionRelativisticElectronic2025}%
  \BibitemOpen
  \bibfield  {author} {\bibinfo {author} {\bibfnamefont {D.~G.}\ \bibnamefont
  {Truhlar}}\ and\ \bibinfo {author} {\bibfnamefont {X.}~\bibnamefont {Li}},\
  }\href {https://doi.org/10.1021/acs.jpca.5c00859} {\bibfield  {journal}
  {\bibinfo  {journal} {J. Phys. Chem. A}\ }\textbf {\bibinfo {volume} {129}},\
  \bibinfo {pages} {4301} (\bibinfo {year} {2025})}\BibitemShut {NoStop}%
\bibitem [{\citenamefont {Liu}(2010)}]{liuIdeasRelativisticQuantum2010}%
  \BibitemOpen
  \bibfield  {author} {\bibinfo {author} {\bibfnamefont {W.}~\bibnamefont
  {Liu}},\ }\href {https://doi.org/10.1080/00268971003781571} {\bibfield
  {journal} {\bibinfo  {journal} {Mol. Phys.}\ }\textbf {\bibinfo {volume}
  {108}},\ \bibinfo {pages} {1679} (\bibinfo {year} {2010})}\BibitemShut
  {NoStop}%
\bibitem [{\citenamefont {Ganyushin}\ and\ \citenamefont
  {Neese}(2013)}]{ganyushinFullyVariationalSpinorbit2013}%
  \BibitemOpen
  \bibfield  {author} {\bibinfo {author} {\bibfnamefont {D.}~\bibnamefont
  {Ganyushin}}\ and\ \bibinfo {author} {\bibfnamefont {F.}~\bibnamefont
  {Neese}},\ }\href {https://doi.org/10.1063/1.4793736} {\bibfield  {journal}
  {\bibinfo  {journal} {J. Chem. Phys.}\ }\textbf {\bibinfo {volume} {138}},\
  \bibinfo {pages} {104113} (\bibinfo {year} {2013})}\BibitemShut {NoStop}%
\bibitem [{\citenamefont {Jenkins}\ \emph {et~al.}(2019)\citenamefont
  {Jenkins}, \citenamefont {Liu}, \citenamefont {Kasper}, \citenamefont
  {Frisch},\ and\ \citenamefont
  {Li}}]{jenkinsVariationalRelativisticTwoComponent2019}%
  \BibitemOpen
  \bibfield  {author} {\bibinfo {author} {\bibfnamefont {A.~J.}\ \bibnamefont
  {Jenkins}}, \bibinfo {author} {\bibfnamefont {H.}~\bibnamefont {Liu}},
  \bibinfo {author} {\bibfnamefont {J.~M.}\ \bibnamefont {Kasper}}, \bibinfo
  {author} {\bibfnamefont {M.~J.}\ \bibnamefont {Frisch}},\ and\ \bibinfo
  {author} {\bibfnamefont {X.}~\bibnamefont {Li}},\ }\href
  {https://doi.org/10.1021/acs.jctc.9b00011} {\bibfield  {journal} {\bibinfo
  {journal} {J. Chem. Theory Comput.}\ }\textbf {\bibinfo {volume} {15}},\
  \bibinfo {pages} {2974} (\bibinfo {year} {2019})}\BibitemShut {NoStop}%
\bibitem [{\citenamefont {Jorgen Aa.~Jensen}\ \emph {et~al.}(1996)\citenamefont
  {Jorgen Aa.~Jensen}, \citenamefont {Dyall}, \citenamefont {Saue},\ and\
  \citenamefont
  {F{\ae}gri}}]{jorgenaa.jensenRelativisticFourcomponentMulticonfigurational1996}%
  \BibitemOpen
  \bibfield  {author} {\bibinfo {author} {\bibfnamefont {H.}~\bibnamefont
  {Jorgen Aa.~Jensen}}, \bibinfo {author} {\bibfnamefont {K.~G.}\ \bibnamefont
  {Dyall}}, \bibinfo {author} {\bibfnamefont {T.}~\bibnamefont {Saue}},\ and\
  \bibinfo {author} {\bibfnamefont {K.}~\bibnamefont {F{\ae}gri}, \bibfnamefont
  {Jr.}},\ }\href {https://doi.org/10.1063/1.471644} {\bibfield  {journal}
  {\bibinfo  {journal} {J. Chem. Phys.}\ }\textbf {\bibinfo {volume} {104}},\
  \bibinfo {pages} {4083} (\bibinfo {year} {1996})}\BibitemShut {NoStop}%
\bibitem [{\citenamefont {Reynolds}\ and\ \citenamefont
  {Shiozaki}(2019)}]{reynoldsZeroFieldSplittingParameters2019}%
  \BibitemOpen
  \bibfield  {author} {\bibinfo {author} {\bibfnamefont {R.~D.}\ \bibnamefont
  {Reynolds}}\ and\ \bibinfo {author} {\bibfnamefont {T.}~\bibnamefont
  {Shiozaki}},\ }\href {https://doi.org/10.1021/acs.jctc.8b00910} {\bibfield
  {journal} {\bibinfo  {journal} {J. Chem. Theory Comput.}\ }\textbf {\bibinfo
  {volume} {15}},\ \bibinfo {pages} {1560} (\bibinfo {year}
  {2019})}\BibitemShut {NoStop}%
\bibitem [{\citenamefont {Shiozaki}\ and\ \citenamefont
  {Mizukami}(2015)}]{shiozakiRelativisticInternallyContracted2015}%
  \BibitemOpen
  \bibfield  {author} {\bibinfo {author} {\bibfnamefont {T.}~\bibnamefont
  {Shiozaki}}\ and\ \bibinfo {author} {\bibfnamefont {W.}~\bibnamefont
  {Mizukami}},\ }\href {https://doi.org/10.1021/acs.jctc.5b00754} {\bibfield
  {journal} {\bibinfo  {journal} {J. Chem. Theory Comput.}\ }\textbf {\bibinfo
  {volume} {11}},\ \bibinfo {pages} {4733} (\bibinfo {year}
  {2015})}\BibitemShut {NoStop}%
\bibitem [{\citenamefont {Tang}, \citenamefont {Sun},\ and\ \citenamefont
  {Li}(2024)}]{tangExactTwoComponentCompleteActive2024}%
  \BibitemOpen
  \bibfield  {author} {\bibinfo {author} {\bibfnamefont {D.}~\bibnamefont
  {Tang}}, \bibinfo {author} {\bibfnamefont {S.}~\bibnamefont {Sun}},\ and\
  \bibinfo {author} {\bibfnamefont {X.}~\bibnamefont {Li}},\ }\href
  {https://doi.org/10.1021/acs.jctc.4c01028} {\bibfield  {journal} {\bibinfo
  {journal} {J. Chem. Theory Comput.}\ }\textbf {\bibinfo {volume} {20}},\
  \bibinfo {pages} {9917} (\bibinfo {year} {2024})}\BibitemShut {NoStop}%
\bibitem [{\citenamefont {Hoyer}\ \emph {et~al.}(2023)\citenamefont {Hoyer},
  \citenamefont {Lu}, \citenamefont {Hu}, \citenamefont {Shumilov},
  \citenamefont {Sun}, \citenamefont {Knecht},\ and\ \citenamefont
  {Li}}]{hoyerCorrelatedDiracCoulomb2023}%
  \BibitemOpen
  \bibfield  {author} {\bibinfo {author} {\bibfnamefont {C.~E.}\ \bibnamefont
  {Hoyer}}, \bibinfo {author} {\bibfnamefont {L.}~\bibnamefont {Lu}}, \bibinfo
  {author} {\bibfnamefont {H.}~\bibnamefont {Hu}}, \bibinfo {author}
  {\bibfnamefont {K.~D.}\ \bibnamefont {Shumilov}}, \bibinfo {author}
  {\bibfnamefont {S.}~\bibnamefont {Sun}}, \bibinfo {author} {\bibfnamefont
  {S.}~\bibnamefont {Knecht}},\ and\ \bibinfo {author} {\bibfnamefont
  {X.}~\bibnamefont {Li}},\ }\href {https://doi.org/10.1063/5.0133741}
  {\bibfield  {journal} {\bibinfo  {journal} {J. Chem. Phys.}\ }\textbf
  {\bibinfo {volume} {158}},\ \bibinfo {pages} {044101} (\bibinfo {year}
  {2023})}\BibitemShut {NoStop}%
\bibitem [{\citenamefont {Liu}\ and\ \citenamefont
  {Cheng}(2021)}]{liuRelativisticCoupledclusterEquationofmotion2021}%
  \BibitemOpen
  \bibfield  {author} {\bibinfo {author} {\bibfnamefont {J.}~\bibnamefont
  {Liu}}\ and\ \bibinfo {author} {\bibfnamefont {L.}~\bibnamefont {Cheng}},\
  }\href {https://doi.org/10.1002/wcms.1536} {\bibfield  {journal} {\bibinfo
  {journal} {WIREs Comput. Mol. Sci.}\ }\textbf {\bibinfo {volume} {11}},\
  \bibinfo {pages} {e1536} (\bibinfo {year} {2021})}\BibitemShut {NoStop}%
\bibitem [{\citenamefont {Wu}\ \emph {et~al.}(2025)\citenamefont {Wu},
  \citenamefont {Lin}, \citenamefont {Wang}, \citenamefont {Wang},
  \citenamefont {Southworth}, \citenamefont {Doumy}, \citenamefont {Young},\
  and\ \citenamefont {Cheng}}]{wuRelativisticCoreValence2025}%
  \BibitemOpen
  \bibfield  {author} {\bibinfo {author} {\bibfnamefont {Y.}~\bibnamefont
  {Wu}}, \bibinfo {author} {\bibfnamefont {Z.}~\bibnamefont {Lin}}, \bibinfo
  {author} {\bibfnamefont {X.}~\bibnamefont {Wang}}, \bibinfo {author}
  {\bibfnamefont {S.}~\bibnamefont {Wang}}, \bibinfo {author} {\bibfnamefont
  {S.~H.}\ \bibnamefont {Southworth}}, \bibinfo {author} {\bibfnamefont
  {G.}~\bibnamefont {Doumy}}, \bibinfo {author} {\bibfnamefont
  {L.}~\bibnamefont {Young}},\ and\ \bibinfo {author} {\bibfnamefont
  {L.}~\bibnamefont {Cheng}},\ }\href {https://doi.org/10.1063/5.0300670}
  {\bibfield  {journal} {\bibinfo  {journal} {J. Chem. Phys.}\ }\textbf
  {\bibinfo {volume} {163}},\ \bibinfo {pages} {244114} (\bibinfo {year}
  {2025})}\BibitemShut {NoStop}%
\bibitem [{\citenamefont {Hu}\ \emph {et~al.}(2020)\citenamefont {Hu},
  \citenamefont {Jenkins}, \citenamefont {Liu}, \citenamefont {Kasper},
  \citenamefont {Frisch},\ and\ \citenamefont
  {Li}}]{huRelativisticTwoComponentMultireference2020}%
  \BibitemOpen
  \bibfield  {author} {\bibinfo {author} {\bibfnamefont {H.}~\bibnamefont
  {Hu}}, \bibinfo {author} {\bibfnamefont {A.~J.}\ \bibnamefont {Jenkins}},
  \bibinfo {author} {\bibfnamefont {H.}~\bibnamefont {Liu}}, \bibinfo {author}
  {\bibfnamefont {J.~M.}\ \bibnamefont {Kasper}}, \bibinfo {author}
  {\bibfnamefont {M.~J.}\ \bibnamefont {Frisch}},\ and\ \bibinfo {author}
  {\bibfnamefont {X.}~\bibnamefont {Li}},\ }\href
  {https://doi.org/10.1021/acs.jctc.9b01290} {\bibfield  {journal} {\bibinfo
  {journal} {J. Chem. Theory Comput.}\ }\textbf {\bibinfo {volume} {16}},\
  \bibinfo {pages} {2975} (\bibinfo {year} {2020})}\BibitemShut {NoStop}%
\bibitem [{\citenamefont {Lu}\ \emph {et~al.}(2022)\citenamefont {Lu},
  \citenamefont {Hu}, \citenamefont {Jenkins},\ and\ \citenamefont
  {Li}}]{luExactTwoComponentRelativisticMultireference2022}%
  \BibitemOpen
  \bibfield  {author} {\bibinfo {author} {\bibfnamefont {L.}~\bibnamefont
  {Lu}}, \bibinfo {author} {\bibfnamefont {H.}~\bibnamefont {Hu}}, \bibinfo
  {author} {\bibfnamefont {A.~J.}\ \bibnamefont {Jenkins}},\ and\ \bibinfo
  {author} {\bibfnamefont {X.}~\bibnamefont {Li}},\ }\href
  {https://doi.org/10.1021/acs.jctc.2c00171} {\bibfield  {journal} {\bibinfo
  {journal} {J. Chem. Theory Comput.}\ }\textbf {\bibinfo {volume} {18}},\
  \bibinfo {pages} {2983} (\bibinfo {year} {2022})}\BibitemShut {NoStop}%
\bibitem [{\citenamefont {Singh}, \citenamefont {Atanasov},\ and\ \citenamefont
  {Neese}(2018)}]{singhChallengesMultireferencePerturbation2018}%
  \BibitemOpen
  \bibfield  {author} {\bibinfo {author} {\bibfnamefont {S.~K.}\ \bibnamefont
  {Singh}}, \bibinfo {author} {\bibfnamefont {M.}~\bibnamefont {Atanasov}},\
  and\ \bibinfo {author} {\bibfnamefont {F.}~\bibnamefont {Neese}},\ }\href
  {https://doi.org/10.1021/acs.jctc.8b00513} {\bibfield  {journal} {\bibinfo
  {journal} {J. Chem. Theory Comput.}\ }\textbf {\bibinfo {volume} {14}},\
  \bibinfo {pages} {4662} (\bibinfo {year} {2018})}\BibitemShut {NoStop}%
\bibitem [{\citenamefont {Tran}\ and\ \citenamefont
  {Neese}(2020)}]{tranDoublehybridDensityFunctional2020}%
  \BibitemOpen
  \bibfield  {author} {\bibinfo {author} {\bibfnamefont {V.~A.}\ \bibnamefont
  {Tran}}\ and\ \bibinfo {author} {\bibfnamefont {F.}~\bibnamefont {Neese}},\
  }\href {https://doi.org/10.1063/5.0013799} {\bibfield  {journal} {\bibinfo
  {journal} {J. Chem. Phys.}\ }\textbf {\bibinfo {volume} {153}},\ \bibinfo
  {pages} {054105} (\bibinfo {year} {2020})}\BibitemShut {NoStop}%
\bibitem [{\citenamefont {Cheng}\ and\ \citenamefont
  {Gauss}(2014)}]{chengPerturbativeTreatmentSpinorbit2014}%
  \BibitemOpen
  \bibfield  {author} {\bibinfo {author} {\bibfnamefont {L.}~\bibnamefont
  {Cheng}}\ and\ \bibinfo {author} {\bibfnamefont {J.}~\bibnamefont {Gauss}},\
  }\href {https://doi.org/10.1063/1.4897254} {\bibfield  {journal} {\bibinfo
  {journal} {J. Chem. Phys.}\ }\textbf {\bibinfo {volume} {141}},\ \bibinfo
  {pages} {164107} (\bibinfo {year} {2014})}\BibitemShut {NoStop}%
\bibitem [{\citenamefont {Cheng}\ \emph {et~al.}(2018)\citenamefont {Cheng},
  \citenamefont {Wang}, \citenamefont {Stanton},\ and\ \citenamefont
  {Gauss}}]{chengPerturbativeTreatmentSpinorbitcoupling2018}%
  \BibitemOpen
  \bibfield  {author} {\bibinfo {author} {\bibfnamefont {L.}~\bibnamefont
  {Cheng}}, \bibinfo {author} {\bibfnamefont {F.}~\bibnamefont {Wang}},
  \bibinfo {author} {\bibfnamefont {J.~F.}\ \bibnamefont {Stanton}},\ and\
  \bibinfo {author} {\bibfnamefont {J.}~\bibnamefont {Gauss}},\ }\href
  {https://doi.org/10.1063/1.5012041} {\bibfield  {journal} {\bibinfo
  {journal} {J. Chem. Phys.}\ }\textbf {\bibinfo {volume} {148}},\ \bibinfo
  {pages} {044108} (\bibinfo {year} {2018})}\BibitemShut {NoStop}%
\bibitem [{\citenamefont {Roos}\ and\ \citenamefont
  {Malmqvist}(2004)}]{roosRelativisticQuantumChemistry2004}%
  \BibitemOpen
  \bibfield  {author} {\bibinfo {author} {\bibfnamefont {B.~O.}\ \bibnamefont
  {Roos}}\ and\ \bibinfo {author} {\bibfnamefont {P.}~\bibnamefont
  {Malmqvist}},\ }\href {https://doi.org/10.1039/B401472N} {\bibfield
  {journal} {\bibinfo  {journal} {Phys. Chem. Chem. Phys.}\ }\textbf {\bibinfo
  {volume} {6}},\ \bibinfo {pages} {2919} (\bibinfo {year} {2004})}\BibitemShut
  {NoStop}%
\bibitem [{\citenamefont {Gerloch}\ and\ \citenamefont
  {McMeeking}(1975)}]{gerlochParamagneticPropertiesUnsymmetricaI1975}%
  \BibitemOpen
  \bibfield  {author} {\bibinfo {author} {\bibfnamefont {M.}~\bibnamefont
  {Gerloch}}\ and\ \bibinfo {author} {\bibfnamefont {R.~F.}\ \bibnamefont
  {McMeeking}},\ }\href@noop {} {\bibfield  {journal} {\bibinfo  {journal} {J.
  Chem. Soc. Dalton Trans.}\ ,\ \bibinfo {pages} {2443}} (\bibinfo {year}
  {1975})}\BibitemShut {NoStop}%
\bibitem [{\citenamefont {Malmqvist}\ and\ \citenamefont
  {Roos}(1989)}]{malmqvistCASSCFStateInteraction1989}%
  \BibitemOpen
  \bibfield  {author} {\bibinfo {author} {\bibfnamefont {P.-{\AA}.}\
  \bibnamefont {Malmqvist}}\ and\ \bibinfo {author} {\bibfnamefont {B.~O.}\
  \bibnamefont {Roos}},\ }\href {https://doi.org/10.1016/0009-2614(89)85347-3}
  {\bibfield  {journal} {\bibinfo  {journal} {Chem. Phys. Lett.}\ }\textbf
  {\bibinfo {volume} {155}},\ \bibinfo {pages} {189} (\bibinfo {year}
  {1989})}\BibitemShut {NoStop}%
\bibitem [{\citenamefont {Malmqvist}, \citenamefont {Roos},\ and\ \citenamefont
  {Schimmelpfennig}(2002)}]{malmqvistRestrictedActiveSpace2002}%
  \BibitemOpen
  \bibfield  {author} {\bibinfo {author} {\bibfnamefont {P.~{\AA}.}\
  \bibnamefont {Malmqvist}}, \bibinfo {author} {\bibfnamefont {B.~O.}\
  \bibnamefont {Roos}},\ and\ \bibinfo {author} {\bibfnamefont
  {B.}~\bibnamefont {Schimmelpfennig}},\ }\href
  {https://doi.org/10.1016/S0009-2614(02)00498-0} {\bibfield  {journal}
  {\bibinfo  {journal} {Chem. Phys. Lett.}\ }\textbf {\bibinfo {volume}
  {357}},\ \bibinfo {pages} {230} (\bibinfo {year} {2002})}\BibitemShut
  {NoStop}%
\bibitem [{\citenamefont
  {Bolvin}(2006)}]{bolvinAlternativeApproachGMatrix2006}%
  \BibitemOpen
  \bibfield  {author} {\bibinfo {author} {\bibfnamefont {H.}~\bibnamefont
  {Bolvin}},\ }\href {https://doi.org/10.1002/cphc.200600051} {\bibfield
  {journal} {\bibinfo  {journal} {ChemPhysChem}\ }\textbf {\bibinfo {volume}
  {7}},\ \bibinfo {pages} {1575} (\bibinfo {year} {2006})}\BibitemShut
  {NoStop}%
\bibitem [{\citenamefont {Tatchen}, \citenamefont {Kleinschmidt},\ and\
  \citenamefont {Marian}(2009)}]{tatchenCalculatingElectronParamagnetic2009}%
  \BibitemOpen
  \bibfield  {author} {\bibinfo {author} {\bibfnamefont {J.}~\bibnamefont
  {Tatchen}}, \bibinfo {author} {\bibfnamefont {M.}~\bibnamefont
  {Kleinschmidt}},\ and\ \bibinfo {author} {\bibfnamefont {C.~M.}\ \bibnamefont
  {Marian}},\ }\href {https://doi.org/10.1063/1.3115965} {\bibfield  {journal}
  {\bibinfo  {journal} {J. Chem. Phys.}\ }\textbf {\bibinfo {volume} {130}},\
  \bibinfo {pages} {154106} (\bibinfo {year} {2009})}\BibitemShut {NoStop}%
\bibitem [{\citenamefont
  {Neese}(2005)}]{neeseEfficientAccurateApproximations2005}%
  \BibitemOpen
  \bibfield  {author} {\bibinfo {author} {\bibfnamefont {F.}~\bibnamefont
  {Neese}},\ }\href {https://doi.org/10.1063/1.1829047} {\bibfield  {journal}
  {\bibinfo  {journal} {J. Chem. Phys.}\ }\textbf {\bibinfo {volume} {122}},\
  \bibinfo {pages} {034107} (\bibinfo {year} {2005})}\BibitemShut {NoStop}%
\bibitem [{\citenamefont {Alessio}\ and\ \citenamefont
  {Krylov}(2021)}]{alessioEquationofMotionCoupledClusterProtocol2021}%
  \BibitemOpen
  \bibfield  {author} {\bibinfo {author} {\bibfnamefont {M.}~\bibnamefont
  {Alessio}}\ and\ \bibinfo {author} {\bibfnamefont {A.~I.}\ \bibnamefont
  {Krylov}},\ }\href {https://doi.org/10.1021/acs.jctc.1c00430} {\bibfield
  {journal} {\bibinfo  {journal} {J. Chem. Theory Comput.}\ }\textbf {\bibinfo
  {volume} {17}},\ \bibinfo {pages} {4225} (\bibinfo {year}
  {2021})}\BibitemShut {NoStop}%
\bibitem [{\citenamefont {Alessio}\ \emph {et~al.}(2023)\citenamefont
  {Alessio}, \citenamefont {Kotaru}, \citenamefont {Giudetti},\ and\
  \citenamefont {Krylov}}]{alessioOriginMagneticAnisotropy2023}%
  \BibitemOpen
  \bibfield  {author} {\bibinfo {author} {\bibfnamefont {M.}~\bibnamefont
  {Alessio}}, \bibinfo {author} {\bibfnamefont {S.}~\bibnamefont {Kotaru}},
  \bibinfo {author} {\bibfnamefont {G.}~\bibnamefont {Giudetti}},\ and\
  \bibinfo {author} {\bibfnamefont {A.~I.}\ \bibnamefont {Krylov}},\ }\href
  {https://doi.org/10.1021/acs.jpcc.2c05940} {\bibfield  {journal} {\bibinfo
  {journal} {J. Phys. Chem. C}\ }\textbf {\bibinfo {volume} {127}},\ \bibinfo
  {pages} {3647} (\bibinfo {year} {2023})}\BibitemShut {NoStop}%
\bibitem [{\citenamefont {Zhou}\ \emph {et~al.}(2021)\citenamefont {Zhou},
  \citenamefont {Wu}, \citenamefont {Gagliardi},\ and\ \citenamefont
  {Truhlar}}]{zhouCalculationZeemanEffect2021}%
  \BibitemOpen
  \bibfield  {author} {\bibinfo {author} {\bibfnamefont {C.}~\bibnamefont
  {Zhou}}, \bibinfo {author} {\bibfnamefont {D.}~\bibnamefont {Wu}}, \bibinfo
  {author} {\bibfnamefont {L.}~\bibnamefont {Gagliardi}},\ and\ \bibinfo
  {author} {\bibfnamefont {D.~G.}\ \bibnamefont {Truhlar}},\ }\href
  {https://doi.org/10.1021/acs.jctc.1c00208} {\bibfield  {journal} {\bibinfo
  {journal} {J. Chem. Theory Comput.}\ }\textbf {\bibinfo {volume} {17}},\
  \bibinfo {pages} {5050} (\bibinfo {year} {2021})}\BibitemShut {NoStop}%
\bibitem [{\citenamefont {{Cebreiro-Gallardo}}\ and\ \citenamefont
  {Casanova}(2025{\natexlab{a}})}]{cebreiro-gallardoStateInteractionApproachGMatrix2025}%
  \BibitemOpen
  \bibfield  {author} {\bibinfo {author} {\bibfnamefont {A.}~\bibnamefont
  {{Cebreiro-Gallardo}}}\ and\ \bibinfo {author} {\bibfnamefont
  {D.}~\bibnamefont {Casanova}},\ }\href
  {https://doi.org/10.1021/acs.jctc.5c00514} {\bibfield  {journal} {\bibinfo
  {journal} {J. Chem. Theory Comput.}\ }\textbf {\bibinfo {volume} {21}},\
  \bibinfo {pages} {6528} (\bibinfo {year} {2025}{\natexlab{a}})}\BibitemShut
  {NoStop}%
\bibitem [{\citenamefont {Jangid}\ \emph {et~al.}(2026)\citenamefont {Jangid},
  \citenamefont {Hennefarth}, \citenamefont {Hermes}, \citenamefont {Truhlar},\
  and\ \citenamefont {Gagliardi}}]{jangidLinearizedPairDensityFunctional2026}%
  \BibitemOpen
  \bibfield  {author} {\bibinfo {author} {\bibfnamefont {B.}~\bibnamefont
  {Jangid}}, \bibinfo {author} {\bibfnamefont {M.~R.}\ \bibnamefont
  {Hennefarth}}, \bibinfo {author} {\bibfnamefont {M.~R.}\ \bibnamefont
  {Hermes}}, \bibinfo {author} {\bibfnamefont {D.~G.}\ \bibnamefont
  {Truhlar}},\ and\ \bibinfo {author} {\bibfnamefont {L.}~\bibnamefont
  {Gagliardi}},\ }\href {https://doi.org/10.1021/acs.jctc.5c01633} {\bibfield
  {journal} {\bibinfo  {journal} {J. Chem. Theory Comput.}\ }\textbf {\bibinfo
  {volume} {22}},\ \bibinfo {pages} {318} (\bibinfo {year} {2026})}\BibitemShut
  {NoStop}%
\bibitem [{\citenamefont {K{\"a}hler}\ \emph {et~al.}(2023)\citenamefont
  {K{\"a}hler}, \citenamefont {{Cebreiro-Gallardo}}, \citenamefont {Pokhilko},
  \citenamefont {Casanova},\ and\ \citenamefont
  {Krylov}}]{kahlerStateInteractionApproachEvaluating2023}%
  \BibitemOpen
  \bibfield  {author} {\bibinfo {author} {\bibfnamefont {S.}~\bibnamefont
  {K{\"a}hler}}, \bibinfo {author} {\bibfnamefont {A.}~\bibnamefont
  {{Cebreiro-Gallardo}}}, \bibinfo {author} {\bibfnamefont {P.}~\bibnamefont
  {Pokhilko}}, \bibinfo {author} {\bibfnamefont {D.}~\bibnamefont {Casanova}},\
  and\ \bibinfo {author} {\bibfnamefont {A.~I.}\ \bibnamefont {Krylov}},\
  }\href {https://doi.org/10.1021/acs.jpca.3c04134} {\bibfield  {journal}
  {\bibinfo  {journal} {J. Phys. Chem. A}\ }\textbf {\bibinfo {volume} {127}},\
  \bibinfo {pages} {8459} (\bibinfo {year} {2023})}\BibitemShut {NoStop}%
\bibitem [{\citenamefont {Vancoillie}, \citenamefont {Malmqvist},\ and\
  \citenamefont {Pierloot}(2007)}]{vancoillieCalculationEPRTensors2007}%
  \BibitemOpen
  \bibfield  {author} {\bibinfo {author} {\bibfnamefont {S.}~\bibnamefont
  {Vancoillie}}, \bibinfo {author} {\bibfnamefont {P.-{\AA}.}\ \bibnamefont
  {Malmqvist}},\ and\ \bibinfo {author} {\bibfnamefont {K.}~\bibnamefont
  {Pierloot}},\ }\href {https://doi.org/10.1002/cphc.200700128} {\bibfield
  {journal} {\bibinfo  {journal} {ChemPhysChem}\ }\textbf {\bibinfo {volume}
  {8}},\ \bibinfo {pages} {1803} (\bibinfo {year} {2007})}\BibitemShut
  {NoStop}%
\bibitem [{\citenamefont {Vancoillie}\ and\ \citenamefont
  {Pierloot}(2008)}]{vancoillieMulticonfigurationalTensorCalculations2008}%
  \BibitemOpen
  \bibfield  {author} {\bibinfo {author} {\bibfnamefont {S.}~\bibnamefont
  {Vancoillie}}\ and\ \bibinfo {author} {\bibfnamefont {K.}~\bibnamefont
  {Pierloot}},\ }\href {https://doi.org/10.1021/jp711345n} {\bibfield
  {journal} {\bibinfo  {journal} {J. Phys. Chem. A}\ }\textbf {\bibinfo
  {volume} {112}},\ \bibinfo {pages} {4011} (\bibinfo {year}
  {2008})}\BibitemShut {NoStop}%
\bibitem [{\citenamefont {Lang}\ and\ \citenamefont
  {Neese}(2019)}]{langSpindependentPropertiesFramework2019}%
  \BibitemOpen
  \bibfield  {author} {\bibinfo {author} {\bibfnamefont {L.}~\bibnamefont
  {Lang}}\ and\ \bibinfo {author} {\bibfnamefont {F.}~\bibnamefont {Neese}},\
  }\href {https://doi.org/10.1063/1.5085203} {\bibfield  {journal} {\bibinfo
  {journal} {J. Chem. Phys.}\ }\textbf {\bibinfo {volume} {150}},\ \bibinfo
  {pages} {104104} (\bibinfo {year} {2019})}\BibitemShut {NoStop}%
\bibitem [{\citenamefont {Fedorov}\ and\ \citenamefont
  {Finley}(2001)}]{fedorovSpinorbitMultireferenceMultistate2001}%
  \BibitemOpen
  \bibfield  {author} {\bibinfo {author} {\bibfnamefont {D.~G.}\ \bibnamefont
  {Fedorov}}\ and\ \bibinfo {author} {\bibfnamefont {J.~P.}\ \bibnamefont
  {Finley}},\ }\href {https://doi.org/10.1103/PhysRevA.64.042502} {\bibfield
  {journal} {\bibinfo  {journal} {Phys. Rev. A}\ }\textbf {\bibinfo {volume}
  {64}},\ \bibinfo {pages} {042502} (\bibinfo {year} {2001})}\BibitemShut
  {NoStop}%
\bibitem [{\citenamefont {Roemelt}(2015)}]{roemeltSpinOrbitCoupling2015}%
  \BibitemOpen
  \bibfield  {author} {\bibinfo {author} {\bibfnamefont {M.}~\bibnamefont
  {Roemelt}},\ }\href {https://doi.org/10.1063/1.4927432} {\bibfield  {journal}
  {\bibinfo  {journal} {J. Chem. Phys.}\ }\textbf {\bibinfo {volume} {143}},\
  \bibinfo {pages} {044112} (\bibinfo {year} {2015})}\BibitemShut {NoStop}%
\bibitem [{\citenamefont {Sayfutyarova}\ and\ \citenamefont
  {Chan}(2018)}]{sayfutyarovaElectronParamagneticResonance2018}%
  \BibitemOpen
  \bibfield  {author} {\bibinfo {author} {\bibfnamefont {E.~R.}\ \bibnamefont
  {Sayfutyarova}}\ and\ \bibinfo {author} {\bibfnamefont {G.~K.-L.}\
  \bibnamefont {Chan}},\ }\href {https://doi.org/10.1063/1.5020079} {\bibfield
  {journal} {\bibinfo  {journal} {J. Chem. Phys.}\ }\textbf {\bibinfo {volume}
  {148}},\ \bibinfo {pages} {184103} (\bibinfo {year} {2018})}\BibitemShut
  {NoStop}%
\bibitem [{\citenamefont {Chibotaru}\ and\ \citenamefont
  {Ungur}(2012)}]{chibotaruInitioCalculationAnisotropic2012}%
  \BibitemOpen
  \bibfield  {author} {\bibinfo {author} {\bibfnamefont {L.~F.}\ \bibnamefont
  {Chibotaru}}\ and\ \bibinfo {author} {\bibfnamefont {L.}~\bibnamefont
  {Ungur}},\ }\href {https://doi.org/10.1063/1.4739763} {\bibfield  {journal}
  {\bibinfo  {journal} {J. Chem. Phys.}\ }\textbf {\bibinfo {volume} {137}},\
  \bibinfo {pages} {064112} (\bibinfo {year} {2012})}\BibitemShut {NoStop}%
\bibitem [{\citenamefont {Vancoillie}\ \emph {et~al.}(2010)\citenamefont
  {Vancoillie}, \citenamefont {Chalupsky}, \citenamefont {Ryde}, \citenamefont
  {Solomon}, \citenamefont {Pierloot}, \citenamefont {Neese},\ and\
  \citenamefont {Rulisek}}]{vancoillieMultireferenceInitioCalculations2010}%
  \BibitemOpen
  \bibfield  {author} {\bibinfo {author} {\bibfnamefont {S.}~\bibnamefont
  {Vancoillie}}, \bibinfo {author} {\bibfnamefont {J.}~\bibnamefont
  {Chalupsky}}, \bibinfo {author} {\bibfnamefont {U.}~\bibnamefont {Ryde}},
  \bibinfo {author} {\bibfnamefont {E.~I.}\ \bibnamefont {Solomon}}, \bibinfo
  {author} {\bibfnamefont {K.}~\bibnamefont {Pierloot}}, \bibinfo {author}
  {\bibfnamefont {F.}~\bibnamefont {Neese}},\ and\ \bibinfo {author}
  {\bibfnamefont {L.}~\bibnamefont {Rulisek}},\ }\href
  {https://doi.org/10.1021/jp103098r} {\bibfield  {journal} {\bibinfo
  {journal} {J. Phys. Chem. B}\ }\textbf {\bibinfo {volume} {114}},\ \bibinfo
  {pages} {7692} (\bibinfo {year} {2010})}\BibitemShut {NoStop}%
\bibitem [{\citenamefont {Bokarev}\ \emph {et~al.}(2014)\citenamefont
  {Bokarev}, \citenamefont {Hollmann}, \citenamefont {Pazidis}, \citenamefont
  {Neubauer}, \citenamefont {Radnik}, \citenamefont {K{\"u}hn}, \citenamefont
  {Lochbrunner}, \citenamefont {Junge}, \citenamefont {Beller},\ and\
  \citenamefont {Br{\"u}ckner}}]{bokarevSpinDensityDistribution2014}%
  \BibitemOpen
  \bibfield  {author} {\bibinfo {author} {\bibfnamefont {S.}~\bibnamefont
  {Bokarev}}, \bibinfo {author} {\bibfnamefont {D.}~\bibnamefont {Hollmann}},
  \bibinfo {author} {\bibfnamefont {A.}~\bibnamefont {Pazidis}}, \bibinfo
  {author} {\bibfnamefont {A.}~\bibnamefont {Neubauer}}, \bibinfo {author}
  {\bibfnamefont {J.}~\bibnamefont {Radnik}}, \bibinfo {author} {\bibfnamefont
  {O.}~\bibnamefont {K{\"u}hn}}, \bibinfo {author} {\bibfnamefont
  {S.}~\bibnamefont {Lochbrunner}}, \bibinfo {author} {\bibfnamefont
  {H.}~\bibnamefont {Junge}}, \bibinfo {author} {\bibfnamefont
  {M.}~\bibnamefont {Beller}},\ and\ \bibinfo {author} {\bibfnamefont
  {A.}~\bibnamefont {Br{\"u}ckner}},\ }\href
  {https://doi.org/10.1039/C3CP54922D} {\bibfield  {journal} {\bibinfo
  {journal} {Phys. Chem. Chem. Phys.}\ }\textbf {\bibinfo {volume} {16}},\
  \bibinfo {pages} {4789} (\bibinfo {year} {2014})}\BibitemShut {NoStop}%
\bibitem [{\citenamefont {{Bad{\'i}a-Romano}}\ \emph
  {et~al.}(2013)\citenamefont {{Bad{\'i}a-Romano}}, \citenamefont
  {Bartolom{\'e}}, \citenamefont {Bartolom{\'e}}, \citenamefont {Luz{\'o}n},
  \citenamefont {Prodius}, \citenamefont {Turta}, \citenamefont {Mereacre},
  \citenamefont {Wilhelm},\ and\ \citenamefont
  {Rogalev}}]{badia-romanoFieldinducedInternalFe2013}%
  \BibitemOpen
  \bibfield  {author} {\bibinfo {author} {\bibfnamefont {L.}~\bibnamefont
  {{Bad{\'i}a-Romano}}}, \bibinfo {author} {\bibfnamefont {F.}~\bibnamefont
  {Bartolom{\'e}}}, \bibinfo {author} {\bibfnamefont {J.}~\bibnamefont
  {Bartolom{\'e}}}, \bibinfo {author} {\bibfnamefont {J.}~\bibnamefont
  {Luz{\'o}n}}, \bibinfo {author} {\bibfnamefont {D.}~\bibnamefont {Prodius}},
  \bibinfo {author} {\bibfnamefont {C.}~\bibnamefont {Turta}}, \bibinfo
  {author} {\bibfnamefont {V.}~\bibnamefont {Mereacre}}, \bibinfo {author}
  {\bibfnamefont {F.}~\bibnamefont {Wilhelm}},\ and\ \bibinfo {author}
  {\bibfnamefont {A.}~\bibnamefont {Rogalev}},\ }\href
  {https://doi.org/10.1103/PhysRevB.87.184403} {\bibfield  {journal} {\bibinfo
  {journal} {Phys. Rev. B}\ }\textbf {\bibinfo {volume} {87}},\ \bibinfo
  {pages} {184403} (\bibinfo {year} {2013})}\BibitemShut {NoStop}%
\bibitem [{\citenamefont {Charron}\ \emph {et~al.}(2016)\citenamefont
  {Charron}, \citenamefont {Malkin}, \citenamefont {Rogez}, \citenamefont
  {Batchelor}, \citenamefont {Mazerat}, \citenamefont {Guillot}, \citenamefont
  {Guih{\'e}ry}, \citenamefont {Barra}, \citenamefont {Mallah},\ and\
  \citenamefont {Bolvin}}]{charronUnravelingEffectsMagnetic2016}%
  \BibitemOpen
  \bibfield  {author} {\bibinfo {author} {\bibfnamefont {G.}~\bibnamefont
  {Charron}}, \bibinfo {author} {\bibfnamefont {E.}~\bibnamefont {Malkin}},
  \bibinfo {author} {\bibfnamefont {G.}~\bibnamefont {Rogez}}, \bibinfo
  {author} {\bibfnamefont {L.~J.}\ \bibnamefont {Batchelor}}, \bibinfo {author}
  {\bibfnamefont {S.}~\bibnamefont {Mazerat}}, \bibinfo {author} {\bibfnamefont
  {R.}~\bibnamefont {Guillot}}, \bibinfo {author} {\bibfnamefont
  {N.}~\bibnamefont {Guih{\'e}ry}}, \bibinfo {author} {\bibfnamefont {A.-L.}\
  \bibnamefont {Barra}}, \bibinfo {author} {\bibfnamefont {T.}~\bibnamefont
  {Mallah}},\ and\ \bibinfo {author} {\bibfnamefont {H.}~\bibnamefont
  {Bolvin}},\ }\href {https://doi.org/10.1002/chem.201602837} {\bibfield
  {journal} {\bibinfo  {journal} {Chem. Eur. J.}\ }\textbf {\bibinfo {volume}
  {22}},\ \bibinfo {pages} {16850} (\bibinfo {year} {2016})}\BibitemShut
  {NoStop}%
\bibitem [{\citenamefont {Pokhilko}\ and\ \citenamefont
  {Pushkar}(2025)}]{pokhilkoMulticonfigurationalElectronicStructure2025}%
  \BibitemOpen
  \bibfield  {author} {\bibinfo {author} {\bibfnamefont {P.}~\bibnamefont
  {Pokhilko}}\ and\ \bibinfo {author} {\bibfnamefont {Y.}~\bibnamefont
  {Pushkar}},\ }\href {https://doi.org/10.1039/D5CP03298A} {\bibfield
  {journal} {\bibinfo  {journal} {Phys. Chem. Chem. Phys.}\ } (\bibinfo {year}
  {2025}),\ 10.1039/D5CP03298A}\BibitemShut {NoStop}%
\bibitem [{\citenamefont {Lunghi}(2019)}]{lunghi2019}%
  \BibitemOpen
  \bibfield  {author} {\bibinfo {author} {\bibfnamefont {A.}~\bibnamefont
  {Lunghi}},\ }\href {https://arxiv.org/abs/1912.04545} {\enquote {\bibinfo
  {title} {Ligand-field contributions to spin-phonon coupling in a family of
  vanadium molecular qubits from multi-reference electronic structure
  theory},}\ } (\bibinfo {year} {2019}),\ \Eprint
  {https://arxiv.org/abs/1912.04545} {arXiv:1912.04545 [cond-mat.mtrl-sci]}
  \BibitemShut {NoStop}%
\bibitem [{\citenamefont {Blackaby}\ \emph {et~al.}(2022)\citenamefont
  {Blackaby}, \citenamefont {Harriman}, \citenamefont {Greer}, \citenamefont
  {Folli}, \citenamefont {Hill}, \citenamefont {Krewald}, \citenamefont
  {Mahon}, \citenamefont {Murphy}, \citenamefont {Murugesu}, \citenamefont
  {Richards}, \citenamefont {Suturina},\ and\ \citenamefont
  {Whittlesey}}]{blackabyExtremeGTensorAnisotropy2022}%
  \BibitemOpen
  \bibfield  {author} {\bibinfo {author} {\bibfnamefont {W.~J.~M.}\
  \bibnamefont {Blackaby}}, \bibinfo {author} {\bibfnamefont {K.~L.~M.}\
  \bibnamefont {Harriman}}, \bibinfo {author} {\bibfnamefont {S.~M.}\
  \bibnamefont {Greer}}, \bibinfo {author} {\bibfnamefont {A.}~\bibnamefont
  {Folli}}, \bibinfo {author} {\bibfnamefont {S.}~\bibnamefont {Hill}},
  \bibinfo {author} {\bibfnamefont {V.}~\bibnamefont {Krewald}}, \bibinfo
  {author} {\bibfnamefont {M.~F.}\ \bibnamefont {Mahon}}, \bibinfo {author}
  {\bibfnamefont {D.~M.}\ \bibnamefont {Murphy}}, \bibinfo {author}
  {\bibfnamefont {M.}~\bibnamefont {Murugesu}}, \bibinfo {author}
  {\bibfnamefont {E.}~\bibnamefont {Richards}}, \bibinfo {author}
  {\bibfnamefont {E.}~\bibnamefont {Suturina}},\ and\ \bibinfo {author}
  {\bibfnamefont {M.~K.}\ \bibnamefont {Whittlesey}},\ }\href
  {https://doi.org/10.1021/acs.inorgchem.1c02413} {\bibfield  {journal}
  {\bibinfo  {journal} {Inorg. Chem.}\ }\textbf {\bibinfo {volume} {61}},\
  \bibinfo {pages} {1308} (\bibinfo {year} {2022})}\BibitemShut {NoStop}%
\bibitem [{\citenamefont {Rouf}, \citenamefont {Mares},\ and\ \citenamefont
  {Vaara}(2017)}]{roufRelativisticApproximationsParamagnetic2017}%
  \BibitemOpen
  \bibfield  {author} {\bibinfo {author} {\bibfnamefont {S.~A.}\ \bibnamefont
  {Rouf}}, \bibinfo {author} {\bibfnamefont {J.}~\bibnamefont {Mares}},\ and\
  \bibinfo {author} {\bibfnamefont {J.}~\bibnamefont {Vaara}},\ }\href
  {https://doi.org/10.1021/acs.jctc.7b00168} {\bibfield  {journal} {\bibinfo
  {journal} {J. Chem. Theory Comput.}\ }\textbf {\bibinfo {volume} {13}},\
  \bibinfo {pages} {3731} (\bibinfo {year} {2017})}\BibitemShut {NoStop}%
\bibitem [{\citenamefont {Sharma}\ \emph {et~al.}(2017)\citenamefont {Sharma},
  \citenamefont {Roemelt}, \citenamefont {Reithofer}, \citenamefont {Schrock},
  \citenamefont {Hoffman},\ and\ \citenamefont
  {Neese}}]{sharmaEPRENDORTheoretical2017}%
  \BibitemOpen
  \bibfield  {author} {\bibinfo {author} {\bibfnamefont {A.}~\bibnamefont
  {Sharma}}, \bibinfo {author} {\bibfnamefont {M.}~\bibnamefont {Roemelt}},
  \bibinfo {author} {\bibfnamefont {M.}~\bibnamefont {Reithofer}}, \bibinfo
  {author} {\bibfnamefont {R.~R.}\ \bibnamefont {Schrock}}, \bibinfo {author}
  {\bibfnamefont {B.~M.}\ \bibnamefont {Hoffman}},\ and\ \bibinfo {author}
  {\bibfnamefont {F.}~\bibnamefont {Neese}},\ }\href
  {https://doi.org/10.1021/acs.inorgchem.7b00364} {\bibfield  {journal}
  {\bibinfo  {journal} {Inorg. Chem.}\ }\textbf {\bibinfo {volume} {56}},\
  \bibinfo {pages} {6906} (\bibinfo {year} {2017})}\BibitemShut {NoStop}%
\bibitem [{\citenamefont {Majumder}\ and\ \citenamefont
  {Sokolov}(2023)}]{majumderSimulatingSpinOrbit2023}%
  \BibitemOpen
  \bibfield  {author} {\bibinfo {author} {\bibfnamefont {R.}~\bibnamefont
  {Majumder}}\ and\ \bibinfo {author} {\bibfnamefont {A.~Y.}\ \bibnamefont
  {Sokolov}},\ }\href {https://doi.org/10.1021/acs.jpca.2c07952} {\bibfield
  {journal} {\bibinfo  {journal} {J. Phys. Chem. A}\ }\textbf {\bibinfo
  {volume} {127}},\ \bibinfo {pages} {546} (\bibinfo {year}
  {2023})}\BibitemShut {NoStop}%
\bibitem [{\citenamefont {Majumder}\ and\ \citenamefont
  {Sokolov}(2024)}]{majumderConsistentSecondOrderTreatment2024}%
  \BibitemOpen
  \bibfield  {author} {\bibinfo {author} {\bibfnamefont {R.}~\bibnamefont
  {Majumder}}\ and\ \bibinfo {author} {\bibfnamefont {A.~Y.}\ \bibnamefont
  {Sokolov}},\ }\href {https://doi.org/10.1021/acs.jctc.4c00458} {\bibfield
  {journal} {\bibinfo  {journal} {J. Chem. Theory Comput.}\ }\textbf {\bibinfo
  {volume} {20}},\ \bibinfo {pages} {4676} (\bibinfo {year}
  {2024})}\BibitemShut {NoStop}%
\bibitem [{\citenamefont {Liu}\ and\ \citenamefont
  {Peng}(2009)}]{liuExactTwocomponentHamiltonians2009}%
  \BibitemOpen
  \bibfield  {author} {\bibinfo {author} {\bibfnamefont {W.}~\bibnamefont
  {Liu}}\ and\ \bibinfo {author} {\bibfnamefont {D.}~\bibnamefont {Peng}},\
  }\href {https://doi.org/10.1063/1.3159445} {\bibfield  {journal} {\bibinfo
  {journal} {J. Chem. Phys.}\ }\textbf {\bibinfo {volume} {131}},\ \bibinfo
  {pages} {031104} (\bibinfo {year} {2009})}\BibitemShut {NoStop}%
\bibitem [{\citenamefont {Li}, \citenamefont {Xiao},\ and\ \citenamefont
  {Liu}(2012)}]{liSpinSeparationAlgebraic2012}%
  \BibitemOpen
  \bibfield  {author} {\bibinfo {author} {\bibfnamefont {Z.}~\bibnamefont
  {Li}}, \bibinfo {author} {\bibfnamefont {Y.}~\bibnamefont {Xiao}},\ and\
  \bibinfo {author} {\bibfnamefont {W.}~\bibnamefont {Liu}},\ }\href
  {https://doi.org/10.1063/1.4758987} {\bibfield  {journal} {\bibinfo
  {journal} {J. Chem. Phys.}\ }\textbf {\bibinfo {volume} {137}},\ \bibinfo
  {pages} {154114} (\bibinfo {year} {2012})}\BibitemShut {NoStop}%
\bibitem [{\citenamefont {Angeli}\ \emph {et~al.}(2004)\citenamefont {Angeli},
  \citenamefont {Borini}, \citenamefont {Cestari},\ and\ \citenamefont
  {Cimiraglia}}]{angeliQuasidegenerateFormulationSecond2004}%
  \BibitemOpen
  \bibfield  {author} {\bibinfo {author} {\bibfnamefont {C.}~\bibnamefont
  {Angeli}}, \bibinfo {author} {\bibfnamefont {S.}~\bibnamefont {Borini}},
  \bibinfo {author} {\bibfnamefont {M.}~\bibnamefont {Cestari}},\ and\ \bibinfo
  {author} {\bibfnamefont {R.}~\bibnamefont {Cimiraglia}},\ }\href
  {https://doi.org/10.1063/1.1778711} {\bibfield  {journal} {\bibinfo
  {journal} {J. Chem. Phys.}\ }\textbf {\bibinfo {volume} {121}},\ \bibinfo
  {pages} {4043} (\bibinfo {year} {2004})}\BibitemShut {NoStop}%
\bibitem [{\citenamefont {Angeli}, \citenamefont {Pastore},\ and\ \citenamefont
  {Cimiraglia}(2007)}]{angeliNewPerspectivesMultireference2007}%
  \BibitemOpen
  \bibfield  {author} {\bibinfo {author} {\bibfnamefont {C.}~\bibnamefont
  {Angeli}}, \bibinfo {author} {\bibfnamefont {M.}~\bibnamefont {Pastore}},\
  and\ \bibinfo {author} {\bibfnamefont {R.}~\bibnamefont {Cimiraglia}},\
  }\href {https://doi.org/10.1007/s00214-006-0207-0} {\bibfield  {journal}
  {\bibinfo  {journal} {Theor. Chem. Acc.}\ }\textbf {\bibinfo {volume}
  {117}},\ \bibinfo {pages} {743} (\bibinfo {year} {2007})}\BibitemShut
  {NoStop}%
\bibitem [{\citenamefont
  {Hinze}(1973)}]{hinzeMCSCFMulticonfigurationSelfconsistentfield1973}%
  \BibitemOpen
  \bibfield  {author} {\bibinfo {author} {\bibfnamefont {J.}~\bibnamefont
  {Hinze}},\ }\href {https://doi.org/10.1063/1.1680022} {\bibfield  {journal}
  {\bibinfo  {journal} {J. Chem. Phys.}\ }\textbf {\bibinfo {volume} {59}},\
  \bibinfo {pages} {6424} (\bibinfo {year} {1973})}\BibitemShut {NoStop}%
\bibitem [{\citenamefont {Roos}, \citenamefont {Taylor},\ and\ \citenamefont
  {Sigbahn}(1980)}]{roosCompleteActiveSpace1980}%
  \BibitemOpen
  \bibfield  {author} {\bibinfo {author} {\bibfnamefont {B.~O.}\ \bibnamefont
  {Roos}}, \bibinfo {author} {\bibfnamefont {P.~R.}\ \bibnamefont {Taylor}},\
  and\ \bibinfo {author} {\bibfnamefont {P.~E.~M.}\ \bibnamefont {Sigbahn}},\
  }\href {https://doi.org/10.1016/0301-0104(80)80045-0} {\bibfield  {journal}
  {\bibinfo  {journal} {Chem. Phys.}\ }\textbf {\bibinfo {volume} {48}},\
  \bibinfo {pages} {157} (\bibinfo {year} {1980})}\BibitemShut {NoStop}%
\bibitem [{\citenamefont {Siegbahn}\ \emph {et~al.}(1981)\citenamefont
  {Siegbahn}, \citenamefont {Alml{\"o}f}, \citenamefont {Heiberg},\ and\
  \citenamefont {Roos}}]{siegbahnCompleteActiveSpace1981}%
  \BibitemOpen
  \bibfield  {author} {\bibinfo {author} {\bibfnamefont {P.~E.~M.}\
  \bibnamefont {Siegbahn}}, \bibinfo {author} {\bibfnamefont {J.}~\bibnamefont
  {Alml{\"o}f}}, \bibinfo {author} {\bibfnamefont {A.}~\bibnamefont
  {Heiberg}},\ and\ \bibinfo {author} {\bibfnamefont {B.~O.}\ \bibnamefont
  {Roos}},\ }\href {https://doi.org/10.1063/1.441359} {\bibfield  {journal}
  {\bibinfo  {journal} {J. Chem. Phys.}\ }\textbf {\bibinfo {volume} {74}},\
  \bibinfo {pages} {2384} (\bibinfo {year} {1981})}\BibitemShut {NoStop}%
\bibitem [{\citenamefont {Werner}\ and\ \citenamefont
  {Meyer}(1981)}]{wernerQuadraticallyConvergentMCSCF1981}%
  \BibitemOpen
  \bibfield  {author} {\bibinfo {author} {\bibfnamefont {H.-J.}\ \bibnamefont
  {Werner}}\ and\ \bibinfo {author} {\bibfnamefont {W.}~\bibnamefont {Meyer}},\
  }\href {https://doi.org/10.1063/1.440892} {\bibfield  {journal} {\bibinfo
  {journal} {J. Chem. Phys.}\ }\textbf {\bibinfo {volume} {74}},\ \bibinfo
  {pages} {5794} (\bibinfo {year} {1981})}\BibitemShut {NoStop}%
\bibitem [{\citenamefont {Werner}\ and\ \citenamefont
  {Knowles}(1985)}]{wernerSecondOrderMulticonfiguration1985}%
  \BibitemOpen
  \bibfield  {author} {\bibinfo {author} {\bibfnamefont {H.-J.}\ \bibnamefont
  {Werner}}\ and\ \bibinfo {author} {\bibfnamefont {P.~J.}\ \bibnamefont
  {Knowles}},\ }\href {https://doi.org/10.1063/1.448627} {\bibfield  {journal}
  {\bibinfo  {journal} {J. Chem. Phys.}\ }\textbf {\bibinfo {volume} {82}},\
  \bibinfo {pages} {5053} (\bibinfo {year} {1985})}\BibitemShut {NoStop}%
\bibitem [{\citenamefont {Angeli}\ \emph {et~al.}(2001)\citenamefont {Angeli},
  \citenamefont {Cimiraglia}, \citenamefont {Evangelisti}, \citenamefont
  {Leininger},\ and\ \citenamefont
  {Malrieu}}]{angeliIntroductionElectronValence2001}%
  \BibitemOpen
  \bibfield  {author} {\bibinfo {author} {\bibfnamefont {C.}~\bibnamefont
  {Angeli}}, \bibinfo {author} {\bibfnamefont {R.}~\bibnamefont {Cimiraglia}},
  \bibinfo {author} {\bibfnamefont {S.}~\bibnamefont {Evangelisti}}, \bibinfo
  {author} {\bibfnamefont {T.}~\bibnamefont {Leininger}},\ and\ \bibinfo
  {author} {\bibfnamefont {J.-P.}\ \bibnamefont {Malrieu}},\ }\href
  {https://doi.org/10.1063/1.1361246} {\bibfield  {journal} {\bibinfo
  {journal} {J. Chem. Phys.}\ }\textbf {\bibinfo {volume} {114}},\ \bibinfo
  {pages} {10252} (\bibinfo {year} {2001})}\BibitemShut {NoStop}%
\bibitem [{\citenamefont {Angeli}, \citenamefont {Cimiraglia},\ and\
  \citenamefont {Malrieu}(2002)}]{angeliElectronValenceState2002}%
  \BibitemOpen
  \bibfield  {author} {\bibinfo {author} {\bibfnamefont {C.}~\bibnamefont
  {Angeli}}, \bibinfo {author} {\bibfnamefont {R.}~\bibnamefont {Cimiraglia}},\
  and\ \bibinfo {author} {\bibfnamefont {J.-P.}\ \bibnamefont {Malrieu}},\
  }\href {https://doi.org/10.1063/1.1515317} {\bibfield  {journal} {\bibinfo
  {journal} {J. Chem. Phys.}\ }\textbf {\bibinfo {volume} {117}},\ \bibinfo
  {pages} {9138} (\bibinfo {year} {2002})}\BibitemShut {NoStop}%
\bibitem [{\citenamefont
  {Dyall}(1995)}]{dyallChoiceZerothorderHamiltonian1995}%
  \BibitemOpen
  \bibfield  {author} {\bibinfo {author} {\bibfnamefont {K.~G.}\ \bibnamefont
  {Dyall}},\ }\href {https://doi.org/10.1063/1.469539} {\bibfield  {journal}
  {\bibinfo  {journal} {J. Chem. Phys.}\ }\textbf {\bibinfo {volume} {102}},\
  \bibinfo {pages} {4909} (\bibinfo {year} {1995})}\BibitemShut {NoStop}%
\bibitem [{\citenamefont
  {Sokolov}(2024)}]{sokolovMultireferencePerturbationTheories2024}%
  \BibitemOpen
  \bibfield  {author} {\bibinfo {author} {\bibfnamefont {A.~Y.}\ \bibnamefont
  {Sokolov}},\ }in\ \href {https://doi.org/10.1016/bs.aiq.2024.04.004} {\emph
  {\bibinfo {booktitle} {Advances in Quantum Chemistry}}},\ Vol.~\bibinfo
  {volume} {90}\ (\bibinfo  {publisher} {Elsevier},\ \bibinfo {year} {2024})\
  pp.\ \bibinfo {pages} {121--155}\BibitemShut {NoStop}%
\bibitem [{\citenamefont {Angeli}\ \emph {et~al.}(2006)\citenamefont {Angeli},
  \citenamefont {Bories}, \citenamefont {Cavallini},\ and\ \citenamefont
  {Cimiraglia}}]{angeliThirdorderMultireferencePerturbation2006}%
  \BibitemOpen
  \bibfield  {author} {\bibinfo {author} {\bibfnamefont {C.}~\bibnamefont
  {Angeli}}, \bibinfo {author} {\bibfnamefont {B.}~\bibnamefont {Bories}},
  \bibinfo {author} {\bibfnamefont {A.}~\bibnamefont {Cavallini}},\ and\
  \bibinfo {author} {\bibfnamefont {R.}~\bibnamefont {Cimiraglia}},\ }\href
  {https://doi.org/10.1063/1.2148946} {\bibfield  {journal} {\bibinfo
  {journal} {J. Chem. Phys.}\ }\textbf {\bibinfo {volume} {124}},\ \bibinfo
  {pages} {054108} (\bibinfo {year} {2006})}\BibitemShut {NoStop}%
\bibitem [{\citenamefont {Kempfer}, \citenamefont {Sivalingam},\ and\
  \citenamefont {Neese}(2025)}]{kempferEfficientImplementationApproximate2025}%
  \BibitemOpen
  \bibfield  {author} {\bibinfo {author} {\bibfnamefont {E.~M.}\ \bibnamefont
  {Kempfer}}, \bibinfo {author} {\bibfnamefont {K.}~\bibnamefont
  {Sivalingam}},\ and\ \bibinfo {author} {\bibfnamefont {F.}~\bibnamefont
  {Neese}},\ }\href {https://doi.org/10.1021/acs.jctc.4c01735} {\bibfield
  {journal} {\bibinfo  {journal} {J. Chem. Theory Comput.}\ }\textbf {\bibinfo
  {volume} {21}},\ \bibinfo {pages} {3953} (\bibinfo {year}
  {2025})}\BibitemShut {NoStop}%
\bibitem [{\citenamefont {Park}(2020)}]{parkAnalyticalGradientTheory2020}%
  \BibitemOpen
  \bibfield  {author} {\bibinfo {author} {\bibfnamefont {J.~W.}\ \bibnamefont
  {Park}},\ }\href {https://doi.org/10.1021/acs.jctc.9b00919} {\bibfield
  {journal} {\bibinfo  {journal} {J. Chem. Theory Comput.}\ }\textbf {\bibinfo
  {volume} {16}},\ \bibinfo {pages} {326} (\bibinfo {year} {2020})}\BibitemShut
  {NoStop}%
\bibitem [{\citenamefont {Sharma}, \citenamefont {Jeanmairet},\ and\
  \citenamefont {Alavi}(2016)}]{sharmaQuasidegeneratePerturbationTheory2016}%
  \BibitemOpen
  \bibfield  {author} {\bibinfo {author} {\bibfnamefont {S.}~\bibnamefont
  {Sharma}}, \bibinfo {author} {\bibfnamefont {G.}~\bibnamefont {Jeanmairet}},\
  and\ \bibinfo {author} {\bibfnamefont {A.}~\bibnamefont {Alavi}},\ }\href
  {https://doi.org/10.1063/1.4939752} {\bibfield  {journal} {\bibinfo
  {journal} {J. Chem. Phys.}\ }\textbf {\bibinfo {volume} {144}},\ \bibinfo
  {pages} {034103} (\bibinfo {year} {2016})}\BibitemShut {NoStop}%
\bibitem [{\citenamefont {Sivalingam}\ \emph {et~al.}(2016)\citenamefont
  {Sivalingam}, \citenamefont {Krupicka}, \citenamefont {Auer},\ and\
  \citenamefont {Neese}}]{sivalingamComparisonFullyInternally2016}%
  \BibitemOpen
  \bibfield  {author} {\bibinfo {author} {\bibfnamefont {K.}~\bibnamefont
  {Sivalingam}}, \bibinfo {author} {\bibfnamefont {M.}~\bibnamefont
  {Krupicka}}, \bibinfo {author} {\bibfnamefont {A.~A.}\ \bibnamefont {Auer}},\
  and\ \bibinfo {author} {\bibfnamefont {F.}~\bibnamefont {Neese}},\ }\href
  {https://doi.org/10.1063/1.4959029} {\bibfield  {journal} {\bibinfo
  {journal} {J. Chem. Phys.}\ }\textbf {\bibinfo {volume} {145}},\ \bibinfo
  {pages} {054104} (\bibinfo {year} {2016})}\BibitemShut {NoStop}%
\bibitem [{\citenamefont {Guo}\ \emph {et~al.}(2016)\citenamefont {Guo},
  \citenamefont {Sivalingam}, \citenamefont {Valeev},\ and\ \citenamefont
  {Neese}}]{guoSparseMapsSystematicInfrastructure2016}%
  \BibitemOpen
  \bibfield  {author} {\bibinfo {author} {\bibfnamefont {Y.}~\bibnamefont
  {Guo}}, \bibinfo {author} {\bibfnamefont {K.}~\bibnamefont {Sivalingam}},
  \bibinfo {author} {\bibfnamefont {E.~F.}\ \bibnamefont {Valeev}},\ and\
  \bibinfo {author} {\bibfnamefont {F.}~\bibnamefont {Neese}},\ }\href
  {https://doi.org/10.1063/1.4942769} {\bibfield  {journal} {\bibinfo
  {journal} {J. Chem. Phys.}\ }\textbf {\bibinfo {volume} {144}},\ \bibinfo
  {pages} {094111} (\bibinfo {year} {2016})}\BibitemShut {NoStop}%
\bibitem [{\citenamefont {Park}(2019)}]{parkAnalyticalGradientTheory2019}%
  \BibitemOpen
  \bibfield  {author} {\bibinfo {author} {\bibfnamefont {J.~W.}\ \bibnamefont
  {Park}},\ }\href {https://doi.org/10.1021/acs.jctc.9b00762} {\bibfield
  {journal} {\bibinfo  {journal} {J. Chem. Theory Comput.}\ }\textbf {\bibinfo
  {volume} {15}},\ \bibinfo {pages} {5417} (\bibinfo {year}
  {2019})}\BibitemShut {NoStop}%
\bibitem [{\citenamefont {Sokolov}\ and\ \citenamefont
  {Chan}(2016)}]{sokolovTimedependentFormulationMultireference2016}%
  \BibitemOpen
  \bibfield  {author} {\bibinfo {author} {\bibfnamefont {A.~Y.}\ \bibnamefont
  {Sokolov}}\ and\ \bibinfo {author} {\bibfnamefont {G.~K.-L.}\ \bibnamefont
  {Chan}},\ }\href {https://doi.org/10.1063/1.4941606} {\bibfield  {journal}
  {\bibinfo  {journal} {J. Chem. Phys.}\ }\textbf {\bibinfo {volume} {144}},\
  \bibinfo {pages} {064102} (\bibinfo {year} {2016})}\BibitemShut {NoStop}%
\bibitem [{\citenamefont {Sokolov}\ \emph {et~al.}(2017)\citenamefont
  {Sokolov}, \citenamefont {Guo}, \citenamefont {Ronca},\ and\ \citenamefont
  {Chan}}]{sokolovTimedependentNelectronValence2017}%
  \BibitemOpen
  \bibfield  {author} {\bibinfo {author} {\bibfnamefont {A.~Y.}\ \bibnamefont
  {Sokolov}}, \bibinfo {author} {\bibfnamefont {S.}~\bibnamefont {Guo}},
  \bibinfo {author} {\bibfnamefont {E.}~\bibnamefont {Ronca}},\ and\ \bibinfo
  {author} {\bibfnamefont {G.~K.-L.}\ \bibnamefont {Chan}},\ }\href
  {https://doi.org/10.1063/1.4986975} {\bibfield  {journal} {\bibinfo
  {journal} {J. Chem. Phys.}\ }\textbf {\bibinfo {volume} {146}},\ \bibinfo
  {pages} {244102} (\bibinfo {year} {2017})}\BibitemShut {NoStop}%
\bibitem [{\citenamefont {Lang}, \citenamefont {Sivalingam},\ and\
  \citenamefont
  {Neese}(2020)}]{langCombinationMultipartitioningHamiltonian2020}%
  \BibitemOpen
  \bibfield  {author} {\bibinfo {author} {\bibfnamefont {L.}~\bibnamefont
  {Lang}}, \bibinfo {author} {\bibfnamefont {K.}~\bibnamefont {Sivalingam}},\
  and\ \bibinfo {author} {\bibfnamefont {F.}~\bibnamefont {Neese}},\ }\href
  {https://doi.org/10.1063/1.5133746} {\bibfield  {journal} {\bibinfo
  {journal} {J. Chem. Phys.}\ }\textbf {\bibinfo {volume} {152}},\ \bibinfo
  {pages} {014109} (\bibinfo {year} {2020})}\BibitemShut {NoStop}%
\bibitem [{\citenamefont {Shavitt}\ and\ \citenamefont
  {Redmon}(1980)}]{shavittQuasidegeneratePerturbationTheories1980}%
  \BibitemOpen
  \bibfield  {author} {\bibinfo {author} {\bibfnamefont {I.}~\bibnamefont
  {Shavitt}}\ and\ \bibinfo {author} {\bibfnamefont {L.~T.}\ \bibnamefont
  {Redmon}},\ }\href {https://doi.org/10.1063/1.440050} {\bibfield  {journal}
  {\bibinfo  {journal} {J. Chem. Phys.}\ }\textbf {\bibinfo {volume} {73}},\
  \bibinfo {pages} {5711} (\bibinfo {year} {1980})}\BibitemShut {NoStop}%
\bibitem [{\citenamefont {Certain}\ and\ \citenamefont
  {Hirschfelder}(1970)}]{certainNewPartitioningPerturbation1970}%
  \BibitemOpen
  \bibfield  {author} {\bibinfo {author} {\bibfnamefont {P.~R.}\ \bibnamefont
  {Certain}}\ and\ \bibinfo {author} {\bibfnamefont {J.~O.}\ \bibnamefont
  {Hirschfelder}},\ }\href {https://doi.org/10.1063/1.1672896} {\bibfield
  {journal} {\bibinfo  {journal} {J. Chem. Phys.}\ }\textbf {\bibinfo {volume}
  {52}},\ \bibinfo {pages} {5977} (\bibinfo {year} {1970})}\BibitemShut
  {NoStop}%
\bibitem [{\citenamefont
  {Kirtman}(1981)}]{kirtmanSimultaneousCalculationSeveral1981}%
  \BibitemOpen
  \bibfield  {author} {\bibinfo {author} {\bibfnamefont {B.}~\bibnamefont
  {Kirtman}},\ }\href {https://doi.org/10.1063/1.442123} {\bibfield  {journal}
  {\bibinfo  {journal} {J. Chem. Phys.}\ }\textbf {\bibinfo {volume} {75}},\
  \bibinfo {pages} {798} (\bibinfo {year} {1981})}\BibitemShut {NoStop}%
\bibitem [{\citenamefont {Zaitsevskii}\ and\ \citenamefont
  {Malrieu}(1995)}]{zaitsevskiiMultipartitioningQuasidegeneratePerturbation1995}%
  \BibitemOpen
  \bibfield  {author} {\bibinfo {author} {\bibfnamefont {A.}~\bibnamefont
  {Zaitsevskii}}\ and\ \bibinfo {author} {\bibfnamefont {J.-P.}\ \bibnamefont
  {Malrieu}},\ }\href {https://doi.org/10.1016/0009-2614(94)01503-N} {\bibfield
   {journal} {\bibinfo  {journal} {Chem. Phys. Lett.}\ }\textbf {\bibinfo
  {volume} {233}},\ \bibinfo {pages} {597} (\bibinfo {year}
  {1995})}\BibitemShut {NoStop}%
\bibitem [{\citenamefont
  {Pyykk{\"o}}(2012)}]{pyykkoRelativisticEffectsChemistry2012}%
  \BibitemOpen
  \bibfield  {author} {\bibinfo {author} {\bibfnamefont {P.}~\bibnamefont
  {Pyykk{\"o}}},\ }\href
  {https://doi.org/10.1146/annurev-physchem-032511-143755} {\bibfield
  {journal} {\bibinfo  {journal} {Annu. Rev. Phys. Chem.}\ }\textbf {\bibinfo
  {volume} {63}},\ \bibinfo {pages} {45} (\bibinfo {year} {2012})}\BibitemShut
  {NoStop}%
\bibitem [{\citenamefont {Liu}(2014)}]{liuAdvancesRelativisticMolecular2014}%
  \BibitemOpen
  \bibfield  {author} {\bibinfo {author} {\bibfnamefont {W.}~\bibnamefont
  {Liu}},\ }\href {https://doi.org/10.1016/j.physrep.2013.11.006} {\bibfield
  {journal} {\bibinfo  {journal} {Phys. Rep.}\ }\textbf {\bibinfo {volume}
  {537}},\ \bibinfo {pages} {59} (\bibinfo {year} {2014})}\BibitemShut
  {NoStop}%
\bibitem [{\citenamefont {Breit}(1932)}]{breitDiracsEquationSpinSpin1932}%
  \BibitemOpen
  \bibfield  {author} {\bibinfo {author} {\bibfnamefont {G.}~\bibnamefont
  {Breit}},\ }\href {https://doi.org/10.1103/PhysRev.39.616} {\bibfield
  {journal} {\bibinfo  {journal} {Phys. Rev.}\ }\textbf {\bibinfo {volume}
  {39}},\ \bibinfo {pages} {616} (\bibinfo {year} {1932})}\BibitemShut
  {NoStop}%
\bibitem [{\citenamefont {Bearpark}\ \emph {et~al.}(1993)\citenamefont
  {Bearpark}, \citenamefont {Handy}, \citenamefont {Palmieri},\ and\
  \citenamefont {Tarroni}}]{bearparkSpinorbitInteractionsSelf1993}%
  \BibitemOpen
  \bibfield  {author} {\bibinfo {author} {\bibfnamefont {M.~J.}\ \bibnamefont
  {Bearpark}}, \bibinfo {author} {\bibfnamefont {N.~C.}\ \bibnamefont {Handy}},
  \bibinfo {author} {\bibfnamefont {P.}~\bibnamefont {Palmieri}},\ and\
  \bibinfo {author} {\bibfnamefont {R.}~\bibnamefont {Tarroni}},\ }\href
  {https://doi.org/10.1080/00268979300102411} {\bibfield  {journal} {\bibinfo
  {journal} {Mol. Phys.}\ }\textbf {\bibinfo {volume} {80}},\ \bibinfo {pages}
  {479} (\bibinfo {year} {1993})}\BibitemShut {NoStop}%
\bibitem [{\citenamefont {Berning}\ \emph {et~al.}(2000)\citenamefont
  {Berning}, \citenamefont {Schweizer}, \citenamefont {Werner}, \citenamefont
  {Knowles},\ and\ \citenamefont
  {Palmieri}}]{berningSpinorbitMatrixElements2000}%
  \BibitemOpen
  \bibfield  {author} {\bibinfo {author} {\bibfnamefont {A.}~\bibnamefont
  {Berning}}, \bibinfo {author} {\bibfnamefont {M.}~\bibnamefont {Schweizer}},
  \bibinfo {author} {\bibfnamefont {H.-J.}\ \bibnamefont {Werner}}, \bibinfo
  {author} {\bibfnamefont {P.~J.}\ \bibnamefont {Knowles}},\ and\ \bibinfo
  {author} {\bibfnamefont {P.}~\bibnamefont {Palmieri}},\ }\href
  {https://doi.org/10.1080/00268970009483386} {\bibfield  {journal} {\bibinfo
  {journal} {Mol. Phys.}\ }\textbf {\bibinfo {volume} {98}},\ \bibinfo {pages}
  {1823} (\bibinfo {year} {2000})}\BibitemShut {NoStop}%
\bibitem [{\citenamefont {He{\ss}}\ \emph {et~al.}(1996)\citenamefont
  {He{\ss}}, \citenamefont {Marian}, \citenamefont {Wahlgren},\ and\
  \citenamefont {Gropen}}]{hessMeanfieldSpinorbitMethod1996}%
  \BibitemOpen
  \bibfield  {author} {\bibinfo {author} {\bibfnamefont {B.~A.}\ \bibnamefont
  {He{\ss}}}, \bibinfo {author} {\bibfnamefont {C.~M.}\ \bibnamefont {Marian}},
  \bibinfo {author} {\bibfnamefont {U.}~\bibnamefont {Wahlgren}},\ and\
  \bibinfo {author} {\bibfnamefont {O.}~\bibnamefont {Gropen}},\ }\href
  {https://doi.org/10.1016/0009-2614(96)00119-4} {\bibfield  {journal}
  {\bibinfo  {journal} {Chem. Phys. Lett.}\ }\textbf {\bibinfo {volume}
  {251}},\ \bibinfo {pages} {365} (\bibinfo {year} {1996})}\BibitemShut
  {NoStop}%
\bibitem [{\citenamefont {Atanasov}\ \emph {et~al.}(2015)\citenamefont
  {Atanasov}, \citenamefont {Aravena}, \citenamefont {Suturina}, \citenamefont
  {Bill}, \citenamefont {Maganas},\ and\ \citenamefont
  {Neese}}]{atanasovFirstPrinciplesApproach2015}%
  \BibitemOpen
  \bibfield  {author} {\bibinfo {author} {\bibfnamefont {M.}~\bibnamefont
  {Atanasov}}, \bibinfo {author} {\bibfnamefont {D.}~\bibnamefont {Aravena}},
  \bibinfo {author} {\bibfnamefont {E.}~\bibnamefont {Suturina}}, \bibinfo
  {author} {\bibfnamefont {E.}~\bibnamefont {Bill}}, \bibinfo {author}
  {\bibfnamefont {D.}~\bibnamefont {Maganas}},\ and\ \bibinfo {author}
  {\bibfnamefont {F.}~\bibnamefont {Neese}},\ }\href
  {https://doi.org/10.1016/j.ccr.2014.10.015} {\bibfield  {journal} {\bibinfo
  {journal} {Coord. Chem. Rev.}\ }\textbf {\bibinfo {volume} {289--290}},\
  \bibinfo {pages} {177} (\bibinfo {year} {2015})}\BibitemShut {NoStop}%
\bibitem [{\citenamefont {Fan}\ \emph {et~al.}(2023)\citenamefont {Fan},
  \citenamefont {Myers}, \citenamefont {Sukra},\ and\ \citenamefont
  {Gabrielse}}]{fanMeasurementElectronMagnetic2023}%
  \BibitemOpen
  \bibfield  {author} {\bibinfo {author} {\bibfnamefont {X.}~\bibnamefont
  {Fan}}, \bibinfo {author} {\bibfnamefont {T.~G.}\ \bibnamefont {Myers}},
  \bibinfo {author} {\bibfnamefont {B.~A.~D.}\ \bibnamefont {Sukra}},\ and\
  \bibinfo {author} {\bibfnamefont {G.}~\bibnamefont {Gabrielse}},\ }\href
  {https://doi.org/10.1103/PhysRevLett.130.071801} {\bibfield  {journal}
  {\bibinfo  {journal} {Phys. Rev. Lett.}\ }\textbf {\bibinfo {volume} {130}},\
  \bibinfo {pages} {071801} (\bibinfo {year} {2023})}\BibitemShut {NoStop}%
\bibitem [{\citenamefont
  {McWeeny}(1965)}]{mcweenyOriginSpinHamiltonianParameters1965}%
  \BibitemOpen
  \bibfield  {author} {\bibinfo {author} {\bibfnamefont {R.}~\bibnamefont
  {McWeeny}},\ }\href {https://doi.org/10.1063/1.1696183} {\bibfield  {journal}
  {\bibinfo  {journal} {J. Chem. Phys.}\ }\textbf {\bibinfo {volume} {42}},\
  \bibinfo {pages} {1717} (\bibinfo {year} {1965})}\BibitemShut {NoStop}%
\bibitem [{\citenamefont {Neese}\ and\ \citenamefont
  {Solomon}(1998)}]{neeseCalculationZeroFieldSplittings1998}%
  \BibitemOpen
  \bibfield  {author} {\bibinfo {author} {\bibfnamefont {F.}~\bibnamefont
  {Neese}}\ and\ \bibinfo {author} {\bibfnamefont {E.~I.}\ \bibnamefont
  {Solomon}},\ }\href {https://doi.org/10.1021/ic980948i} {\bibfield  {journal}
  {\bibinfo  {journal} {Inorg. Chem.}\ }\textbf {\bibinfo {volume} {37}},\
  \bibinfo {pages} {6568} (\bibinfo {year} {1998})}\BibitemShut {NoStop}%
\bibitem [{\citenamefont {{Cebreiro-Gallardo}}\ and\ \citenamefont
  {Casanova}(2025{\natexlab{b}})}]{cebreiro-gallardoEfficientStateinteractionApproach2025}%
  \BibitemOpen
  \bibfield  {author} {\bibinfo {author} {\bibfnamefont {A.}~\bibnamefont
  {{Cebreiro-Gallardo}}}\ and\ \bibinfo {author} {\bibfnamefont
  {D.}~\bibnamefont {Casanova}},\ }\href {https://doi.org/10.1039/D4CP04511D}
  {\bibfield  {journal} {\bibinfo  {journal} {Phys. Chem. Chem. Phys.}\
  }\textbf {\bibinfo {volume} {27}},\ \bibinfo {pages} {7093} (\bibinfo {year}
  {2025}{\natexlab{b}})}\BibitemShut {NoStop}%
\bibitem [{\citenamefont {Lan}, \citenamefont {Chalupsk{\'y}},\ and\
  \citenamefont {Yanai}(2015)}]{lanMolecularTensorsAnalytical2015}%
  \BibitemOpen
  \bibfield  {author} {\bibinfo {author} {\bibfnamefont {T.~N.}\ \bibnamefont
  {Lan}}, \bibinfo {author} {\bibfnamefont {J.}~\bibnamefont {Chalupsk{\'y}}},\
  and\ \bibinfo {author} {\bibfnamefont {T.}~\bibnamefont {Yanai}},\ }\href
  {https://doi.org/10.1080/00268976.2015.1012128} {\bibfield  {journal}
  {\bibinfo  {journal} {Mol. Phys.}\ }\textbf {\bibinfo {volume} {113}},\
  \bibinfo {pages} {1750} (\bibinfo {year} {2015})}\BibitemShut {NoStop}%
\bibitem [{\citenamefont {Weltner}(1983)}]{weltnerMagneticAtomsMolecules1983}%
  \BibitemOpen
  \bibfield  {author} {\bibinfo {author} {\bibfnamefont {W.~{\relax Jr}.}\
  \bibnamefont {Weltner}},\ }\href@noop {} {\emph {\bibinfo {title} {Magnetic
  Atoms and Molecules}}}\ (\bibinfo  {publisher} {Van Nostrand Reinhold},\
  \bibinfo {address} {New York},\ \bibinfo {year} {1983})\BibitemShut {NoStop}%
\bibitem [{\citenamefont {Choe}\ \emph {et~al.}(2001)\citenamefont {Choe},
  \citenamefont {Witek}, \citenamefont {Finley},\ and\ \citenamefont
  {Hirao}}]{choeIdentifyingRemovingIntruder2001}%
  \BibitemOpen
  \bibfield  {author} {\bibinfo {author} {\bibfnamefont {Y.-K.}\ \bibnamefont
  {Choe}}, \bibinfo {author} {\bibfnamefont {H.~A.}\ \bibnamefont {Witek}},
  \bibinfo {author} {\bibfnamefont {J.~P.}\ \bibnamefont {Finley}},\ and\
  \bibinfo {author} {\bibfnamefont {K.}~\bibnamefont {Hirao}},\ }\href
  {https://doi.org/10.1063/1.1345510} {\bibfield  {journal} {\bibinfo
  {journal} {J. Chem. Phys.}\ }\textbf {\bibinfo {volume} {114}},\ \bibinfo
  {pages} {3913} (\bibinfo {year} {2001})}\BibitemShut {NoStop}%
\bibitem [{\citenamefont {Forsberg}\ and\ \citenamefont
  {Malmqvist}(1997)}]{forsbergMulticonfigurationPerturbationTheory1997}%
  \BibitemOpen
  \bibfield  {author} {\bibinfo {author} {\bibfnamefont {N.}~\bibnamefont
  {Forsberg}}\ and\ \bibinfo {author} {\bibfnamefont {P.-{\AA}.}\ \bibnamefont
  {Malmqvist}},\ }\href {https://doi.org/10.1016/S0009-2614(97)00669-6}
  {\bibfield  {journal} {\bibinfo  {journal} {Chem. Phys. Lett.}\ }\textbf
  {\bibinfo {volume} {274}},\ \bibinfo {pages} {196} (\bibinfo {year}
  {1997})}\BibitemShut {NoStop}%
\bibitem [{\citenamefont {Hayashi}\ \emph {et~al.}(2024)\citenamefont
  {Hayashi}, \citenamefont {Saitow}, \citenamefont {Uemura},\ and\
  \citenamefont {Yanai}}]{hayashiQuasidegenerateExtensionLocal2024}%
  \BibitemOpen
  \bibfield  {author} {\bibinfo {author} {\bibfnamefont {M.}~\bibnamefont
  {Hayashi}}, \bibinfo {author} {\bibfnamefont {M.}~\bibnamefont {Saitow}},
  \bibinfo {author} {\bibfnamefont {K.}~\bibnamefont {Uemura}},\ and\ \bibinfo
  {author} {\bibfnamefont {T.}~\bibnamefont {Yanai}},\ }\href
  {https://doi.org/10.1063/5.0204419} {\bibfield  {journal} {\bibinfo
  {journal} {J. Chem. Phys.}\ }\textbf {\bibinfo {volume} {160}},\ \bibinfo
  {pages} {194105} (\bibinfo {year} {2024})}\BibitemShut {NoStop}%
\bibitem [{\citenamefont {Li}\ \emph {et~al.}(2025)\citenamefont {Li},
  \citenamefont {Luo}, \citenamefont {Wu},\ and\ \citenamefont
  {Lei}}]{liApplicationDyallHamiltonianbased2025}%
  \BibitemOpen
  \bibfield  {author} {\bibinfo {author} {\bibfnamefont {Y.}~\bibnamefont
  {Li}}, \bibinfo {author} {\bibfnamefont {S.}~\bibnamefont {Luo}}, \bibinfo
  {author} {\bibfnamefont {P.}~\bibnamefont {Wu}},\ and\ \bibinfo {author}
  {\bibfnamefont {Y.}~\bibnamefont {Lei}},\ }\href
  {https://doi.org/10.1016/j.cplett.2025.142096} {\bibfield  {journal}
  {\bibinfo  {journal} {Chem. Phys. Lett.}\ }\textbf {\bibinfo {volume}
  {871}},\ \bibinfo {pages} {142096} (\bibinfo {year} {2025})}\BibitemShut
  {NoStop}%
\bibitem [{\citenamefont {Chang}\ and\ \citenamefont
  {Witek}(2012)}]{changChoiceOptimalShift2012}%
  \BibitemOpen
  \bibfield  {author} {\bibinfo {author} {\bibfnamefont {S.-W.}\ \bibnamefont
  {Chang}}\ and\ \bibinfo {author} {\bibfnamefont {H.~A.}\ \bibnamefont
  {Witek}},\ }\href {https://doi.org/10.1021/ct2006924} {\bibfield  {journal}
  {\bibinfo  {journal} {J. Chem. Theory Comput.}\ }\textbf {\bibinfo {volume}
  {8}},\ \bibinfo {pages} {4053} (\bibinfo {year} {2012})}\BibitemShut
  {NoStop}%
\bibitem [{\citenamefont {Camacho}, \citenamefont {Witek},\ and\ \citenamefont
  {Yamamoto}(2009)}]{camachoIntruderStatesMultireference2009}%
  \BibitemOpen
  \bibfield  {author} {\bibinfo {author} {\bibfnamefont {C.}~\bibnamefont
  {Camacho}}, \bibinfo {author} {\bibfnamefont {H.~A.}\ \bibnamefont {Witek}},\
  and\ \bibinfo {author} {\bibfnamefont {S.}~\bibnamefont {Yamamoto}},\ }\href
  {https://doi.org/10.1002/jcc.21074} {\bibfield  {journal} {\bibinfo
  {journal} {J. Comput. Chem.}\ }\textbf {\bibinfo {volume} {30}},\ \bibinfo
  {pages} {468} (\bibinfo {year} {2009})}\BibitemShut {NoStop}%
\bibitem [{\citenamefont {Roos}\ and\ \citenamefont
  {Andersson}(1995)}]{roosMulticonfigurationalPerturbationTheory1995}%
  \BibitemOpen
  \bibfield  {author} {\bibinfo {author} {\bibfnamefont {B.~O.}\ \bibnamefont
  {Roos}}\ and\ \bibinfo {author} {\bibfnamefont {K.}~\bibnamefont
  {Andersson}},\ }\href {https://doi.org/10.1016/0009-2614(95)01010-7}
  {\bibfield  {journal} {\bibinfo  {journal} {Chem. Phys. Lett.}\ }\textbf
  {\bibinfo {volume} {245}},\ \bibinfo {pages} {215} (\bibinfo {year}
  {1995})}\BibitemShut {NoStop}%
\bibitem [{\citenamefont {Park}\ and\ \citenamefont
  {Shiozaki}(2017)}]{parkAnalyticalDerivativeCoupling2017}%
  \BibitemOpen
  \bibfield  {author} {\bibinfo {author} {\bibfnamefont {J.~W.}\ \bibnamefont
  {Park}}\ and\ \bibinfo {author} {\bibfnamefont {T.}~\bibnamefont
  {Shiozaki}},\ }\href {https://doi.org/10.1021/acs.jctc.7b00018} {\bibfield
  {journal} {\bibinfo  {journal} {J. Chem. Theory Comput.}\ }\textbf {\bibinfo
  {volume} {13}},\ \bibinfo {pages} {2561} (\bibinfo {year}
  {2017})}\BibitemShut {NoStop}%
\bibitem [{\citenamefont {Park}\ \emph {et~al.}(2019)\citenamefont {Park},
  \citenamefont {{Al-Saadon}}, \citenamefont {Strand},\ and\ \citenamefont
  {Shiozaki}}]{parkImaginaryShiftCASPT22019}%
  \BibitemOpen
  \bibfield  {author} {\bibinfo {author} {\bibfnamefont {J.~W.}\ \bibnamefont
  {Park}}, \bibinfo {author} {\bibfnamefont {R.}~\bibnamefont {{Al-Saadon}}},
  \bibinfo {author} {\bibfnamefont {N.~E.}\ \bibnamefont {Strand}},\ and\
  \bibinfo {author} {\bibfnamefont {T.}~\bibnamefont {Shiozaki}},\ }\href
  {https://doi.org/10.1021/acs.jctc.9b00368} {\bibfield  {journal} {\bibinfo
  {journal} {J. Chem. Theory Comput.}\ }\textbf {\bibinfo {volume} {15}},\
  \bibinfo {pages} {4088} (\bibinfo {year} {2019})}\BibitemShut {NoStop}%
\bibitem [{\citenamefont {Li}\ and\ \citenamefont
  {Evangelista}(2017)}]{liDrivenSimilarityRenormalization2017}%
  \BibitemOpen
  \bibfield  {author} {\bibinfo {author} {\bibfnamefont {C.}~\bibnamefont
  {Li}}\ and\ \bibinfo {author} {\bibfnamefont {F.~A.}\ \bibnamefont
  {Evangelista}},\ }\href {https://doi.org/10.1063/1.4979016} {\bibfield
  {journal} {\bibinfo  {journal} {J. Chem. Phys.}\ }\textbf {\bibinfo {volume}
  {146}},\ \bibinfo {pages} {124132} (\bibinfo {year} {2017})}\BibitemShut
  {NoStop}%
\bibitem [{\citenamefont {Radzig}\ and\ \citenamefont
  {Smirnov}(1985)}]{radzigReferenceDataAtoms1985}%
  \BibitemOpen
  \bibfield  {author} {\bibinfo {author} {\bibfnamefont {A.~A.}\ \bibnamefont
  {Radzig}}\ and\ \bibinfo {author} {\bibfnamefont {B.~M.}\ \bibnamefont
  {Smirnov}},\ }\href {https://doi.org/10.1007/978-3-642-82048-9} {\emph
  {\bibinfo {title} {Reference Data on Atoms, Molecules, and Ions}}},\ edited
  by\ \bibinfo {editor} {\bibfnamefont {V.~I.}\ \bibnamefont {Goldanskii}},
  \bibinfo {editor} {\bibfnamefont {R.}~\bibnamefont {Gomer}}, \bibinfo
  {editor} {\bibfnamefont {F.~P.}\ \bibnamefont {Sch{\"a}fer}},\ and\ \bibinfo
  {editor} {\bibfnamefont {J.~P.}\ \bibnamefont {Toennies}},\ Vol.~\bibinfo
  {volume} {31}\ (\bibinfo  {publisher} {Springer},\ \bibinfo {address}
  {Berlin, Heidelberg},\ \bibinfo {year} {1985})\BibitemShut {NoStop}%
\bibitem [{\citenamefont {Belanzoni}, \citenamefont {Van~Lenthe},\ and\
  \citenamefont {Baerends}(2001)}]{belanzoniEvaluationDensityFunctional2001}%
  \BibitemOpen
  \bibfield  {author} {\bibinfo {author} {\bibfnamefont {P.}~\bibnamefont
  {Belanzoni}}, \bibinfo {author} {\bibfnamefont {E.}~\bibnamefont
  {Van~Lenthe}},\ and\ \bibinfo {author} {\bibfnamefont {E.~J.}\ \bibnamefont
  {Baerends}},\ }\href {https://doi.org/10.1063/1.1345509} {\bibfield
  {journal} {\bibinfo  {journal} {J. Chem. Phys.}\ }\textbf {\bibinfo {volume}
  {114}},\ \bibinfo {pages} {4421} (\bibinfo {year} {2001})}\BibitemShut
  {NoStop}%
\bibitem [{\citenamefont {Patchkovskii}\ and\ \citenamefont
  {Ziegler}(2001)}]{patchkovskiiCalculationEPRGTensors2001}%
  \BibitemOpen
  \bibfield  {author} {\bibinfo {author} {\bibfnamefont {S.}~\bibnamefont
  {Patchkovskii}}\ and\ \bibinfo {author} {\bibfnamefont {T.}~\bibnamefont
  {Ziegler}},\ }\href {https://doi.org/10.1021/jp010457a} {\bibfield  {journal}
  {\bibinfo  {journal} {J. Phys. Chem. A}\ }\textbf {\bibinfo {volume} {105}},\
  \bibinfo {pages} {5490} (\bibinfo {year} {2001})}\BibitemShut {NoStop}%
\bibitem [{\citenamefont {Linstrom}\ and\ \citenamefont
  {Mallard}(2024)}]{NISTWebBook}%
  \BibitemOpen
  \bibfield  {author} {\bibinfo {author} {\bibfnamefont {P.~J.}\ \bibnamefont
  {Linstrom}}\ and\ \bibinfo {author} {\bibfnamefont {W.~G.}\ \bibnamefont
  {Mallard}},\ }\href {https://webbook.nist.gov/} {\enquote {\bibinfo {title}
  {{NIST Chemistry WebBook: NIST Standard Reference Database Number 69}},}\ }
  (\bibinfo {year} {2024}),\ \bibinfo {note} {accessed 2025}\BibitemShut
  {NoStop}%
\bibitem [{\citenamefont {Ram}\ and\ \citenamefont
  {Bernath}(2013)}]{RAM201318}%
  \BibitemOpen
  \bibfield  {author} {\bibinfo {author} {\bibfnamefont {R.}~\bibnamefont
  {Ram}}\ and\ \bibinfo {author} {\bibfnamefont {P.}~\bibnamefont {Bernath}},\
  }\href {https://doi.org/10.1016/j.jms.2012.12.004} {\bibfield  {journal}
  {\bibinfo  {journal} {J. Mol. Spectrosc.}\ }\textbf {\bibinfo {volume}
  {283}},\ \bibinfo {pages} {18} (\bibinfo {year} {2013})}\BibitemShut
  {NoStop}%
\bibitem [{\citenamefont {Balasubramanian}\ and\ \citenamefont
  {Liao}(1988)}]{balasubramanianElectronicStatesPotential1988}%
  \BibitemOpen
  \bibfield  {author} {\bibinfo {author} {\bibfnamefont {K.}~\bibnamefont
  {Balasubramanian}}\ and\ \bibinfo {author} {\bibfnamefont {D.~W.}\
  \bibnamefont {Liao}},\ }\href {https://doi.org/10.1021/j100333a018}
  {\bibfield  {journal} {\bibinfo  {journal} {J. Phys. Chem.}\ }\textbf
  {\bibinfo {volume} {92}},\ \bibinfo {pages} {6259} (\bibinfo {year}
  {1988})}\BibitemShut {NoStop}%
\bibitem [{\citenamefont {Balasubramanian}\ and\ \citenamefont
  {Dai}(1990)}]{balasubramanianPotentialEnergySurfaces1990}%
  \BibitemOpen
  \bibfield  {author} {\bibinfo {author} {\bibfnamefont {K.}~\bibnamefont
  {Balasubramanian}}\ and\ \bibinfo {author} {\bibfnamefont {D.}~\bibnamefont
  {Dai}},\ }\href {https://doi.org/10.1063/1.459447} {\bibfield  {journal}
  {\bibinfo  {journal} {J. Chem. Phys.}\ }\textbf {\bibinfo {volume} {93}},\
  \bibinfo {pages} {7243} (\bibinfo {year} {1990})}\BibitemShut {NoStop}%
\bibitem [{\citenamefont {Van~Zee}\ \emph {et~al.}(1992)\citenamefont
  {Van~Zee}, \citenamefont {Li}, \citenamefont {Hamrick},\ and\ \citenamefont
  {Weltner}}]{vanzeeElectronspinResonanceCo1992}%
  \BibitemOpen
  \bibfield  {author} {\bibinfo {author} {\bibfnamefont {R.~J.}\ \bibnamefont
  {Van~Zee}}, \bibinfo {author} {\bibfnamefont {S.}~\bibnamefont {Li}},
  \bibinfo {author} {\bibfnamefont {Y.~M.}\ \bibnamefont {Hamrick}},\ and\
  \bibinfo {author} {\bibfnamefont {W.}~\bibnamefont {Weltner}, \bibfnamefont
  {Jr.}},\ }\href {https://doi.org/10.1063/1.463433} {\bibfield  {journal}
  {\bibinfo  {journal} {J. Chem. Phys.}\ }\textbf {\bibinfo {volume} {97}},\
  \bibinfo {pages} {8123} (\bibinfo {year} {1992})}\BibitemShut {NoStop}%
\bibitem [{\citenamefont {Chow}, \citenamefont {Chang},\ and\ \citenamefont
  {Willett}(1973)}]{chowElectronSpinResonance1973}%
  \BibitemOpen
  \bibfield  {author} {\bibinfo {author} {\bibfnamefont {C.}~\bibnamefont
  {Chow}}, \bibinfo {author} {\bibfnamefont {K.}~\bibnamefont {Chang}},\ and\
  \bibinfo {author} {\bibfnamefont {R.~D.}\ \bibnamefont {Willett}},\ }\href
  {https://doi.org/10.1063/1.1680380} {\bibfield  {journal} {\bibinfo
  {journal} {J. Chem. Phys.}\ }\textbf {\bibinfo {volume} {59}},\ \bibinfo
  {pages} {2629} (\bibinfo {year} {1973})}\BibitemShut {NoStop}%
\bibitem [{\citenamefont {Scholl}\ and\ \citenamefont
  {Huettermann}(1992)}]{schollESRENDORCopperII1992}%
  \BibitemOpen
  \bibfield  {author} {\bibinfo {author} {\bibfnamefont {H.~J.}\ \bibnamefont
  {Scholl}}\ and\ \bibinfo {author} {\bibfnamefont {J.}~\bibnamefont
  {Huettermann}},\ }\href {https://doi.org/10.1021/j100203a023} {\bibfield
  {journal} {\bibinfo  {journal} {J. Phys. Chem.}\ }\textbf {\bibinfo {volume}
  {96}},\ \bibinfo {pages} {9684} (\bibinfo {year} {1992})}\BibitemShut
  {NoStop}%
\bibitem [{\citenamefont {De~Vore}\ and\ \citenamefont
  {Weltner}(1977)}]{devoreTitaniumDifluorideTitanium1977}%
  \BibitemOpen
  \bibfield  {author} {\bibinfo {author} {\bibfnamefont {T.~C.}\ \bibnamefont
  {De~Vore}}\ and\ \bibinfo {author} {\bibfnamefont {W.~{\relax Jr}.}\
  \bibnamefont {Weltner}},\ }\href {https://doi.org/10.1021/ja00456a028}
  {\bibfield  {journal} {\bibinfo  {journal} {J. Am. Chem. Soc.}\ }\textbf
  {\bibinfo {volume} {99}},\ \bibinfo {pages} {4700} (\bibinfo {year}
  {1977})}\BibitemShut {NoStop}%
\bibitem [{\citenamefont {de~Moura}\ and\ \citenamefont
  {Sokolov}(2025)}]{MouraSokolov2025Prism}%
  \BibitemOpen
  \bibfield  {author} {\bibinfo {author} {\bibfnamefont {C.~E.~V.}\
  \bibnamefont {de~Moura}}\ and\ \bibinfo {author} {\bibfnamefont {A.~Y.}\
  \bibnamefont {Sokolov}},\ }\href@noop {} {\enquote {\bibinfo {title} {{Prism}
  (github repository)},}\ }\bibinfo {howpublished}
  {https://github.com/sokolov-group/prism} (\bibinfo {year} {2025})\BibitemShut
  {NoStop}%
\bibitem [{\citenamefont {Sun}\ \emph {et~al.}(2020)\citenamefont {Sun},
  \citenamefont {Zhang}, \citenamefont {Banerjee}, \citenamefont {Bao},
  \citenamefont {Barbry}, \citenamefont {Blunt}, \citenamefont {Bogdanov},
  \citenamefont {Booth}, \citenamefont {Chen}, \citenamefont {Cui},
  \citenamefont {Eriksen}, \citenamefont {Gao}, \citenamefont {Guo},
  \citenamefont {Hermann}, \citenamefont {Hermes}, \citenamefont {Koh},
  \citenamefont {Koval}, \citenamefont {Lehtola}, \citenamefont {Li},
  \citenamefont {Liu}, \citenamefont {Mardirossian}, \citenamefont {McClain},
  \citenamefont {Motta}, \citenamefont {Mussard}, \citenamefont {Pham},
  \citenamefont {Pulkin}, \citenamefont {Purwanto}, \citenamefont {Robinson},
  \citenamefont {Ronca}, \citenamefont {Sayfutyarova}, \citenamefont
  {Scheurer}, \citenamefont {Schurkus}, \citenamefont {Smith}, \citenamefont
  {Sun}, \citenamefont {Sun}, \citenamefont {Upadhyay}, \citenamefont {Wagner},
  \citenamefont {Wang}, \citenamefont {White}, \citenamefont {Whitfield},
  \citenamefont {Williamson}, \citenamefont {Wouters}, \citenamefont {Yang},
  \citenamefont {Yu}, \citenamefont {Zhu}, \citenamefont {Berkelbach},
  \citenamefont {Sharma}, \citenamefont {Sokolov},\ and\ \citenamefont
  {Chan}}]{pyscf}%
  \BibitemOpen
  \bibfield  {author} {\bibinfo {author} {\bibfnamefont {Q.}~\bibnamefont
  {Sun}}, \bibinfo {author} {\bibfnamefont {X.}~\bibnamefont {Zhang}}, \bibinfo
  {author} {\bibfnamefont {S.}~\bibnamefont {Banerjee}}, \bibinfo {author}
  {\bibfnamefont {P.}~\bibnamefont {Bao}}, \bibinfo {author} {\bibfnamefont
  {M.}~\bibnamefont {Barbry}}, \bibinfo {author} {\bibfnamefont {N.~S.}\
  \bibnamefont {Blunt}}, \bibinfo {author} {\bibfnamefont {N.~A.}\ \bibnamefont
  {Bogdanov}}, \bibinfo {author} {\bibfnamefont {G.~H.}\ \bibnamefont {Booth}},
  \bibinfo {author} {\bibfnamefont {J.}~\bibnamefont {Chen}}, \bibinfo {author}
  {\bibfnamefont {Z.-H.}\ \bibnamefont {Cui}}, \bibinfo {author} {\bibfnamefont
  {J.~J.}\ \bibnamefont {Eriksen}}, \bibinfo {author} {\bibfnamefont
  {Y.}~\bibnamefont {Gao}}, \bibinfo {author} {\bibfnamefont {S.}~\bibnamefont
  {Guo}}, \bibinfo {author} {\bibfnamefont {J.}~\bibnamefont {Hermann}},
  \bibinfo {author} {\bibfnamefont {M.~R.}\ \bibnamefont {Hermes}}, \bibinfo
  {author} {\bibfnamefont {K.}~\bibnamefont {Koh}}, \bibinfo {author}
  {\bibfnamefont {P.}~\bibnamefont {Koval}}, \bibinfo {author} {\bibfnamefont
  {S.}~\bibnamefont {Lehtola}}, \bibinfo {author} {\bibfnamefont
  {Z.}~\bibnamefont {Li}}, \bibinfo {author} {\bibfnamefont {J.}~\bibnamefont
  {Liu}}, \bibinfo {author} {\bibfnamefont {N.}~\bibnamefont {Mardirossian}},
  \bibinfo {author} {\bibfnamefont {J.~D.}\ \bibnamefont {McClain}}, \bibinfo
  {author} {\bibfnamefont {M.}~\bibnamefont {Motta}}, \bibinfo {author}
  {\bibfnamefont {B.}~\bibnamefont {Mussard}}, \bibinfo {author} {\bibfnamefont
  {H.~Q.}\ \bibnamefont {Pham}}, \bibinfo {author} {\bibfnamefont
  {A.}~\bibnamefont {Pulkin}}, \bibinfo {author} {\bibfnamefont
  {W.}~\bibnamefont {Purwanto}}, \bibinfo {author} {\bibfnamefont {P.~J.}\
  \bibnamefont {Robinson}}, \bibinfo {author} {\bibfnamefont {E.}~\bibnamefont
  {Ronca}}, \bibinfo {author} {\bibfnamefont {E.~R.}\ \bibnamefont
  {Sayfutyarova}}, \bibinfo {author} {\bibfnamefont {M.}~\bibnamefont
  {Scheurer}}, \bibinfo {author} {\bibfnamefont {H.~F.}\ \bibnamefont
  {Schurkus}}, \bibinfo {author} {\bibfnamefont {J.~E.~T.}\ \bibnamefont
  {Smith}}, \bibinfo {author} {\bibfnamefont {C.}~\bibnamefont {Sun}}, \bibinfo
  {author} {\bibfnamefont {S.-N.}\ \bibnamefont {Sun}}, \bibinfo {author}
  {\bibfnamefont {S.}~\bibnamefont {Upadhyay}}, \bibinfo {author}
  {\bibfnamefont {L.~K.}\ \bibnamefont {Wagner}}, \bibinfo {author}
  {\bibfnamefont {X.}~\bibnamefont {Wang}}, \bibinfo {author} {\bibfnamefont
  {A.}~\bibnamefont {White}}, \bibinfo {author} {\bibfnamefont {J.~D.}\
  \bibnamefont {Whitfield}}, \bibinfo {author} {\bibfnamefont {M.~J.}\
  \bibnamefont {Williamson}}, \bibinfo {author} {\bibfnamefont
  {S.}~\bibnamefont {Wouters}}, \bibinfo {author} {\bibfnamefont
  {J.}~\bibnamefont {Yang}}, \bibinfo {author} {\bibfnamefont {J.~M.}\
  \bibnamefont {Yu}}, \bibinfo {author} {\bibfnamefont {T.}~\bibnamefont
  {Zhu}}, \bibinfo {author} {\bibfnamefont {T.~C.}\ \bibnamefont {Berkelbach}},
  \bibinfo {author} {\bibfnamefont {S.}~\bibnamefont {Sharma}}, \bibinfo
  {author} {\bibfnamefont {A.~Y.}\ \bibnamefont {Sokolov}},\ and\ \bibinfo
  {author} {\bibfnamefont {G.~K.-L.}\ \bibnamefont {Chan}},\ }\href
  {https://doi.org/10.1063/5.0006074} {\bibfield  {journal} {\bibinfo
  {journal} {J. Chem. Phys.}\ }\textbf {\bibinfo {volume} {153}},\ \bibinfo
  {pages} {024109} (\bibinfo {year} {2020})}\BibitemShut {NoStop}%
\bibitem [{\citenamefont {Wang}(2022)}]{Wang2022Socutils}%
  \BibitemOpen
  \bibfield  {author} {\bibinfo {author} {\bibfnamefont {X.}~\bibnamefont
  {Wang}},\ }\href@noop {} {\enquote {\bibinfo {title} {{socutils} (github
  repository)},}\ }\bibinfo {howpublished} {https://github.com/xubwa/socutils}
  (\bibinfo {year} {2022})\BibitemShut {NoStop}%
\bibitem [{\citenamefont
  {Perdew}(1986)}]{perdewDensityfunctionalApproximationCorrelation1986}%
  \BibitemOpen
  \bibfield  {author} {\bibinfo {author} {\bibfnamefont {J.~P.}\ \bibnamefont
  {Perdew}},\ }\href {https://doi.org/10.1103/PhysRevB.33.8822} {\bibfield
  {journal} {\bibinfo  {journal} {Phys. Rev. B}\ }\textbf {\bibinfo {volume}
  {33}},\ \bibinfo {pages} {8822} (\bibinfo {year} {1986})}\BibitemShut
  {NoStop}%
\bibitem [{\citenamefont
  {Becke}(1988)}]{beckeDensityfunctionalExchangeenergyApproximation1988}%
  \BibitemOpen
  \bibfield  {author} {\bibinfo {author} {\bibfnamefont {A.~D.}\ \bibnamefont
  {Becke}},\ }\href {https://doi.org/10.1103/PhysRevA.38.3098} {\bibfield
  {journal} {\bibinfo  {journal} {Phys. Rev. A}\ }\textbf {\bibinfo {volume}
  {38}},\ \bibinfo {pages} {3098} (\bibinfo {year} {1988})}\BibitemShut
  {NoStop}%
\bibitem [{\citenamefont {Roos}\ \emph {et~al.}(2004)\citenamefont {Roos},
  \citenamefont {Lindh}, \citenamefont {Malmqvist}, \citenamefont {Veryazov},\
  and\ \citenamefont {Widmark}}]{roosMainGroupAtoms2004}%
  \BibitemOpen
  \bibfield  {author} {\bibinfo {author} {\bibfnamefont {B.~O.}\ \bibnamefont
  {Roos}}, \bibinfo {author} {\bibfnamefont {R.}~\bibnamefont {Lindh}},
  \bibinfo {author} {\bibfnamefont {P.-{\AA}.}\ \bibnamefont {Malmqvist}},
  \bibinfo {author} {\bibfnamefont {V.}~\bibnamefont {Veryazov}},\ and\
  \bibinfo {author} {\bibfnamefont {P.-O.}\ \bibnamefont {Widmark}},\ }\href
  {https://doi.org/10.1021/jp031064+} {\bibfield  {journal} {\bibinfo
  {journal} {J. Phys. Chem. A}\ }\textbf {\bibinfo {volume} {108}},\ \bibinfo
  {pages} {2851} (\bibinfo {year} {2004})}\BibitemShut {NoStop}%
\bibitem [{\citenamefont {Weigend}\ and\ \citenamefont
  {Ahlrichs}(2005)}]{weigendBalancedBasisSets2005}%
  \BibitemOpen
  \bibfield  {author} {\bibinfo {author} {\bibfnamefont {F.}~\bibnamefont
  {Weigend}}\ and\ \bibinfo {author} {\bibfnamefont {R.}~\bibnamefont
  {Ahlrichs}},\ }\href {https://doi.org/10.1039/B508541A} {\bibfield  {journal}
  {\bibinfo  {journal} {Phys. Chem. Chem. Phys.}\ }\textbf {\bibinfo {volume}
  {7}},\ \bibinfo {pages} {3297} (\bibinfo {year} {2005})}\BibitemShut
  {NoStop}%
\bibitem [{\citenamefont {Sharma}\ and\ \citenamefont
  {DiVincenzo}(2024)}]{sharmaGfactorSymmetryTopology2024}%
  \BibitemOpen
  \bibfield  {author} {\bibinfo {author} {\bibfnamefont {M.}~\bibnamefont
  {Sharma}}\ and\ \bibinfo {author} {\bibfnamefont {D.~P.}\ \bibnamefont
  {DiVincenzo}},\ }\href {https://doi.org/10.1073/pnas.2404298121} {\bibfield
  {journal} {\bibinfo  {journal} {Proc. Natl. Acad. Sci.}\ }\textbf {\bibinfo
  {volume} {121}},\ \bibinfo {pages} {e2404298121} (\bibinfo {year}
  {2024})}\BibitemShut {NoStop}%
\bibitem [{\citenamefont {Luzanov}, \citenamefont {Babich},\ and\ \citenamefont
  {Ivanov}(1994)}]{luzanovGaugeinvariantCalculationsMagnetic1994}%
  \BibitemOpen
  \bibfield  {author} {\bibinfo {author} {\bibfnamefont {A.~V.}\ \bibnamefont
  {Luzanov}}, \bibinfo {author} {\bibfnamefont {E.~N.}\ \bibnamefont
  {Babich}},\ and\ \bibinfo {author} {\bibfnamefont {V.~V.}\ \bibnamefont
  {Ivanov}},\ }\href {https://doi.org/10.1016/S0166-1280(09)80059-6} {\bibfield
   {journal} {\bibinfo  {journal} {J. Mol. Struct. THEOCHEM}\ }\textbf
  {\bibinfo {volume} {311}},\ \bibinfo {pages} {211} (\bibinfo {year}
  {1994})}\BibitemShut {NoStop}%
\bibitem [{osc(1987)}]{osc1987}%
  \BibitemOpen
  \href {http://osc.edu/ark:/19495/f5s1ph73} {\enquote {\bibinfo {title} {Ohio
  supercomputer center},}\ } (\bibinfo {year} {1987}),\ \bibinfo {note}
  {accessed Oct 2, 2025}\BibitemShut {NoStop}%
\end{thebibliography}

%

\end{document}